\documentclass[a4paper,12pt]{article}

\RequirePackage[T1]{fontenc}

\RequirePackage{graphicx}
\RequirePackage{mathptmx}      
\RequirePackage{flushend}
\RequirePackage[numbers,sort&compress]{natbib}
\RequirePackage[colorlinks,citecolor=blue,urlcolor=blue,linkcolor=blue]{hyperref}
\usepackage{amssymb}
\usepackage{graphicx}
\usepackage{float}
\usepackage{stfloats}
\usepackage{multicol}
\usepackage{afterpage}

\usepackage{dcolumn}
\usepackage{bm}
\usepackage{amssymb}
\usepackage{amsmath}
\usepackage{hyperref}
\usepackage{float}
\usepackage{caption}

\usepackage{natbib}
\usepackage{color,soul}
\usepackage{multirow}
\usepackage{setspace}
\usepackage[shortlabels]{enumitem}
\usepackage{mathtools, nccmath}

\usepackage[T1]{fontenc}
\usepackage[english]{babel}
\usepackage[utf8]{inputenc}
\usepackage{authblk}

\begin{document} 

\title{Chiral QCD phase in equilibrium with Hadron Gas \\
and the location of the critical point}
\author{N.~G.~Antoniou\thanks{Dedicated to the memory of N.~G.~Antoniou (1939-2020)}}
\author{F.~K.~Diakonos\thanks{email: fdiakono@phys.uoa.gr}}
\author{A.~S.~Kapoyannis\thanks{email: akapog@phys.uoa.gr}}
\affil{Nuclear and Particle Physics Section, Faculty of Physics, \\
National and Kapodistrian University of Athens, GR-15784 Greece}
\date{}

\maketitle

\begin{abstract}
We develop a description of the equation of state of QCD matter with restored chiral symmetry, 
which is in thermal and
chemical equilibrium with the hadronic phase. The hadron gas is described with thermodynamically consistent volume corrections. The chiral
phase is composed of a set of few quark condensates, each of which corresponds to a family of 
hadrons with specific quark
content. On the boundary between the two phases we apply the requirement 
of conservation of particle numbers per family. We use lattice calculations for temperatures 
below the transition curve to determine hadronic volumes.
We find that the pion system plays decisive role in the shift of the transition from higher 
order (crossover) to first order. 
For four volume models and five hadron sets we calculate the location of the critical point as function of critical 
temperature $T_c$ at vanishing baryon density. Among other acceptable solutions, if we additionally impose the equality between the densities of quarks contained in mesons 
and baryons, we find a critical point residing in the interval of
baryon chemical potential $\mu_B \simeq$ 234-279 MeV and of
temperature $T \simeq$ 148-156 MeV. We propose as signal of approach to the critical point
the determination of the chiral pion mass through
experimentally measured pion multiplicities in heavy-ion collisions.
\end{abstract}


\section{Introduction} \label{sec:intro}
The QCD phase diagram is a subject of intense research during the last years. This phase diagram 
describes the transition to hadronic matter at lower temperatures, a phase with equation of 
state (EoS) well described by different versions of the Hadron Gas (HG) model \cite{HG}. However, 
knowledge of the EoS of the phase at higher temperatures is accessible only through lattice 
calculations that are limited mainly to the zero baryon density regime \cite{lat2+1}. 
The ideal 
quark-gluon plasma can, only, be reached at extremely high temperatures. The region which 
resides at temperatures higher than those determining the boundaries of the hadronic phase, is 
proposed to be associated with several exotic phases, like chiral-density waves, crystalline 
colour 
superconductivity,  gluonic phase and quarkyonic matter \cite{QCDphd}. 
The knowledge about this region is still developing and our work aims at this direction. We will 
focus on describing a model for the EoS at the high temperature region just above the transition 
to HG curve. The derivation of our model is based on a few useful, to our opinion, observations 
which are listed in the following:

\begin{enumerate}[(a)]
\item The chiral condensate, which serves as the order parameter for chiral restoration, 
changes sharply 
with temperature. In contrast, deconfinement, related to the renormalized Polyakov loop, occurs over 
a broader temperature range \cite{ch-d}. 
This suggests chiral restoration may complete before quarks fully 
``dissolve'' into independent entities. Our argument is based on the width of the deconfinement
transition, not its mean temperature. Due to finite size, there are no strict transition points but 
transition regions for deconfinement and chiral transitions. These regions may lay around the same  
temperature \cite{Polyakov}, but they can differ in extent. 
The transition points, defined for the infinite systems, should lie within these transition regions
and can coincide.
However, we assume the deconfinement transition is broader than the chiral
transition, potentially allowing quarks to organize into hadron-like particles.

\item In \cite{particleQCD} an equivalent description of the lattice results at vanishing 
baryonic density has been carried 
out, using one particle states with mass and number of states calculated from the fit to these 
results.
It was found that the quark-gluon interacting system, when crossing the critical temperature 
$T_c$, exhibits an abrupt decrease in the number of states and, also, in the respective mass. 
Moreover, the minimum value of the degeneracy factor, reached just above $T_c$, cannot be 
attributed to deconfined quark degrees of freedom, thus, leading to the conclusion that {\em 
these states should be still quark condensates}. Based on this observation, as well as the 
previous one (a), we will adopt in the present work the viewpoint that the quarks remain 
confined in hadronic type condensates even at temperatures above the QCD transition line.

\item In \cite{paritydoubling1} a model has been developed which simulates the parity doubling 
of the nucleon. In this model chiral
symmetry and an explicit mass term for the nucleon can coexist without contradiction. The 
nucleon has a partner of opposite parity. These particles have different masses in the broken 
symmetry phase (HG). However in the restored chiral phase these masses become degenerate. In 
\cite{paritydoubling2}, with the use of lattice simulations, a similar behaviour is found for 
opposite parity pairs of $N$, $\Delta$ and $\Omega$ baryons.

\item One of the proposed exotic phases, situated at high density and relatively low 
temperatures is the quarkyonic phase, which is valid for a system with a large number of colours 
$N_c$ \cite{quarkyonic}. In the quarkyonic matter there are quarks forming a Fermi sea, but, 
also, baryons which occupy a Fermi surface. These baryonic particles are connected with 
phenomena like superconductivity and superfluidity. 
Likewise, here we assume that hadronic states can exist at densities 
and temperatures higher than those characterizing the hadronic phase.

\item An important issue when crossing the transition into the hadron gas (HG) phase is the 
conservation of quantum numbers (baryon number, strangeness, charge, etc.) carried by quarks. This 
conservation is an additional constraint, along with the continuity of chemical potentials and 
pressure \cite{Gibbs}, assuming 
thermal and chemical equilibrium between the phases. However, these conditions alone cannot 
determine how quarks above the transition group into hadrons below it. For instance, conserving net
baryon density does not guarantee continuity in the proton and antiproton multiplicities. This can 
happen with the creation or annihilation of quark-antiquark pairs across the transition, altering
quark and antiquark numbers without changing net baryon density—though such a scenario is excluded 
in a smooth crossover.

\end{enumerate}

Having presented the key observations which inspire our approach, we now proceed to suggest our 
working hypothesis in this paper: 

{\it A result of the partial chiral restoration is that the resonances with different masses and 
the same quark content, occurring below the critical temperature, become quasi-particle states 
(condensates), which are degenerate in mass, above the critical temperature.}
 
The emerging picture we adopt is that the quark masses are reduced with the increase of 
temperature, while they are still 
interacting to form ``hadronic-like'' particles. This leads to an equivalent reformulation of 
the above hypothesis: {\it All the resonances with the same quark content at lower temperatures 
originate from the same quasi-particle state at higher temperatures}. As a consequence, since 
chiral symmetry breaking causes increase in the mass, we cannot allow the mass of a ``hadronic-like'' 
quasi-particle in the chiral phase to be greater than the mass of the lightest hadron 
in the broken symmetry phase. With this hypothesis we are able to solve the issue raised in 
remark (e) in a seamless way.

Another important point we are confronted with in this work is the fact that the exact position 
of the critical point depends crucially on the size of the hadronic volumes which encode the 
repulsive part of the strong interaction. To handle this uncertainty we estimate the hadronic 
volumes by fitting the Hadron Gas model with volume corrections for temperatures lower than the critical temperature to the lattice pressure with 
(2+1) flavours for zero \cite{lat2+1} or finite \cite{lat_muB1} baryon chemical potential. These, 
lattice QCD-based hadron volume determinations, together with the hypothesis of mass degeneration presented 
in the previous paragraph, constitute the backbone of the model we employ here to explore the 
QCD phase diagram region close to the boundary of the HG phase.

Therefore, in this work we will construct a model for the transition between
chiral and HG phase and seek solutions for the position of the critical point.
The free parameters we will be concerned with, will be:
i) The value $T_c$ at which the freeze-out curve crosses the temperature axis.
 We will equate this
with the lattice QCD transition temperature at zero baryon chemical potential.
ii) The hadron eigenvolumes that represent the repulsive hadronic interactions in the
HG phase. We will use lattice QCD data to have an estimation of these eigenvolumes
in four volume models.
The model choices controlling its construction are: i) Hadron eigenvolumes that
remain constant or depend on temperature, baryon chemical potential or both.
ii) The hadronic states that compose the spectrum of the HG phase. 
We will use five data sets. The first contains only
the verified states and the rest additional ones which are partially verified or
predicted from other studies or a state with different mass.

The structure of the remaining part of our paper is as follows. In Section \ref{sec:constr} we 
present in more detail the main ideas of our model. In Section \ref{sec:mech} we use them to 
propose a procedure for the determination of the critical point. In Section \ref{sec:constvary} 
we investigate the effect on HG of volumes depending on
temperature or chemical potentials and present results
for hadron volumes calculated from
fits to the lattice pressure at zero baryon density. In Section 
\ref{sec:loccp} we use the results of Sections \ref{sec:constr}, \ref{sec:mech} and 
\ref{sec:constvary} to propose four different models for the description of the volume 
corrections leading to the determination of the critical point location in terms of $T_c$ 
(critical temperature at zero baryonic density from lattice QCD). The first 
model assumes a common radius for all hadrons which is 
independent from temperature and chemical potential.
The second model assumes a common hadronic radius for all species which 
is varying with temperature. The third model assumes different hadronic radius for mesons which vary with temperature and for
baryons which vary with baryon chemical potential. The fourth model utilises again a common
hadron radius which depends simultaneously on both temperature and baryon chemical potential. 
In Section \ref{sec:crit} we investigate an additional condition for the determination of the 
critical point which ensures equality of the density of the quarks which are enclosed in 
mesons and baryons of the hadron gas. 
In Section \ref{sec:FD-ET} we discuss further development of our model and propose 
experimental tests for its validity and the approach to the critical point.
Finally, in Section \ref{sec:conclu} we summarise our findings 
and record our concluding remarks.

\section{Modelling the hadron-quark condensate transition} \label{sec:constr}

The HG phase, in its simplest form, can be described as point particles in the Boltzmann 
approximation 
with total particle density:
\begin{equation} \label{eq:HGptBo}
n^{pt}_{HG,Bo} =\frac{T}{2\pi^2} \sum_i \lambda_i \sum_j g_{ij} m_{ij}^2 K_2 \left[\frac{m_{ij}}
{T} \right]\;\;,
\end{equation}
where $\lambda_i$ is the fugacity corresponding to a specific group of hadrons that have the 
same quark content and which will be called as ``family''. Also, 
$\lambda_i=\lambda^k_B\lambda^l_S\lambda^n_Q$ if the particles of the family carry baryon number $B$ equal to $k$, strangeness $S$ equal to $l$ and electric charge $Q$ equal to $n$. The index 
$j$ runs over all resonances of the $i$ family. Since $i$ runs over particles and 
antiparticles, $g_{ij}$ is the degeneracy factor due only to the spin of the particle and 
$m_{ij}$ is the respective mass. Equation (\ref{eq:HGptBo}) easily shows that the HG partition 
function is a sum of terms, each of which corresponds to a specific fugacity. 

Next we turn to the description of the QCD phase at temperatures just above the transition 
curve. In this phase the constituent
quark masses are reduced due to the partial chiral restoration and in effect the mass of the 
corresponding quasi-particle is also reduced. Therefore, we shall call this phase in the 
following as ``chiral'' phase. We will, also, use the ``tilde'' ($\Large \tilde{\;\;}$) symbol 
to denote all quantities associated with this phase. We consider it to be a sum of quark 
condensates which represent quasi-particles carrying fugacities $\lambda_i$ and degeneracy 
factors $\tilde{g}_i$, where the index ``$i$'' runs over the families. Then the total particle 
density in the Boltzmann approximation for the chiral phase reads:
\begin{equation} \label{eq:chiBo}
\tilde{n}_{Bo} =\frac{T}{2\pi^2} \sum_i \lambda_i \tilde{g}_i \tilde{m}_i^2 K_2 
\left[\frac{\tilde{m}_i}{T} \right]\;\;.
\end{equation}
If the condensates in the chiral phase have the same quark content as the particles of the HG 
phase, then the particle number conservation at an arbitrary point $(T,\lambda_i)$ of the 
transition curve is converted to the conservation of the number of particles per family. For $N$ 
existing families in the HG, this accounts to $N$ equations, each denoted by the index $i$, of 
the form:
\begin{equation} \label{eq:HGchiptBo}
V_{HG}\sum_j g_{ij} m_{ij}^2 K_2 \left[\frac{m_{ij}}{T} \right]=
\tilde{V} \tilde{g}_{i} \tilde{m}_{i}(T)^2 K_2 
\left[\frac{\tilde{m}_{i}(T)}{T} \right]\;\;,
\end{equation}
where $V_{HG}$ $(\tilde{V})$ is the system volume in the HG (chiral) phase. Due to continuity 
criteria, in the crossover $V_{HG}=\tilde{V}$, while in the 1st order region $V_{HG} \neq 
\tilde{V}$. In order to fulfil eqs.~(\ref{eq:HGchiptBo}), we allow the masses in the chiral 
phase to be functions of temperature, $\tilde{m}_{i}(T)$. In principle, the $N$ equations in 
(\ref{eq:HGchiptBo}), since they do not depend on the fugacities, can be fulfilled at any point of a
transition curve which may depend on fugacities.

Since chiral restoration is associated with the reduction of the quark masses, we expect the 
quasi-particle, which contains the relevant quarks, to reduce its mass. So we can associate the 
chiral state of each family with the ground mass state of the HG phase. This HG state will attribute 
the degeneracy factor $g_{i1}=\tilde{g}_i$ to the respective chiral phase state. 
However, one can easily see that eqs.~(\ref{eq:HGchiptBo}) are impossible to be fulfilled if 
$\tilde{m}_{i}(T)=m_{i1}$ since the sum of several positive terms cannot be equal to the first term only. 
To overcome this inconsistency one should require:
\begin{equation} \label{eq:mchi}
0< \tilde{m}_{i}(T) \leq m_{i1}.
\end{equation}
Then, eqs.~(\ref{eq:HGchiptBo}) could in principle be fulfilled, since the functions 
$K_2 \left[\frac{\tilde{m}_{i}(T)}{T} \right]$ 
are sensitive enough increasing exponentially with decreasing 
$\tilde{m}_{i}(T)$. The constraint (\ref{eq:mchi}) has a 
straightforward physical interpretation: the quasi-particle existing in the chiral restored 
phase will produce in the chiral broken (HG) phase, among other resonances, the ground state 
particle with the minimum mass among all resonances, $m_{i1}$. Since the breaking of chiral 
symmetry increases mass, the chiral mass cannot be greater than $m_{i1}$. So, we will set as the 
upper limit for the masses $\tilde{m}_{i}(T)$ the respective mass $m_{i1}$ of the lightest 
hadron in the $i$ family.

It can be checked that for some families eqs.~(\ref{eq:HGchiptBo}) cannot be fulfilled for any 
value of mass $\tilde{m}_{i}(T)$ in the range of relation (\ref{eq:mchi}). This fact forces us 
to turn to a more realistic description of the HG phase. We have to take into account the hadron 
volume corrections which will effectively reduce all the particle number densities. For 
consistency reasons we will, also, use the correct 
quantum statistics (Bose-Einstein or Fermi-Dirac). The volume corrections in the HG have been 
accounted for in a thermodynamically consistent way which avoids negative contributions to the 
system volume \cite{volume_cor}. 
The partial pressure of a point particle $j$ belonging to the $i$ family using the correct 
quantum statistics is given by:
\begin{equation} \label{eq:HGppptBF}
P^{pt}_{ij}(T, \mu_{i})=
\frac{g_{ij}}{2\pi^2} \int_0^{\infty} dk \frac{k^4}{(k^2+m_{ij}^2)^{1/2}} 
f(k;m_{ij},\mu_{i},\alpha)\;,
\end{equation}
\vspace{-0.5cm}
\begin{equation} \label{eq:fBF}
f(k;m_{ij},\mu_{i},\alpha) \equiv
\left[\exp\left( \frac{{(k^2+m_{ij}^2)^{1/2}} -\mu_{i}}{T}\right) +\alpha  \right]^{-1} \;\;,
\end{equation}
where $\alpha=-1$ ($+1$) for bosons (fermions). If each particle carries volume $v_{ij}$, then 
the total HG pressure can be calculated from the partial pressures of the point particles. 
The relevant 
chemical potentials $\mu_{i}$ have to be replaced by the chemical potentials 
$\mathord{\buildrel{\lower3pt\hbox{$\scriptscriptstyle\frown$}} 
 \over \mu}_i $ \cite{volume_cor}:
\begin{equation} \label{eq:HGtpvBF}
P_{HG}(T, \ldots,\mu_{i}, \ldots)=
P_{HG}^{pt}(T, \ldots,\mathord{\buildrel{\lower3pt\hbox{$\scriptscriptstyle\frown$}} 
 \over \mu}_i, \ldots)=
\sum_i \sum_j P^{pt}_{ij}(T, \mathord{\buildrel{\lower3pt\hbox{$\scriptscriptstyle\frown$}} 
 \over \mu}_i) \;\;,
\end{equation}

\begin{equation} \label{eq:tildemu}
\mathord{\buildrel{\lower3pt\hbox{$\scriptscriptstyle\frown$}} 
 \over \mu}_i=\mu_{i}-v_{ij} P_{HG}(T, \ldots,\mu_{i}, \ldots) \;\;.
\end{equation}
Eq.~(\ref{eq:HGtpvBF}) has to be solved numerically for the system pressure $p_{HG}$. The 
density for point particles in the Bose/Fermi statistics reads:
\begin{equation} \label{eq:HGnptBF}
n^{pt}_{ij}(T, \mu_{i})=
\frac{g_{ij}}{2\pi^2} \int_0^{\infty} dk k^2 f(k;m_{ij},\mu_{i},\alpha)\;.
\end{equation}
Then we can proceed with the calculation of the particle densities with volume corrections:
\begin{equation} \label{eq:HGnvBF}
n_{HG,ij}(T, \ldots,\mu_{i}, \ldots)=
\frac{n^{pt}_{ij}(T, \mathord{\buildrel{\lower3pt\hbox{$\scriptscriptstyle\frown$}} 
 \over \mu}_{i})}
{1+\sum_r \sum_s v_{rs} n^{pt}_{rs}(T, \mathord{\buildrel{\lower3pt\hbox{$\scriptscriptstyle\frown$}} 
 \over \mu}_{r})}
 \;\;,
\end{equation}
where the volume of the $ij$ particle is connected to the corresponding radius \cite{volume_cor} 
through:
\vspace{-0.1cm}
\begin{equation} \label{eq:v-r}
v_{ij}=4(4\pi/3)r_{ij}^3 \;.
\end{equation}
With the use of the volume corrections and the Bose/Fermi statistics, eqs.~(\ref{eq:HGchiptBo}), 
for the equality
of particle densities between the HG and the chiral phase, are replaced by:
\begin{equation} \label{eq:HGchnBF}
V_{HG}\sum_j n_{HG,ij}(T, \mu_{i})=\tilde{V}\tilde{n}_{i}(T, \mu_{i}) \Rightarrow
V_{HG}\frac{n^{pt}_{ij}(T, \mathord{\buildrel{\lower3pt\hbox{$\scriptscriptstyle\frown$}} 
 \over \mu}_{i})}
{1+\sum_r \sum_s v_{rs} n^{pt}_{rs}(T, \mathord{\buildrel{\lower3pt\hbox{$\scriptscriptstyle\frown$}} 
 \over \mu}_{r})}=
\tilde{V}\frac{\tilde{g}_{i}}{2\pi^2} \int_0^{\infty} dk k^2 
f(k;\tilde{m}_{i}(T),\mu_{i},\alpha)
\;\;.
\end{equation}

The fugacities in the densities do not factorize when the Bose/Fermi statistics are used. 
As a consequence, the solution for the chiral state mass, 
$\tilde{m}_{i}(T)$, in
eq.~(\ref{eq:HGchnBF}) is different between particles and 
antiparticles. This dependence is not 
surprising since we are considering matter at positive baryon
densities, and we are allowing the masses of baryons and antibaryons to change. Also, while in 
the Boltzmann approximation particles and antiparticles can be grouped together in the same 
family, in the Bose/Fermi statistics the particles have to form
separate family from the respective antiparticles. However, as it will become evident in the 
next section, the mass difference between particles and antiparticles is low. Also, we do not allow dependence of the
chiral masses on the chemical potentials. In this way the
 densities of the particles at the chiral phase do not depend
on derivatives of the masses with respect to chemical potentials
and so we are able to write down eq.~(\ref{eq:HGchnBF}).
On the other hand we have to keep in mind that eq.~(\ref{eq:HGchnBF}) holds on a set of
points which form a curve in the space of the parameters 
$(T,\{\mu\})$, after the imposition of additional constraints and which is identified as the phase 
transition curve. Chiral masses depend only on 
temperature, because it parametrises smoothly this curve.

In the treatment of two QCD phases in equilibrium it is needed to apply the condition of the 
conservation of quantum numbers carried by the quarks \cite{Gibbs}. In \cite{Gibbs} in order to conserve quantum numbers, like baryon number, strangeness, etc. on the transition curve it was necessary to introduce an additional thermodynamic variable for each quantum number, namely the partial equilibrium fugacity $\gamma$. In the absence of such fugacities it is not
possible to conserve the quantum numbers on a transition curve between two phases, one of which is composed of hadrons and the
other of quarks. Here we give a different approach to the problem and, additionally, we impose a more 
strict condition of the equality of the particle numbers per family. The solution is given by our assumption that quarks have 
already been grouped in condensates at the chiral phase near the transition curve, which, after 
chiral breaking, evolve to the hadronic spectrum of the specific family.
The conservation of these particle numbers automatically insures the conservation of quantum 
numbers, e.g. for the baryon number $B$:
\begin{equation} \label{eq:eqqn}
N_{B,HG}=\sum_i k_i \sum_j V_{HG} n_{HG,ij}(T, \mu_{i})=
\sum_i k_i \tilde{V} \tilde{n}_i(T, \mu_{i})=\tilde{N}_B \;\;,
\end{equation}
where $k_i$ equals to the baryon number carried in the particles of the $i$ family, common to 
both the HG and the chiral phase and $n$ refers to the
particle number density.

To complete the set of requirements for equilibrium between the two phases we have to 
demand for the equality of pressures. In the chiral phase, apart from the quark condensates 
which contribute partial pressure $\tilde{P}_q$, also the
gluons show considerable pressure, $\tilde{P}_g$, for $T>T_c$ \cite{SU(3),SU(3)_}. Thus:
\begin{equation} \label{eq:pg}
P_{HG}=\tilde{P}_q + \tilde{P}_g - B_v \;\;,
\end{equation}
\noindent
where $B_v$ accounts for the possible influence of the vacuum.

\begin{table}
\vspace{-0.2cm}
\begin{center}
\renewcommand{\arraystretch}{1.3}
\begin{tabular}{|c|c|c|c|c|c|c|c|c|} \hline
\centering
    &        &               &                 & Ground &          &          &             &            \\ 
$i$ & Sym-   & Name          & Quark Content   & Mass   & $m_{i1}$ & $g_{i1}$ & $\lambda_i$ & $\alpha$    \\
    &  bol   &               &                 & State  & (MeV)    &          &             &             \\ \hline \hline
1  & $\pi$  & L.U.M., $I=1$ & $u\bar{d},(u\bar{u}-d\bar{d})/\sqrt{2},d\bar{u}$ & $\pi^{+,0,-}$ & 138,03919 & 3 & 1 & -1 \\
   &        &               & (flavour-                           &               &           &   &   &    \\ 
   &        &               & antisymmetric)                           &               &           &   &   &    \\ \hline
2a & $\eta$ & L.U.M., $I=0$ & $(u\bar{u}+d\bar{d})/\sqrt{2}$  & $\eta$  & 547,862   & 1 & 1 & -1 \\
   &        &               & (flavour-symmetric,             &         &           &   &   &    \\ 
   &        &               & no $s$-quark)                   &         &           &   &   &    \\ \hline
2b & $\eta'$& L.U.M., $I=0$ & $c_1(u\bar{u}+d\bar{d})+c_2 s\bar{s}$& $\eta'(958)$  & 957,78 & 1 & 1 & -1 \\
   &        &               & (flavour-symmetric,                  &               &        &   &   &    \\
   &        &               & with $s$-quark)                      &               &        &   &   &    \\ \hline
3  & $N$       & $N$ Baryons     & $uud,udd$                    & $p,n$   & 938,918754  & 4 & $\lambda_B$ & +1 \\
   &           & ($I=1/2$)       &                              &         &             &   &             &    \\ \hline
-3 & $\bar{N}$ & $N$ Antibaryons & $\bar{u}\bar{u}\bar{d},\bar{u}\bar{d}\bar{d}$& $\bar{p},\bar{n}$& 938,918754  & 4 & 
$\lambda_B^{-1}$ & +1 \\
   &           & ($I=1/2$)       &                              &         &             &   &             &    \\ \hline
4  & $\Delta$       & $\Delta$ Baryons     & $uuu,uud,udd,ddd$                    & $\Delta^{++,+,0,-}$   & 1232  & 16 & $\lambda_B$ & +1 \\
   &                & ($I=3/2$)       &   (flavour-symmetric)      &         &             &   &             &    \\ \hline
-4 & $\bar{\Delta}$ & $\Delta$ Antibaryons & $\bar{u}\bar{u}\bar{u},\bar{u}\bar{u}\bar{d},\bar{u}\bar{d}\bar{d},\bar{d}\bar{d}\bar{d}$& $\bar{\Delta}^{--,-,0,+}$& 1232  & 16 & 
$\lambda_B^{-1}$ & +1 \\
   &           & ($I=3/2$)       &   (flavour-symmetric)     &         &             &   &             &    \\ \hline
5  & $K$       & Strange Mesons     & $u\bar{s},d\bar{s}$ & $K^+,K^0$      & 495,644 & 2 & $\lambda_S$     & -1 
\\ \hline
-5 & $\bar{K}$ & Antistrange & $\bar{d}s,\bar{u}s$ & $\bar{K}^0,K^-$& 495,644 & 2 & $\lambda_S^{-1}$& -1 \\ 
   &           & Mesons      &                     &                &         &   &                 &    \\ \hline                   
6  & $\Lambda$ & $\Lambda$ Baryons & $uds$ & $\Lambda$      & 1115,683 & 2 & $\lambda_B\lambda_S^{-1}$  & +1 \\
   &           & ($I=0$)           &       &                &          &   &                            &   \\ \hline   
-6 & $\bar{\Lambda}$&$\Lambda$ Antibaryons &$\bar{u}\bar{d}\bar{s}$& $\bar{\Lambda}$& 1115,683 & 2 &$\lambda_B^{-1}\lambda_S$& +1 \\
   &                & ($I=0$)              &                       &                &          &   &                         &    \\ \hline
7  & $\Sigma$ & $\Sigma$ Baryons & $uus,uds,dds$ & $\Sigma^{+,0,-}$ & 1193,154 & 6 & $\lambda_B\lambda_S^{-1}$  & +1 \\
   &          & ($I=1$)          &               &                  &          &   &                            &    \\ \hline
-7 & $\bar{\Sigma}$ & $\Sigma$ Antibaryons & $\bar{d}\bar{d}\bar{s},\bar{u}\bar{d}\bar{s},\bar{u}\bar{u}\bar{s}$ & 
$\bar{\Sigma}^{+,0,-}$ & 1193,154 & 6 & $\lambda_B^{-1}\lambda_S$  & +1 \\
   &                & ($I=1$)              &               &                  &          &   &                            &    \\ \hline
8  & $\Xi$ & $\Xi$ Baryons & $uss,dss$ & $\Xi^{0,-}$ & 1318,285 & 4 & $\lambda_B\lambda_S^{-2}$  & +1 \\ \hline
-8 & $\bar{\Xi}$ & $\Xi$ Antibaryons & $\bar{d}\bar{s}\bar{s},\bar{u}\bar{s}\bar{s}$ & $\bar{\Xi}^{+,0}$ & 1318,285 & 4 & 
$\lambda_B^{-1}\lambda_S^2$  & +1 \\ \hline
9  & $\Omega$ & $\Omega$ Baryons & $sss$ & $\Omega^-$ & 1672,45 & 4 & $\lambda_B\lambda_S^{-3}$  & +1 \\ \hline
-9 & $\bar{\Omega}$ & $\Omega$ Antibaryons & $\bar{s}\bar{s}\bar{s}$ & $\bar{\Omega}^+$ & 1672,45 & 4 & 
$\lambda_B^{-1}\lambda_S^{-3}$  & +1 \\ \hline
\end{tabular}
\renewcommand{\arraystretch}{1}
\end{center}
\setcounter{table}{0}
\vspace{-0.2cm}
\caption{\label{tab:families} {\small The families where the hadrons are grouped when the 
Bose/Fermi statistics
are used. We list the difference in the quark content and the difference in the flavour wave 
function where it
exists. Also, we list the lowest mass hadron in the HG phase, which is the upper limit for 
the respective mass in
the chiral phase, according to eq.~(\ref{eq:mchi}).}}
\end{table}



\noindent
Eqs.~(\ref{eq:HGchnBF}) and (\ref{eq:pg}) constitute the minimum requirements for two
phases in 
equilibrium. Since the two phases are described by different partition functions, 
continuation 
of first and higher order derivatives cannot be fulfilled by these equations
alone. We use, though, these equations assuming that the region for the transition between 
the 
HG states and the chiral states is small. This allows to consider the transition curve of 
the 
phase diagram as points of equilibrium between the HG and chiral phases.
Indeed, in a more complete description the quantities $\tilde{P}_g$ and $B_v$ should be 
taken into account. Also, smoothing functions, like in \cite{matching}, would have to be 
introduced in the crossover region to ensure continuation of the first and higher order 
derivatives. However, this procedure would increase the free parameters of the model in cost 
of physical insight.

To complete this section we refer to the families in which the hadronic particles have been 
grouped. Related information is given in Table \ref{tab:families}. 
We use hadronic states 
listed in tables of ``Particle Data Group'' \cite{PDG2}\footnote{We include all particle 
states
containing only $u$, $d$ and $s$ quarks.}. 
To explore the effect of possible missing (not yet 
discovered) hadrons, we perform our calculations using five sets of 
hadronic states. The first 
includes all the
fully verified states of \cite{PDG2}, listed in the summary 
tables and it is denoted as (vh). In this set we, also, include the particle state of 
$f_0(500)$ state (also known as $\sigma$) with mass equal to 600 
MeV. Throughout our paper, wherever no description is given
for the used states it is implied that this is the set of states 
which is used. The second set includes, apart from the exact 
states of the first set, all the partially verified states in
\cite{PDG2}. This set will be denoted in the paper with a simple 
asterisk, $(*)$. The third set is exactly the same as the second
one, apart from the replacement of the mass value of the 
$f_0(500)$ state with 475 MeV. We choose to explore the effect of a different mass
for the $f_0(500)$ hadron because there is great uncertainty for the exact value of this 
mass and because it affects considerably the HG pressure, since this mass is relatively low. 
This third set will 
be denoted with a double asterisk, $(**)$.
In the third set $f_0(500)$ becomes the ground state of the family $2a$, since its mass is 
reduced below the $\eta$ mass.
However, as both $\eta$ and $f_0(500)$ share the same spin
degeneracy factor, $g$, the solution of eq.~(\ref{eq:HGchnBF}) for the chiral mass of the 
family
remains the same. Of course, we have to ensure that this chiral mass does not exceed the
reduced value of $f_0(500)$.
We, also, explore the effect of hadron states and resonances which
are not known experimentally, but they are predicted from quark models and lattice
QCD studies \cite{lp,lp1,lp2,qm}.
These, are grouped in $N$-$\Delta$ baryonic states, strange baryonic states and few meson 
states
(see Fig.~1 in \cite{lp}).
In \cite{lp} it is found that the additional $N$-$\Delta$ states, if they exist, are not 
realized in nature as independent resonances and their inclusion, if it is left uncorrected,
leads to 
deterioration of thermal fits to experimental particle yields.
On the contrary, most of the known resonances of excited strange baryons have rather small 
widths and, hence, need smaller corrections \cite{lp}.
For these reasons we form a fourth hadron set, denoted as (lps), where we include the states 
of the hadron set $(**)$, as well as predicted strange and meson states hadron states 
recorded in \cite{lp}.
Finally in a fifth hadron set, denoted as (lp) set, we include the states of the (lps) set, 
as well
as the $N$-$\Delta$ states of \cite{lp}. The findings of this set should be treated 
cautiously, since
the predicted non-strange baryonic states are included in our study as independent 
resonances without
any correction which according to \cite{lp} may be problematic.
From set one to set five, apart from other effects, we have
gradually increasing pressure in the hadronic phase (at the same
thermodynamic conditions). 
So, for the set including only experimentally verified states, the HG pressure 
is the lowest one.
For the (lp) set we can, also, argue 
that, with the available information, the HG pressure probably approaches the higher
possible value (all possible states have been included and  $f_0(500)$ retains a
lower mass value). Therefore, with the use of the five hadrons sets we completely explore
the effect of hadron states on our calculations\footnote{The higher hadron state mass in the 
(vh) set we use is $\sim$2600 MeV,
while in the $(*)$, $(**)$ and (lp) states it is $\sim$3170 MeV.}.

For simplicity we have neglected the dependence of the particles 
on the electric charge ($\lambda_Q=1$). For this
reason, when the lightest HG states in a family are nearly 
degenerate we consider as lowest mass 
their average value 
(e.g. $m_{31}=(m_p+m_n)/2$). Notice that some information for 
the lowest mass states presented 
in Table \ref{tab:families} have been taken from the Quark Model 
classification scheme in 
\cite{PDG1} (see the relevant review ``Quark Model'' therein).

There is not enough information of how to distribute the 
Light Unflavoured Mesons (L.U.M.) with Isotopic Spin $I=0$ 
among families 2a and 2b. For this 
reason we consider the  
conservation of the inclusive particle numbers of both families, according to the formula:
\begin{equation} \label{eq:HGchN23}
V_{HG}\left[\sum_j n_{HG,2aj}(T)+\sum_j n_{HG,2bj}(T)\right]=
\tilde{V}\left[\tilde{n}_{2a}(T)+\tilde{n}_{2b}(T)\right]\;\;.
\end{equation}
Last equation does not contain enough information for fixing both  chiral masses. For this 
reason we assume the ratio of the chiral masses to be equal to the respective ratio of the 
lowest mass HG  states:
\begin{equation} \label{eq:HGchm23}
\frac{\tilde{m}_{2a}(T)}{\tilde{m}_{2b}(T)}=\frac{m_{2a,1}}{m_{2b,1}}\;.
\end{equation}
This particular choice influences only the calculations concerning families 2a and 2b. In 
the following we shall consider families 2a and 2b as a common family 2. 

An additional comment is 
here in order: for the $\Omega$ family there are listed 
in \cite{PDG2} only a few states. Thus, there are 
probably still missing states in this 
family and this may influence family specific properties. 
In order to reduce the impact of this 
missing information in our analysis we will occasionally 
neglect the, anyway very small, 
contribution of this family (as well as $\bar{\Omega}$), 
when referring to family specific 
results.

\section{A Procedure to locate the Critical Point} \label{sec:mech}

We first need a transition curve upon which we shall impose the equality of particle 
multiplicities of the two phases in equilibrium. Lattice results for the QCD pressure 
 do not fix the transition temperature $T_c$ at zero baryon chemical potential. 
The value of $T_c=$154 MeV 
is suggested by \cite{lat2+1}. A chemical 
freeze-out curve evaluated from heavy-ion 
collisions is suggested to end at $\mu_B=0$ at 
a temperature $T_{ch,0} \simeq$166 MeV in \cite{freeze}. It is natural to assume that the HG 
state (in which the hadron particles exist) cannot be retained in higher temperatures than 
$T_c$, thus
\begin{equation} \label{eq:Tch<=Tc}
T_{ch,0} \le T_c \;.
\end{equation}
On the other hand, the results in \cite{lat2+1} and \cite{freeze} obey $T_{ch,0} > T_c$ which 
can be compatible with the relation (\ref{eq:Tch<=Tc}) only if
\begin{equation} \label{eq:Tch=Tc}
T_{ch,0} \simeq T_c \;.
\end{equation}

In fact such a condition is also compatible with the analysis 
presented in \cite{highTc} ($T_c = 157.5 \pm 1.5 $ MeV) and \cite{lowTch} 
($T_{ch,0}$ $ \simeq 157$ MeV),
which suggest that the critical QCD temperature 
and the freeze-out temperature at zero baryon chemical potential are very close, if not 
identical. We will adopt this scenario in the following, assuming that the chemical 
freeze-out curve \cite{freeze} coincides with the chiral transition curve. This curve is 
parametrised \cite{freeze} as:
\begin{equation} \label{eq:freeze}
T(\mu_B)= 166 - 0.139 \cdot 10^{-3} \mu_B^2 -0.053 \cdot 10^{-9} \mu_B^4\;\;,
\end{equation}
with $T$ and $\mu_B$ measured in MeV. Here, we will use a more general parametrization 
of (\ref{eq:freeze}) in order to implement the fact that $T_c$ is not uniquely defined 
and to fulfil $T_{ch,0}=T_c$:
\begin{equation} \label{eq:freezeout}
T=T_c-A(T_c) \mu_B^2-B(T_c) \mu_B^4 \Rightarrow T=f_{fr} (\mu_B;T_c),
\end{equation}
\[
A(T_c)=3.25\cdot 10^{-6}  \;{\rm MeV}^{-2} \cdot T_c
- 4.005\cdot 10^{-4}
\;{\rm MeV}^{-1},\;
\]
\[
B(T_c)=(2.5 \cdot 10^{-3}\; 
{\rm MeV}^{-4} \cdot T_c
- 0.468 \;{\rm MeV}^{-3})\cdot 10^{-9}\;.
\]
This parametrization leads to the 
freeze-out curve of \cite{freeze} for $T_{ch,0}=$166 MeV. The curves of eq.~(\ref{eq:freezeout}) 
approach the curve in \cite{freeze} at high values of baryon chemical potential, while they 
differ from it at low values of 
baryon chemical potential (each approaches a different value of $T_{ch,0}$ at zero density). In 
that sense, $T_c$ remains in eq.~(\ref{eq:freezeout}) a free parameter. Upon this curve we shall 
apply the conservation of particle numbers between the HG and the chiral phase. In fact, any 
choice for a transition curve is able to accommodate the constraints of particle 
conservation. However, the numerical results for the densities and chiral masses will depend on 
the particular choice.

We, also, impose the strangeness neutrality
\begin{equation} \label{eq:<S>=0}
<S>_{HG}=0 \;.
\end{equation}

Equations (\ref{eq:freezeout}), (\ref{eq:<S>=0}) can be solved for a given value of $T_c$ and so 
the thermodynamic parameters of temperature $T$ and strange quark chemical potential $\mu_s$ can 
be expressed as a function of the baryon chemical potential $\mu_B$, $T=T(\mu_B;T_c)$, 
$\mu_s=\mu_s(\mu_B;T_c)$. Following this procedure, in the next equations 
of this section there will be no explicit dependence on the parameters $T$ and $\mu_s$.

We will now investigate whether solutions of eqs.~(\ref{eq:HGchnBF}) for given values of 
$T_c,~\mu_B$ exist. 
In the crossover region, the equality of volumes, $V_{HG}=\tilde{V}$, leads to the equality of 
densities for the $i$ family:
\[
\sum_j n_{HG,ij}(\mu_B;T_c)=\tilde{n}_i(\mu_B;T_c) \Rightarrow
f_{vc,i}(\mu_B;T_c) \sum_j n^{pt}_{HG,ij}(\mu_B;T_c)=\tilde{n}_i(\mu_B;T_c) \Rightarrow
\]
\vspace{-0.2cm}
\begin{equation} \label{eq:fvc}
\Rightarrow
f_{vc,i}(\mu_B;T_c) = \frac{\tilde{n}_i(\mu_B;T_c)}{\sum_j n^{pt}_{HG,ij}(\mu_B;T_c)} \;\;,
\end{equation}
where we have written the HG particle density with volume corrections as the product of the 
corresponding point particle density times a volume correction factor $f_{vc,i}$:
\begin{equation} \label{eq:fvc1}
f_{vc,i}(\mu_B;T_c)\equiv \frac{\sum_j n_{HG,ij}(\mu_B;T_c)}{\sum_j n^{pt}_{HG,ij}(\mu_B;T_c)}
\;\;,
\end{equation}
which represents the ratio of the density
with volume corrections
of the family $i$ to the corresponding
density for point particles at specific
conditions.

Eq.~(\ref{eq:fvc}) should hold in the crossover region and, according to eq.~(\ref{eq:mchi}), it 
can be fulfilled as long as the chiral mass for each family is less than the maximum 
allowed value. So the limit of the crossover region (critical point) is reached if for some 
family $i$:
\begin{equation} \label{eq:mi1}
\tilde{m}_i=m_{i1} \;.
\end{equation}
When eq.~(\ref{eq:mi1}) holds, the right hand side of eq.~(\ref{eq:fvc}) becomes:
\begin{equation} \label{eq:Ri}
R_i(\mu_B;T_c) \equiv \frac{\tilde{n}_{i,min}(\mu_B;T_c)}{\sum_j n^{pt}_{HG,ij}(\mu_B;T_c)} =
\frac{n^{pt}_{HG,i1}(\mu_B;T_c)}{\sum_j n^{pt}_{HG,ij}(\mu_B;T_c)} \;\;.
\end{equation}
Here as $R_i$ we define the ratio of the density of the least mass particle to the
density of the whole family $i$ for specific
conditions. The nominator of this
ratio equals the minimum allowed density at
the chiral phase at these conditions, which is achieved for the maximum allowed mass.
With this definition we are led to 
\begin{equation} \label{eq:fvccpi}
f_{vc,i}(\mu_B;T_c)=R_i(\mu_B;T_c) \Rightarrow
 \sum_j n_{HG,ij}(\mu_B;T_c)=
n^{pt}_{HG,i1}(\mu_B;T_c) \;\;,
\end{equation} 
i.e. {\bf the critical point is reached at the
conditions where the density of a family
(with volume corrections) equals
the point particle density of the family with least mass}.

The quantities $f_{vc,i}$ and $R_i$ in the above equation, also, depend on $\mu_s$ which is 
calculated by solving eq.~(\ref{eq:<S>=0}). Actually, $\mu_s$ is almost independent of the 
particle volumes in the Bose/Fermi statistics, while it is completely independent in the 
Boltzmann 
approximation, since in this case the strangeness neutrality condition reduces to 
$<S>^{pt}=0$. 
To show that this dependence is negligible we depict in Fig.~\ref{fig:msmbr0} the solution 
for 
the strange quark chemical potential $\mu_s=\mu_B/3-\mu_S$ for the point particle case 
($r_0=0$), 
as well as, for different values of the hadron radius $r_0$ and we observe that the results 
almost coincide. We present calculations for a high and a low value of $T_c$.

The only dependence of ratios $R_i$ on particle volumes comes through $\mu_s$. In turn, we have
shown that $\mu_s$ weakly depends on particle volumes.
Thus, we can draw our conclusions using $R_i$ calculated for the $r_0=0$ case. The 
$f_{vc,i}$ we have defined, carry, then, almost all the dependence on the particle volumes 
corrections. Also, these volume correction factors are practically the same for all families, 
provided that the same radius $r_0$ is used for all hadrons. This property, which is 
approximate 
in the Bose/Fermi statistics, holds exactly in the Boltzmann description. Indeed, for equal 
hadron volumes, $v_0$, eq.~(\ref{eq:fvc1}) reduces to:
\begin{equation} \label{eq:fvc1bo}
f_{vc,i}(\mu_B;T_c)=
 \frac{\exp[-v_0 P]}{1+ v_0 \exp[-v_0 P] \sum_i \sum_j n_{HG,ij}^{pt}}\;\;.
\end{equation}
To verify that the dependence of the correction factor on the family is weak, we plot in 
Fig.~\ref{fig:fvc} the factor 
$f_{vc}$ in the Bose/Fermi statistics for the different families
$i$ and for two hadron volumes (which correspond to the radii $r_0=0.285$ fm and $r_0=0.25$ fm 
respectively). It is evident from the graph that the values of this ratio are very close for all 
the families. We have to point out, though,
that if different radii $r_i$ are used for different 
families, then the corresponding factors $f_{vc,i}$ will differ.

After these considerations, we plot in Fig.~\ref{fig:RiTc} the values 
of the 
ratios $R_i$ for each family, calculated along the transition
curves which correspond to the two extreme values of 
$T_c \simeq$166 MeV in (i) and $T_c \simeq$154 MeV in (ii). 
In this figure we see, firstly, that the differences between families which contain particles 
and antiparticles are small. 
Secondly, we notice that the ratio for the $\Omega,\bar{\Omega}$ families is artificially
high as a consequence of unknown resonances in these families. Neglecting this discrepancy,
the ratio $R_i$ for the $\pi$ family for certain value of baryon chemical potential has 
the highest value of all the families. Thus, if we apply the volume corrections 
for $r_0$ equal for 
all hadrons, the $\pi$'s will be the first family (i.e. for the lower value of $\mu_B$) which 
satisfies the condition that its chiral mass acquires the maximum allowed value:

\begin{figure}[H]
\centering
\vspace{-0.4cm}
\includegraphics[scale=0.5,trim=2.in 0.3in 2.in 0.2in,angle=0]{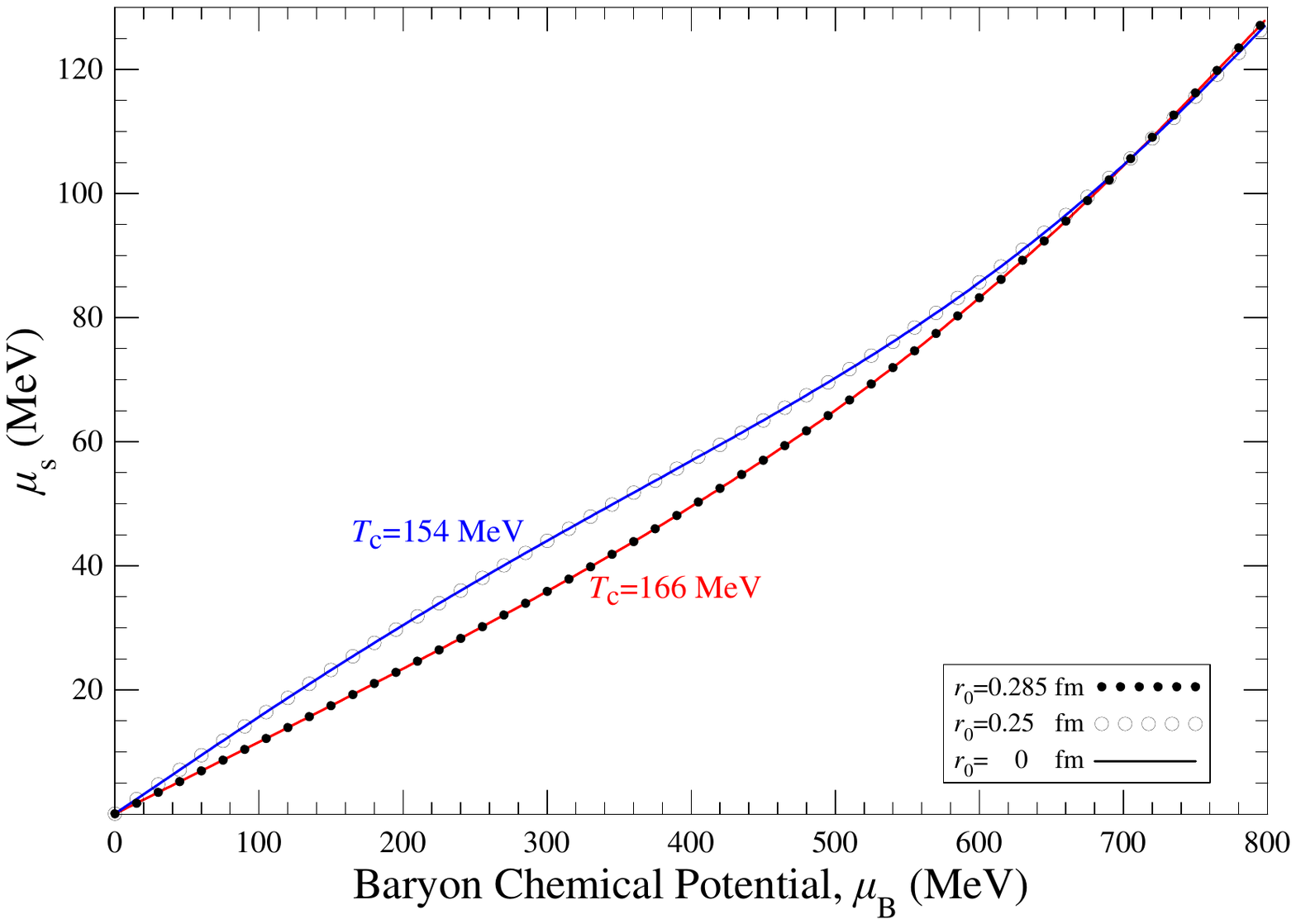}
\caption{\label{fig:msmbr0} {\small The dependence of the strange quark chemical potential 
$\mu_s$, which solves the $<S>=0$ equation for given hadronic radius $r_0$, on the baryon 
chemical potential $\mu_B$. Shown are two cases which correspond to a high value of 
$T_c=$166 MeV (lower curves) and a low value of $T_c=$154 MeV (upper curves) respectively. 
Continuous lines correspond to $r_0$=0 fm (point particle case), solid circles to $r_0$=0.285 fm 
and open circles to $r_0$=0.25 fm. It is evident that volume corrections practically do not 
affect the calculated value of $\mu_s$.}}
\end{figure}

\begin{figure}[ht]
\centering
\vspace{-0.4cm}
\includegraphics[scale=0.55,trim=1.5in 0.8in 2in 0.5in,angle=0]{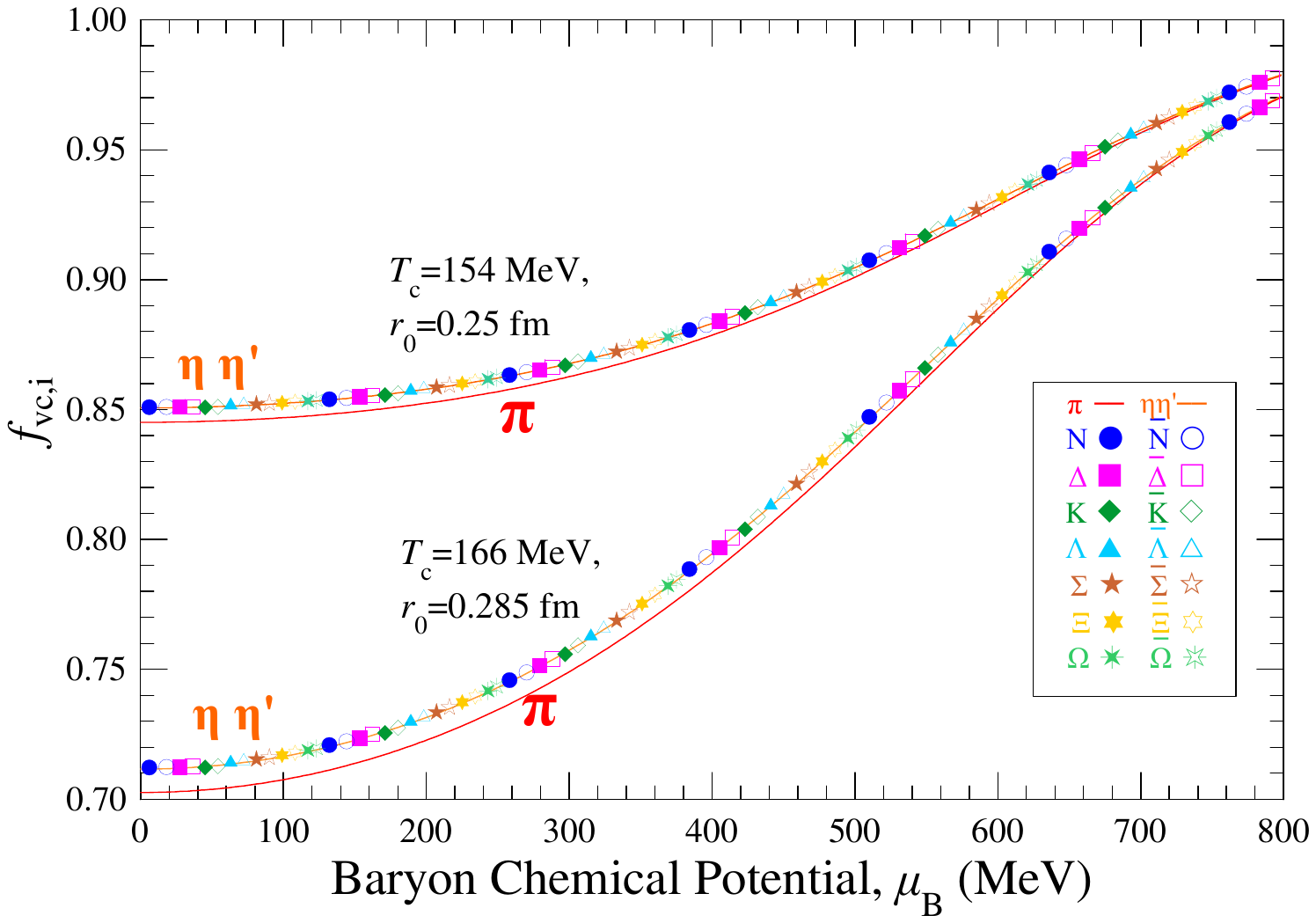}
\vspace{-0.2cm}
\caption{\label{fig:fvc} {\small The volume correction factor $f_{vc}$ for the hadronic 
families, with Bose/Fermi statistics for $T_c=$166 MeV and radius $r_0=0.285$ fm (lower curves)
and for $T_c=$154 MeV and radius $r_0=0.25$ fm (upper curves),
as function of the baryon chemical potential $\mu_B$. Since the curves for all families except 
the pions are close, we depict, instead of almost coinciding curves, calculations at distinct points of non-overlapping lattices.}}
\end{figure}


\begin{figure}
\vspace{-0.2cm}
\centering
(i)\hspace{0.5cm}
\includegraphics[scale=0.5,trim=2.5in 0.3in 2in 0.5in,angle=0]{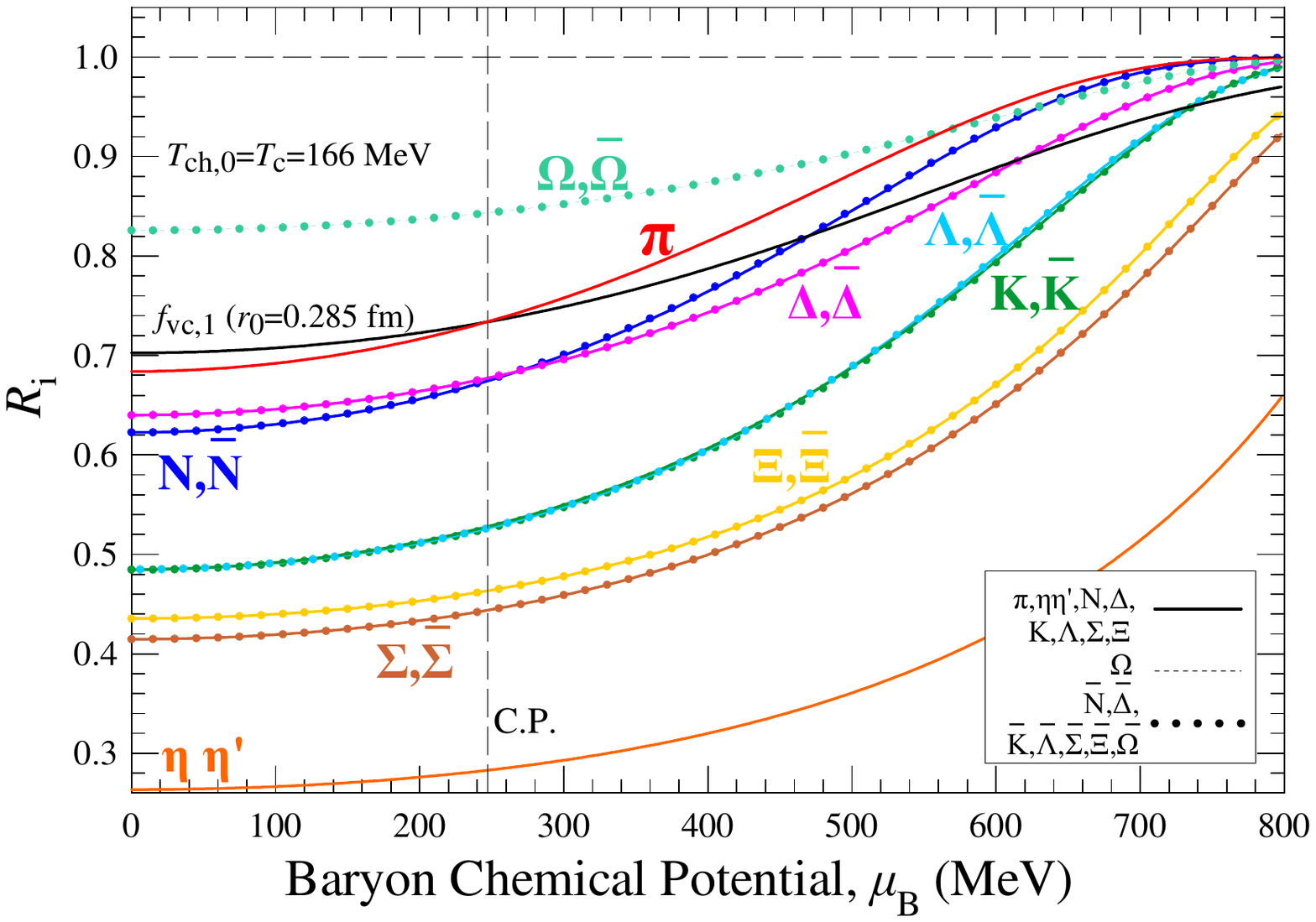}\\
\vspace{0.5cm}
(ii)\hspace{0.5cm}
\includegraphics[scale=0.5,trim=2.5in 0.3in 2in 0.5in,angle=0]{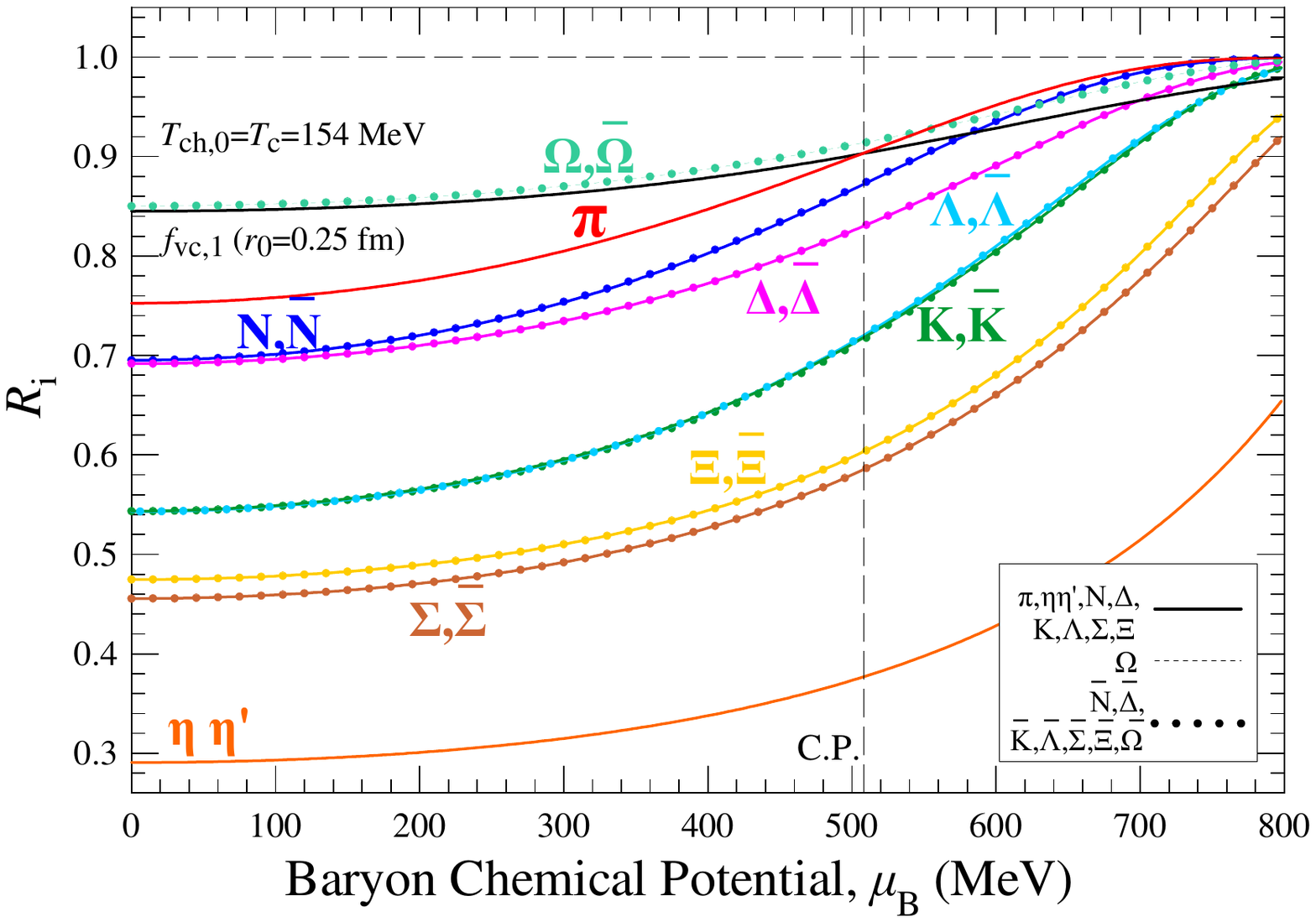}
\caption{\label{fig:RiTc} {\small 
The ratios $R_i$ along the transition curve for the families $i$ as function of $\mu_B$. These 
ratios are calculated for point particles and are equal to the density of the lowest mass 
particles of each family (equal to the minimum density of the chiral density at the same 
conditions) divided by the total particle density of the family. The ratios $R_i$ here are 
calculated in the Bose/Fermi statistics and without volume corrections, since $\mu_s$ is 
practically unaffected by the hadron volumes, according to Fig.~\ref{fig:msmbr0}. In (i) we show 
calculations for a high value of $T_c=$166 MeV. Also, it is shown the volume correction 
factor for the pion family, $f_{vc,1}$ for $r_0=$0.285 fm. The intersection of $f_{vc,1}$ with 
$R_1$ determines the position of the critical point (C.P.). In (ii) we show analogues 
calculations for a low value of $T_c=$154 MeV and for $f_{vc,1}$ for $r_0=$0.25 fm. The 
pion family retains for all values of $T_c$ the higher value of ratio $R$. The second important 
family is the nucleon family which, almost everywhere, retains, the second higher value of ratio 
$R$.
The ratio for the $\Omega$ family is shown with dotted line to represent the
artificially high value of the $\Omega,\bar{\Omega}$ families which are due to the absence of 
unknown resonances.
}}
\end{figure}

\noindent
\begin{equation} \label{eq:mchcp}
\tilde{m}_1(\mu_{Bcr}) = m_{\pi} = m_{11} \;\;.
\end{equation}
Equivalently, last equation can be written as:
\begin{equation} \label{eq:fvccp}
f_{vc,1}(\mu_{Bcr})=R_1(\mu_{Bcr}) \Rightarrow
\sum_j n_{HG,1j}(\mu_{Bcr}) = n^{pt}_{HG,11}(\mu_{Bcr}) \;\;.
\end{equation}
Eqs.~(\ref{eq:mchcp}), (\ref{eq:fvccp}) enable us to calculate the critical point position at 
baryon chemical potential $\mu_{Bcr}$.

For values $\mu_B \leq \mu_{Bcr}$ the crossover conditions (\ref{eq:fvc}) can be fulfilled not 
only for the $\pi$ family, but also for all the families (since $R_i(\mu_B)<R_1(\mu_B),i\neq$ 
1,9,-9).
For $\mu_B > \mu_{Bcr}$ the chiral mass of the pion family remains equal to the maximum allowed 
pion mass. The conservation of the particle number of the pion family can no longer be fulfilled 
through the reduction of the mass and, thus, it has to be fulfilled through the 
alteration of the system volume, $V_{HG} > \tilde{V}$. So we are in the region of the 1st order 
transition. To ensure the conservation of particle number for the pion family we impose the 
condition:
\begin{equation} \label{eq:pc11o}
V_{HG}\sum_j n_{HG,1j}(\mu_{B}) = \tilde{V}n^{pt}_{HG,1}(\mu_{B}) \Rightarrow
v_{er} \equiv \frac{V_{HG}}{\tilde{V}} = \frac {n^{pt}_{HG,1}(\mu_{B})}{\sum_j n_{HG,1j}
(\mu_{B})}  \;\;.
\end{equation}
We have defined as $v_{er}$ the volume expansion ratio. This ratio is determined by 
eq.~(\ref{eq:pc11o}) for $\mu_B>\mu_{Bcr}$, while it remains equal to $1$ in the crossover. 
Consequently, $v_{er}$ is inherited to the rest of the families. The conservation of the
particle number in these families is ensured by:
\begin{equation} \label{eq:pci1o}
v_{er} \sum_j n_{HG,ij}(\mu_B)=\tilde{n}_{i}(\mu_B) \;, i\neq 1 \;.
\end{equation}
The last equation determines the chiral masses $\tilde{m}_i(\mu_B)$ for the particle families 
different than the pion family.
However, it has to be checked for consistency reasons that the condition 
\begin{equation} \label{eq:milm1}
\tilde{m}_i(\mu_B) \leq m_{i1} \;, i\neq 1 \;,
\end{equation}
continues to hold in the region of the 1st-order transition, as well.

It is evident from this description that the pions with the associated spectrum of resonances 
play fundamental role.
The position of the critical point is determined by the knowledge of the pion family particle 
spectrum. The particle spectrum of the rest of the families does not influence the critical 
point position. The addition of currently unknown resonances in the
spectrum of these families will further reduce the ratios $R_i$, depicted in Figs.~\ref{fig:RiTc}.

In Figs.~\ref{fig:RiTc} we, also, present graphical solutions of eq.~(\ref{eq:fvccp}) -- for the 
transition curves which correspond to $T_c \simeq$ 166 and 154 MeV -- determining the 
position of the critical point. To this end we plot the volume correction factors for the 
relevant value of hadron radii $r_0$. The intersection of these factors with the pion ratios, 
$R_1$, determine the critical baryon chemical potential, $\mu_{Bcr}$. 
The factors 
$f_{vc,i}$ are almost the same for all the families $i$ for equal hadron volumes, according to 
Fig.~\ref{fig:fvc}. Thus, the fact that the ratio $R_1$ retains the higher value among all $R_i$ 
ensures that the intersection of the curves
$f_{vc,1}$ and $R_1$ will occur first (at lower $\mu_B$). Also, it is evident from 
Fig.~\ref{fig:fvc} that $f_{vc,1}$ is slightly 
lower than the rest of the volume correction factors. This further confirms our conclusion that 
the pion family will reach first the relative maximum mass at the chiral phase. However, we have 
to point out that the factors $f_{vc,i}$ will differ if different particle volumes are used.

This description is consistent with the fact that in the low baryon density area the HG state is 
dominated mainly by mesons
and the pion is the meson with the lower mass. Thus, its production is favoured at this 
territory, which leads to enhanced
pion multiplicities. Also, we observe that the family with $R_i$ closest to the pion family is 
the nucleon family $N$. Indeed, the nucleon family contains the nucleons, protons and neutrons, 
which are the lightest hadrons carrying baryon number. Thus, these particles, with their family, 
should play important role as the baryonic density increases and their multiplicity becomes 
enhanced. So, the next family to the pions which may play important role are the nucleons. 

In this section for simplicity we have used volume correction factors for the case of 
constant volumes equal for
all hadrons. If different volumes are introduced for specific hadron species these factors 
will differ among these species. Additionally, for volumes depending on chemical potentials
the volume correction factors may differ between particles and antiparticles 
(according to eq.~(\ref{eq:n_HG})). In each case we have to ensure that the pion family is the family for which the volume 
correction factor meets the corresponding ratio $R_i$ at 
the lowest value of baryon chemical potential.


\section{Constrains of lattice QCD on hadron volumes - Thermodynamic implications} \label{sec:constvary}

The hadron densities in the HG state depend on the eigenvolume of each particle, which 
incorporate the effect of the 
repulsive part of the strong interaction. 
In the previous section, to show the effect of the hadron volumes on our model, we used arbitrary values, which
where not subject to any constraint.
In this 
section we take a step further and we try to acquire knowledge of the values of these particle volumes by a fit on the lattice QCD 
results. We consider here the QCD pressure $P_L$ as a function of temperature for vanishing baryon 
chemical potential which is calculated in \cite{lat2+1}. Our results will depend on the critical 
temperature $T_c$ for zero density. Below $T_c$ the system exists in the Hadron Gas (HG) state. 
We will consider both scenarios: (i) constant eigenvolumes (independent of temperature and baryon 
chemical potential),  as well as, (ii) eigenvolumes depending on temperature and baryon chemical potential.

\subsection{Constant Eigenvolumes} \label{subsec:const}

\begin{figure}[H]
\centering
\vspace{-0.3cm}
\includegraphics[scale=0.45,trim=0.3in 0.3in 0.in 0.2in,angle=0]{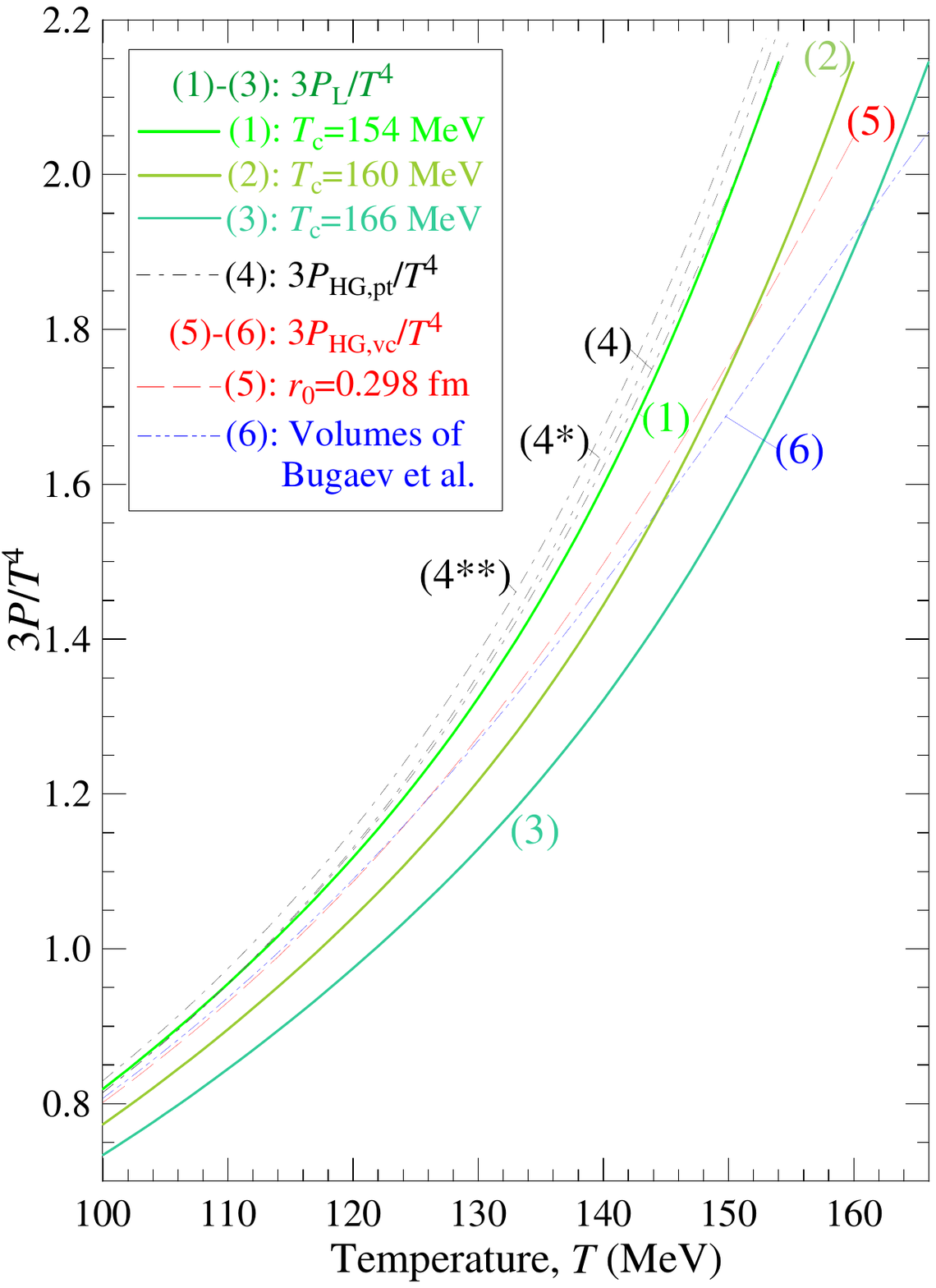}
\caption{\label{fig:PLHGB_} {\small 
Normalised pressures, $3P/T^4$, as function of temperature, $T$: 
Lattice QCD of \cite{lat2+1} (continuous lines) corresponding to $T_c=$ 154 MeV (line (1)), 
160 MeV (line (2)) and 166 MeV (line (3)). 
Hadron Gas for point particles with Bose/Fermi statistics for 3 hadron sets (slashed-dotted lines (4)).
Hadron Gas for extended particles, with common radius $r_0=0.298$ fm (slashed line (5)) and with radii fixed
at the values cited in \cite{hadvol} (slashed-double dotted line (6)).}}
\vspace{-0.3cm}
\end{figure}

We start our investigation with the scenario of constant eigenvolumes. In Fig.~\ref{fig:PLHGB_} 
we present the normalised lattice pressure $3P_L/T^4$ as function of temperature for three 
values of $T_c$: 154 MeV (which is the case in \cite{lat2+1}), 160 MeV and 166 MeV  
(lines, (1)-(3)). 
We focus on the temperature region $T \geq$ 100 MeV, since the lattice results may not be 
accurate for $T \ll T_c$. As it is evident 
from Fig.~\ref{fig:PLHGB_}, the increase of $T_c$ 
causes the lattice pressure to decrease for the same temperature. With line (4) we 
depict the normalised HG pressure for point particles with Bose/Fermi 
statistics. However, 
line (4) lacks the additional requirements for a correct HG description, that is 
volume corrections. 
It is necessary that this HG pressure curve lies above the relevant lattice pressure curve. This is due to the decrease of the HG
pressure caused by the introduction of volume corrections.
We observe that this requirement for lattice critical temperature $T_c$=154 MeV is fulfilled for the most part of the temperature interval (100-154) MeV for the set of 
(vh) states (line (4)) and for the set (*) (line (4*)). It is, though, fulfilled in the whole 
temperature interval for the 
set of states (**) (line (4**)). 
The increase of $T_c$ above 154 MeV moves the lattice pressure curve to lower values for fixed temperature. So the HG pressure will be compatible with these
lattice results, as well.

\begin{figure}[H]
\vspace{-0.8cm}
\centering
\includegraphics[scale=0.52,trim=0.3in 0.8in 0.3in 0.in, angle=0]{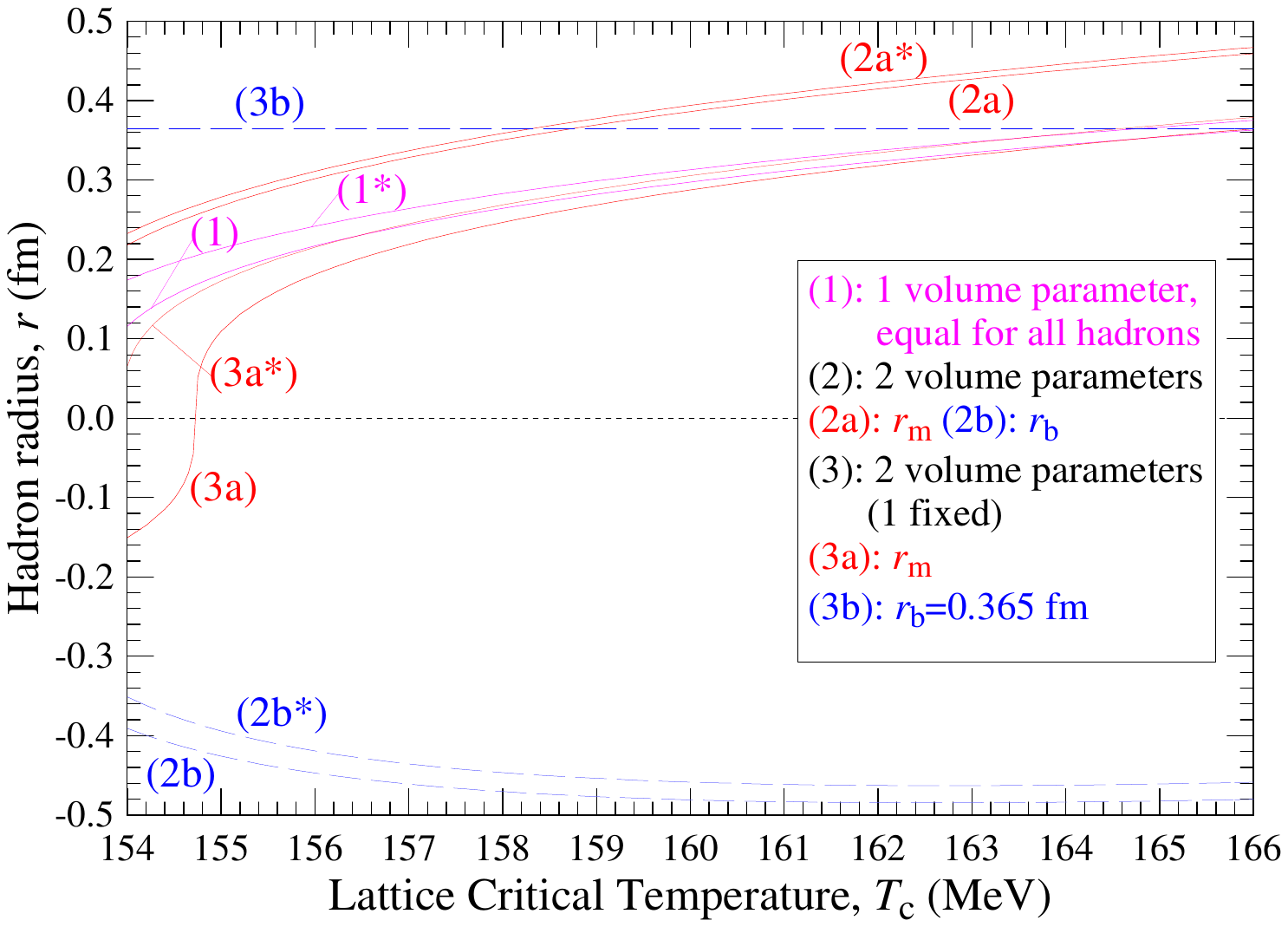}
\caption{\label{fig:PLHGR0} {\small 
The constant hadron radii in HG model which best fit the lattice QCD pressure, as function of the critical
QCD temperature, $T_c$. 
Line (1): a common radius for all hadrons.
Lines (2): Continuous line (2a): the radius for all mesons, $r_m$. 
Slashed line (2b): the radius for all baryons, $r_b$ (lying completely to the unphysical negative territory).
Lines (3): Continuous line (3a): the radius for all mesons, $r_m$. 
Slashed line (3b): the radius for all baryons, held fixed at $r_b=0.365$ fm (as in \cite{hadvol}).
}}
\end{figure}

We, then, attempt to fit the lattice pressure with constant values of volumes. First we employ 
the same volume parameter, 
$v_0$, for all the hadrons. We determine this volume by a fit on the 
lattice pressure \cite{lat2+1}, $P_L$, so that the quantity
\begin{equation} \label{eq:chi2}
\chi^2=\sum^N_{i=1}\left\{\frac{3}{T_i^4}\frac{[P_L(T_i;T_c)-P_{HG}(T_i,\{\mu\}=0;v_0)]}{s_i}
\right\}^2
\end{equation}
is minimized. The $N=20$ temperature points, $T_i$, are taken at equal intervals in the range 
$(100 {\rm \;MeV} - T_c)$, while the errors are all taken equal $s_i=1$. The degrees of freedom 
are $dof=N-k$, where $k$ is the number of the volume parameters we shall determine by the fit. 
The value of $s_i$ certainly affects any conclusion about the quality of the fit, however we are 
only interested in the relative success of the different fits we shall perform.
In Fig.~\ref{fig:PLHGR0} we show the results of this fit in 
line (1), which represents the radius $r_0$, common to all hadrons, as a function of $T_c$. 
We find for the whole range of values of $T_c$, depicted in the graph,
a physical  positive value for the radius $r_0$.
We observe a 
tendency for saturation as the maximum value of $T_c=166$ MeV is approached.

\begin{figure}[H]
\centering
\includegraphics[scale=0.52,trim=0.3in 0.6in 0.in 0.3in, angle=0]{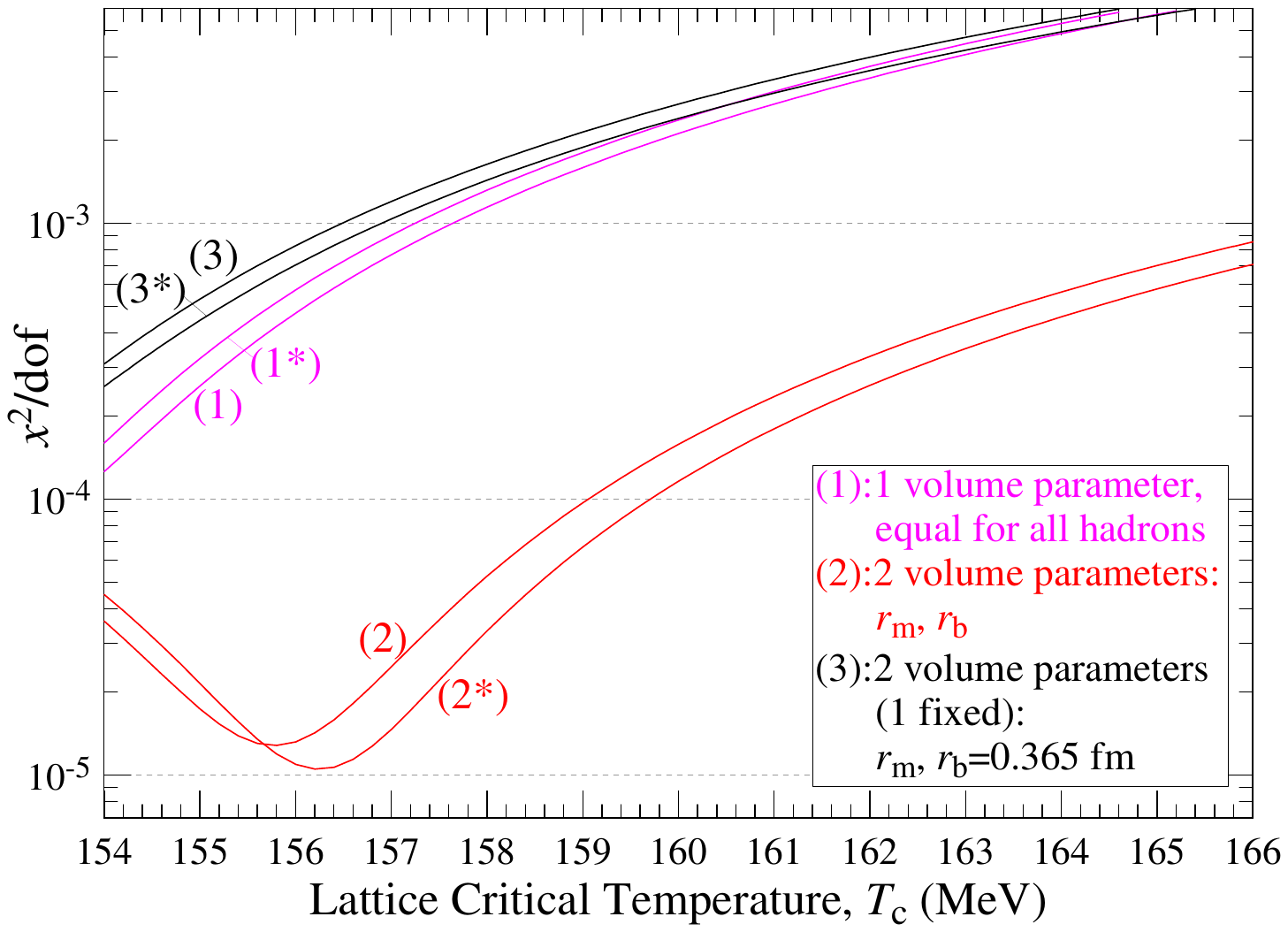}
\vspace{-0.2cm}
\caption{\label{fig:PLHGX2} {\small 
The quality, $\chi^2/dof$, of the fits performed with constant hadron radii in HG model 
in order to fit the lattice QCD pressure, as 
function of the critical QCD temperature $T_c$. 
Since the errors are taken {\it ad hoc} $s_i=1$, the graph serves to compare the relevant quality 
between the different cases.
Line (1): a common radius for all hadrons determined by the fit.
Line (2): two radii determined by the fit one for mesons, $r_m$, and one for baryons, $r_b$.
Line (3): the radius for baryons is held fixed at $r_b=0.365$ fm (as in \cite{hadvol}) and 
only the meson radius, $r_m$, is determined by the fit.}}
\end{figure}

In Fig.~\ref{fig:PLHGX2} we present results of the quality of the fit. From this graph, it is 
evident that the quality worsens as $T_c$ increases.
Returning to Fig.~\ref{fig:PLHGB_}, where we have plotted the normalised HG pressures for the 
fitted radius value which corresponds to $T_c=$160 MeV, we observe that the lattice pressure is 
fitted poorly by the one volume parameter (line (5)). 
For comparison we have also plotted the HG 
pressure for the hadron volumes of \cite{hadvol} (line (6)). It is evident that it does not fit 
well the lattice pressure.

We then attempt to fit the lattice pressure with two volume parameters, one for the mesons, 
$v_m$ and one for the baryons, $v_b$, which are both left free to be determined by the fit. 
The procedure entails to determine these two parameters by minimizing the quantity $\chi^2$
of eq.~(\ref{eq:chi2}), which now involves HG pressure depending on $v_m,v_b$,
$P_{HG} (T_i,\{\mu\}=0;v_m,v_b)$.
The results for the fitted hadron radii are shown in 
Fig.~\ref{fig:PLHGR0} with lines (2a) and 
(2b) and in Fig.~\ref{fig:PLHGX2} with line (2). This fit achieves better quality than the one 
parameter case, as it is shown in Fig.~\ref{fig:PLHGX2}. This is expected due to the use of an 
extra parameter. However, the baryon volumes are shifted, completely or partly, to unphysical 
negative volumes for all values of $T_c$.
This is also understandable, because for zero baryon density the HG is composed mainly by 
mesons, while the baryon content remains
relatively small.
Thus, small variations of the mesonic radius produce 



\vspace{-1cm}

\noindent
large variations to the 
meson multiplicities. 
On the contrary, to obtain comparable variations to the baryon 
multiplicities the respective radius has to vary considerably.  
Since the parameters in the 
performed fit are treated without constraints, the baryon radius is easily shifted to negative 
values whenever there is a tendency for the reduction of this parameter.
To avoid this 
behaviour, while keeping two volume parameters in the HG description, we decided to fit the 
lattice pressure varying only the mesonic radius $r_m$. During the fitting procedure the baryon 
radius remains fixed to the value $r_b=0.365$ fm given in \cite{hadvol}. Our results for $r_m$ 
are shown in Fig.~\ref{fig:PLHGR0} with lines (3a) and (3b) and in
Fig.~\ref{fig:PLHGX2} with line (3). From Fig.~\ref{fig:PLHGR0} we
find that $r_m$ acquires physical values for $T_c \geq 154.7$ MeV. The quality of the fit 
remains worst than the one volume parameter case (Fig.~\ref{fig:PLHGX2}, line (3)) for all 
$T_c$.

\begin{figure}
\vspace{-0.1cm}
\centering
\hspace{-0.1cm}(i)\hspace{-0.1cm}
\includegraphics[scale=0.4,trim=0.3in 0.2in 0.in 0.3in, angle=0]{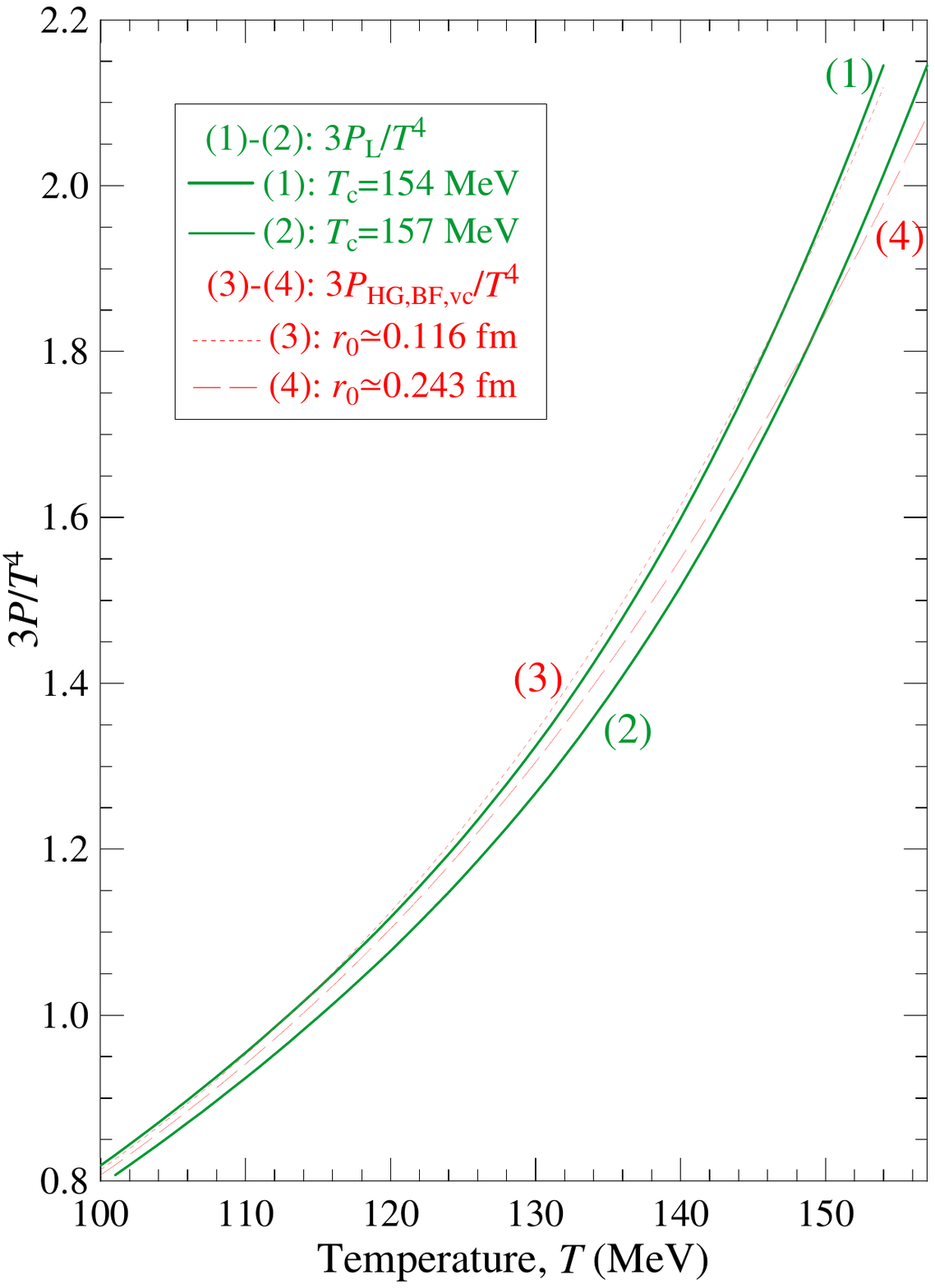}
\hspace{0.5cm}(ii)\hspace{-0.1cm}
\includegraphics[scale=0.4,trim=0.3in 0.2in 0.3in 0.3in, angle=0]{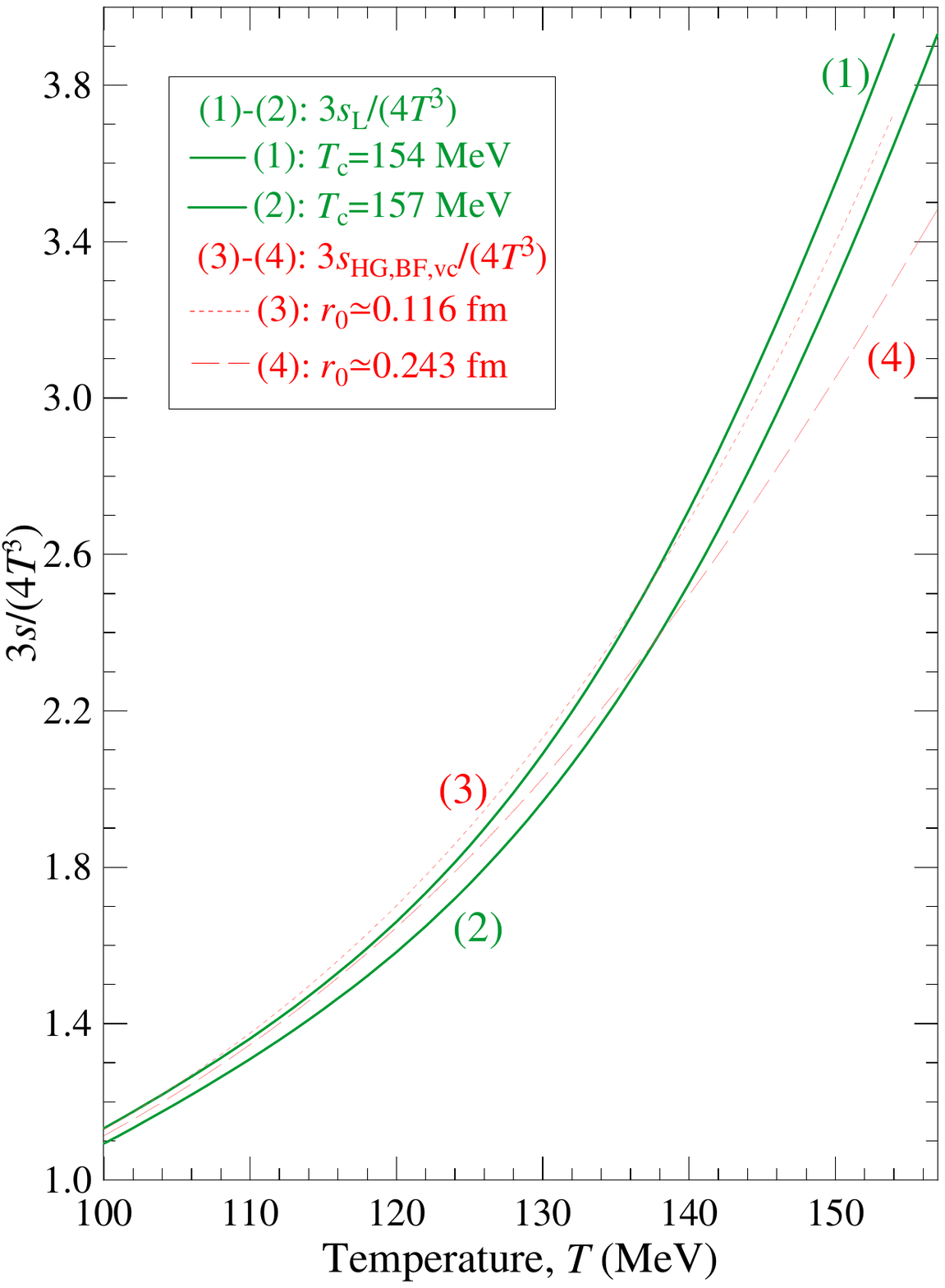}\\
\vspace{-0.cm}
\caption{\label{fig:PLHGC_} {\small 
(i) Comparison of lattice normalised pressures with HG normalised pressures, as function of 
temperature, for two values of $T_c=$154 MeV and $T_c=$157 MeV. The relevant HG pressures correspond 
to common radius determined by the fit, $r_0=$0.116 fm and 0.243 fm, respectively. 
(ii) Comparison of lattice normalised entropy densities with HG normalised entropy densities, as 
function of temperature, for the parameters in (i). 
Although, the pressures are fitted quite well in (i), the entropy densities 
here fail to do so.
}}
\end{figure}

We end this subsection by showing representative fits with common particle eigenvolumes for both 
mesons and baryons. We concentrate on this case since the fit with the two free volume 
parameters leads to unphysical results and the one parameter fit has better quality than the one 
with the fixed baryon volume. Thus, we present in Fig.~\ref{fig:PLHGC_}(i) results of the HG 
pressure corresponding to $T_c=$154 and 157 MeV. The best fit among the two
is at $T_c=$154 MeV, while the fit starts to deteriorate at 157 MeV. In any case, 
the HG model fails to fit sufficiently the lattice entropy density, which is the temperature 
derivative of the pressure. In Fig.~\ref{fig:PLHGC_}(ii) we compare the entropy densities of 
lattice and HG for the parameters of Fig.~\ref{fig:PLHGC_}(i). It is clear that the best 
constant HG volume parameters, which fit quite well
the lattice pressure, are not adequate to fit the lattice entropy density.

\subsection{Varying hadron eigenvolumes} \label{subsec:vary}

Since the constant particle eigenvolumes are inadequate to fit the lattice results (pressure and 
entropy density) we turn now to the use of volumes which depend on temperature and chemical 
potentials. However, this approach affects certain thermodynamic quantities.
The total HG pressure can be calculated by eqs.~(\ref{eq:HGtpvBF})-(\ref{eq:tildemu}). 

For simplicity we use, here, only one index $i$ for a specific particle, without reference to 
its family. If the particle volumes remain fixed, then the particle densities and 
the system entropy density read, respectively:
\begin{equation} 
n^v_{HG,j}(T, \{\mu\})=
\left.\frac{\partial P_{HG}(T,\{\mu\})}{\partial \mu_j}\right|_{T,\{\mu\}_j} \;,
\end{equation}
\begin{equation}
s^v_{HG}(T, \{\mu\})=
\left.\frac{\partial P_{HG}(T,\{\mu\})}{\partial T}\right|_{\{\mu\}}\;,
\end{equation}
where the upper index $v$ denotes constant particle volumes $ v_1,v_2, \ldots \equiv \{v\}$ and 
we have defined $ \mu_1,\ldots ,\mu_{j-1},\mu_{j+1},\ldots \equiv \{\mu\}_j$, i.e. the group of 
all the items without the $j$-th one.

The last two equations can be worked out to give the known results \cite{volume_cor},
with the upper index ``{\it pt}'' denoting the point-particle case:
\begin{equation} \label{eq:nv_HG}
n^v_{HG,j}(T, \{\mu\})=
\frac{n^{pt}_{j}(T, \mathord{\buildrel{\lower3pt\hbox{$\scriptscriptstyle\frown$}} 
 \over \mu}_{j})}
{1+\sum_i v_{i} n^{pt}_{i}(T,\mathord{\buildrel{\lower3pt\hbox{$\scriptscriptstyle\frown$}} 
 \over \mu}_{i})} \;\;,
\end{equation}
\vspace{-0.3cm}
\begin{equation} \label{eq:sv_HG}
s^v_{HG}(T, \{\mu\})=
\frac{s^{pt}(T, \mathord{\buildrel{\lower3pt\hbox{$\scriptscriptstyle\frown$}} 
 \over \mu})}
{1+\sum_i v_{i} n^{pt}_{i}(T, \mathord{\buildrel{\lower3pt\hbox{$\scriptscriptstyle\frown$}} 
 \over \mu}_{i})} \;\;.
\end{equation}

Then we examine the effect of having volumes which are {\it not constant} with respect of 
temperature and/or chemical potentials.
We use the notation $\ldots ,v_{j-1},v_{j+1},\ldots \equiv \{v\}_j $. Then we calculate:
\[
\left.\frac{\partial P_{HG}(T,\{\mu\};\{v\})}{\partial v_j}\right|_{T,\{\mu\},\{v\}_j}=
\sum_i \frac{\partial P^{pt}_i}{\partial \mathord{\buildrel{\lower3pt\hbox{$\scriptscriptstyle\frown$}} 
 \over \mu}_i} \frac{\partial \mathord{\buildrel{\lower3pt\hbox{$\scriptscriptstyle\frown$}} 
 \over \mu}_i}{\partial v_j}=
\sum_i \frac{\partial P^{pt}_i}{\partial \mathord{\buildrel{\lower3pt\hbox{$\scriptscriptstyle\frown$}} 
 \over \mu}_i} 
\left(-\delta_{ij} P_{HG}-v_{i} \frac {\partial P_{HG}} {\partial v_j}\right)=
\]
\[
=-P_{HG} \frac{\partial P^{pt}_j}{\partial \mathord{\buildrel{\lower3pt\hbox{$\scriptscriptstyle\frown$}} 
 \over \mu}_j}-
\frac {\partial P_{HG}} {\partial v_j}
\sum_i v_i \frac{\partial P^{pt}_i}{\partial \mathord{\buildrel{\lower3pt\hbox{$\scriptscriptstyle\frown$}} 
 \over \mu}_i} \Rightarrow
\frac{\partial P_{HG}}{\partial v_j}=
\frac{-P_{HG} \frac{\partial P^{pt}_j}{\partial \mathord{\buildrel{\lower3pt\hbox{$\scriptscriptstyle\frown$}} 
 \over \mu}_j}}{1+\sum_i v_i \frac{\partial P^{pt}_i}
{\partial \mathord{\buildrel{\lower3pt\hbox{$\scriptscriptstyle\frown$}} 
 \over \mu}_i}} =
\frac{-P_{HG} n^{pt}_j(T,\mathord{\buildrel{\lower3pt\hbox{$\scriptscriptstyle\frown$}} 
 \over \mu}_j)}{1+\sum_i v_i n^{pt}_i(T,\mathord{\buildrel{\lower3pt\hbox{$\scriptscriptstyle\frown$}} 
 \over \mu}_i)} \Rightarrow
\]
\vspace{-0cm}
\begin{equation} \label{eq:dPdv}
\hspace{0cm} \Rightarrow \frac {\partial P_{HG}} {\partial v_j}=
-P_{HG} n^v_{HG,j}\;\;.
\end{equation}

If the particle volumes depend on temperature, $v_i=v_i(T)$, then, using eq.~(\ref{eq:dPdv}), 
we determine the entropy density:
\[
s_{HG}(T, \{\mu\};\{v\})=
\left.\frac{\partial P_{HG}(T,\{\mu\};\{v\})}{\partial T}\right|_{\{\mu\}}=
\left.\frac{\partial P_{HG}(T,\{\mu\};\{v\})}{\partial T}\right|_{\{\mu\},\{v\}}+
\sum_i \frac{\partial P_{HG}}{\partial v_i} \frac{\partial v_i}{\partial T}\Rightarrow
\]
\vspace{-0.cm}
\begin{equation} \label{eq:s_HG}
\Rightarrow s_{HG}=s^v_{HG}-P_{HG} \sum_i n^v_{HG,i} \frac{\partial v_i}{\partial T}\;,
\end{equation}
where with $s^v_{HG}$ we denote the part of the HG entropy density which is calculated with 
constant volumes, according to eq.~(\ref{eq:sv_HG}).

If the particle volumes depend on chemical potentials, $v_i=v_i(\{\mu\})$, then, using 
eq.~(\ref{eq:dPdv}), we calculate the particle densities:
\[
n_{HG,j}(T, \{\mu\};\{v\})=
\left.\frac{\partial P_{HG}(T,\{\mu\};\{v\})}{\partial \mu_j}\right|_{T,\{\mu\}_j}=
\left.\frac{\partial P_{HG}(T,\{\mu\};\{v\})}{\partial \mu_j}\right|_{T,\{\mu\}_j,\{v\}}+
\sum_i \frac{\partial P_{HG}}{\partial v_i} \frac{\partial v_i}{\partial \mu_j}\Rightarrow
\]
\vspace{-0.1cm}
\begin{equation} \label{eq:n_HG_ini}
\Rightarrow n_{HG,j}=n^v_{HG,j}-P_{HG} \sum_i n^v_{HG,i} \frac{\partial v_i}{\partial \mu_j}\;,
\end{equation}
\noindent where with $n^v_{HG,j}$ we denote the part of the HG  density of the particle $j$, 
which is calculated with constant volumes, according to eq.~(\ref{eq:nv_HG}). 

At this point we want to have the same result if we evaluate the density for two particle
species $i$, $j$ and then add them and if we evaluate the density of the two particles together:
\begin{equation} \label{eq:n_ij}
n_{HG,j}+n_{HG,k}=n_{HG,j+k}\;.
\end{equation}
For example we can imagine that we have a family with two
particle species $j$ and $k$ and common volume $v$ and we evaluate each density
separately with chemical potentials $\mu_j$, $\mu_k$.
Then we evaluate the density for the whole family $j+k$
with chemical potential $\mu_j=\mu_k$. If the index $i$
in eq.~(\ref{eq:n_HG_ini}) runs to all the hadrons of the
system, then eqs.~(\ref{eq:n_HG_ini}) and (\ref{eq:n_ij})
cannot hold simultaneously. So we make the 
{\it choice} that the dependence of a hadron volume on chemical potentials is restricted to the chemical potential of the specific hadron:
\begin{equation} \label{eq:dvdmu}
\frac{\partial v_i}{\partial \mu_j}=\delta_{ij} \frac{\partial v_j}{\partial \mu_j} 
\;.
\end{equation}
Then eq.~(\ref{eq:n_HG_ini}) results in
\vspace{-0.2cm}
\begin{equation} \label{eq:n_HG}
n_{HG,j}=n^v_{HG,j}-P_{HG} n^v_{HG,j} \frac{\partial v_j}{\partial \mu_j}
=n^v_{HG,j}\left(1-P_{HG} \frac{\partial v_j}{\partial \mu_j} \right)
\;.
\end{equation}

After evaluating the effect of the varying volumes on entropy density and particle densities, we 
shall use this type of volumes to fit the lattice results \cite{lat2+1}\footnote{We use as a
reproduction of the lattice results the ansatz given in eq.~(16) and Table II in \cite{lat2+1} which is valid for $T\ge$100 MeV.}. Obviously, the lattice pressure at 
vanishing baryon density depends only on temperature. Thus, the hadron volumes we will use in 
the fitting procedure depend only on temperature, as well. This, indeed, will allow us 
to produce a HG curve which completely coincides with the lattice pressure
curve. Also, it will be possible to reproduce exactly the lattice entropy density curve with the 
use of eq.~(\ref{eq:s_HG}).

Our first attempt (case A) is to use one volume parameter, $v_0(T)$, for all hadrons. This 
volume is determined for every $T$ if we solve the equation:
\begin{equation}
\frac{3P_L(T;T_c)}{T^4}-\frac{3P_{HG}(T,\{\mu\}=0;v_0(T))}{T^4}=0\;\;.
\end{equation}
\noindent
Then, the corresponding HG entropy density can be calculated by eq.~(\ref{eq:s_HG}) with only 
one volume parameter available:
\vspace{-0.2cm}
\begin{equation} \label{eq:s_HGder}
s_{HG}=s^v_{HG}-P_{HG} \frac{v_0(T+\delta T)-v_0(T-\delta T)}{2\delta T} \sum_i n^v_{HG,i} \;,
\end{equation}
\noindent
where $\delta T$ is infinitesimal, $i$ runs over all hadrons and in the approximate calculation of 
the derivative $dv_0(T)/dT$ we have used terms up to the order of $\delta T^2$.

\begin{figure}[H]
\centering
\vspace{-0.5cm}
\includegraphics[scale=0.52,trim=0.5in 0.0in 0.5in 0.in,angle=0]{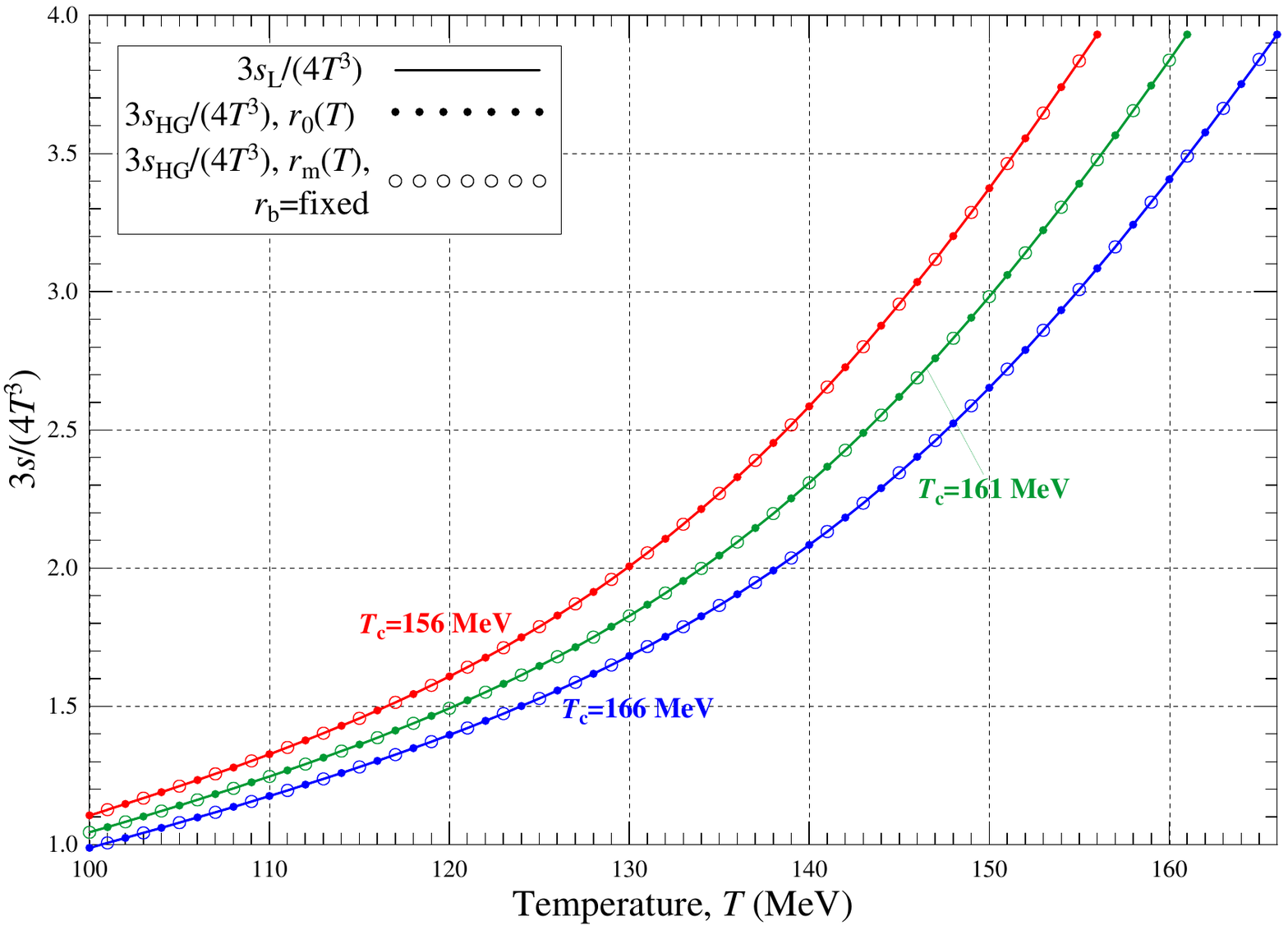}
\vspace{-0.3cm}
\caption{\label{fig:sL-sHG} {\small 
Comparison between the normalised entropy density, $3s/(4T^3)$, from lattice (continuous lines) and from Hadron Gas
(symbols). The entropy density of the Hadron Gas has been calculated by eq.~(\ref{eq:s_HG}) with a radius
dependent on temperature which is calculated by a fit on the lattice pressure. With filled circles is shown the case of a unique
radius, $r_0(T)$, for all hadrons. With open circles we depict the case of a radius $r_m(T)$ for mesons, while the radius for baryons is
held fixed ($r_b$=0.20 fm for $T_c$=156 MeV and $r_b$=0.365 fm for $T_c$=161 and 166 MeV). It is evident that the parameters
which fit the lattice pressure, also fit exactly the lattice entropy density.
}}
\end{figure}


The HG entropy density of eq.~(\ref{eq:s_HGder}) fits exactly $s_L$, as it is evident from 
Fig.~\ref{fig:sL-sHG}. Also, because $s_L > s^v_{HG}$, $P_{HG}>0$ and $n^v_{HG,i}>0$, we must 
have $dv_0(T)/dT<0$. Thus, the hadron volume decreases with the increase of temperature. In 
Fig.~\ref{fig:R0_Tc=154,157,161,166} we present the results for the hadronic radii calculated 
from
the fit to the lattice results for specific critical temperatures $T_c$: in (i) $T_c$=154 MeV, 
in (ii) $T_c$=157 MeV, in (iii) $T_c$=161 MeV and in (iv) $T_c$=166 MeV. In each graph the 
result corresponding to a common radius for all hadrons is shown with
line (1).$\;$

Our second try (case B) utilises two separate volume parameters, one for mesons $v_m(T)$ and 
one for baryons $v_b(T)$. We determine these two parameters for every $T$, so that the HG 
pressure and entropy density equals the relevant quantities of the lattice. Thus, we 
simultaneously solve the equations:
\begin{equation}
\frac{3P_L(T;T_c)}{T^4}-\frac{3P_{HG}(T,\{\mu\}=0;v_m(T),v_b(T))}{T^4}=0\;\;,
\end{equation}
\vspace{-0.5cm}
\begin{equation}
\frac{3s_L(T;T_c)}{4T^3}-\frac{3s^v_{HG}(T,\{\mu\}=0;v_m(T),v_b(T))}{4T^3}=0\;\;.
\end{equation}

This case is interesting because the HG entropy density for these specific volumes has the same 
functional dependence on the hadron eigenvolumes as the HG entropy density with constant 
volumes,
that is 
\[
s^v_{HG}(T,\{\mu\}=0;v_m(T),v_b(T))=
s_{HG}(T,\{\mu\}=0;v_m(T),v_b(T)).
\]
\vspace{-0.cm}
\noindent
Consequently, in view of eq.~(\ref{eq:s_HG}), this leads to:
\begin{equation} \label{eq:vmvbders}
n^v_{HG,m} \frac{d v_m}{d T}+n^v_{HG,b} \frac{d v_b}{d T}=0\;,
\end{equation}

\begin{figure}[H]
\centering
(i) \includegraphics[scale=0.6,trim=0.8in 0.8in 0.8in 0.5in,angle=0]{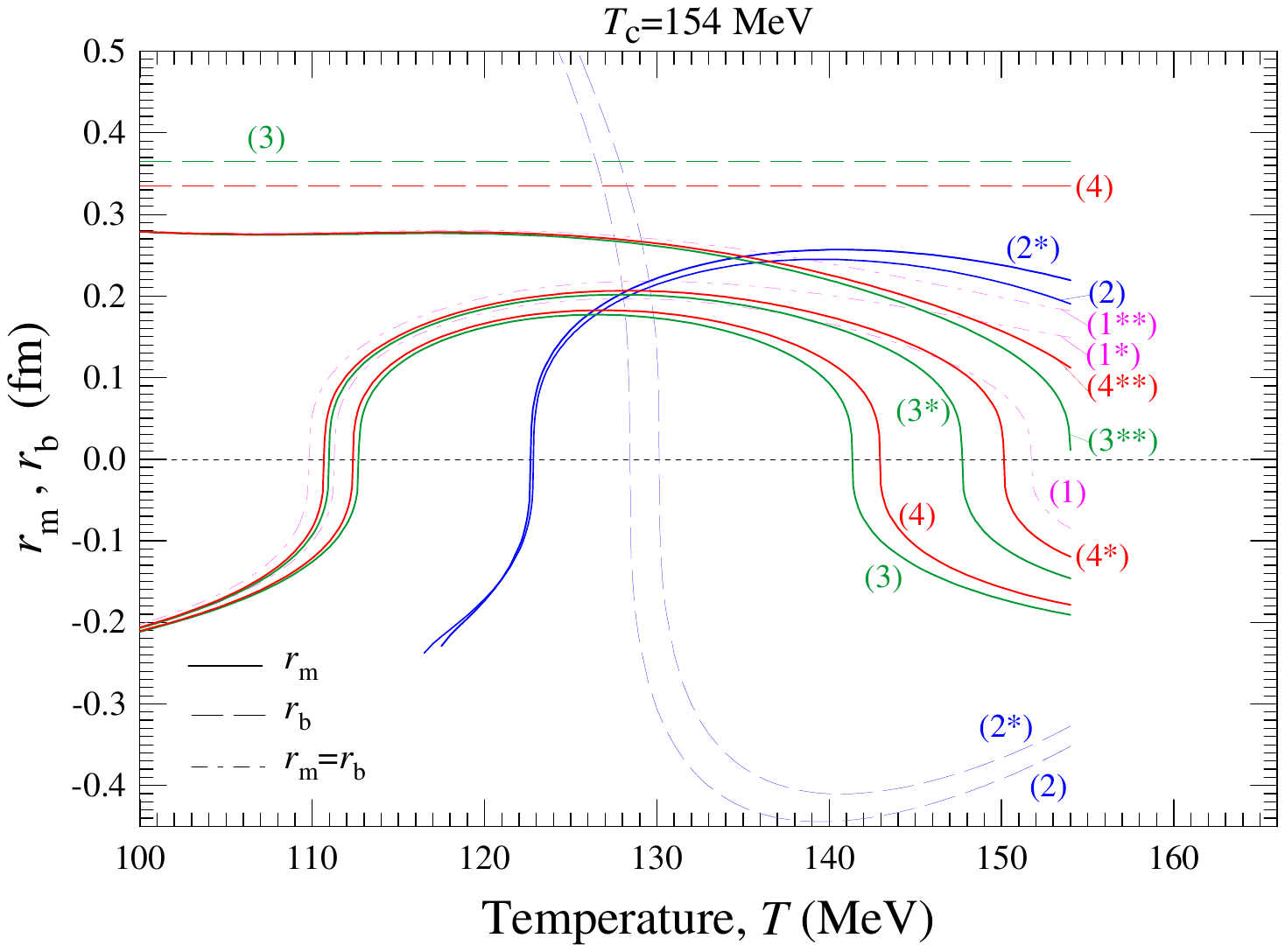}\\
(ii) \includegraphics[scale=0.6,trim=0.8in 0.8in 0.8in 0.in,angle=0]{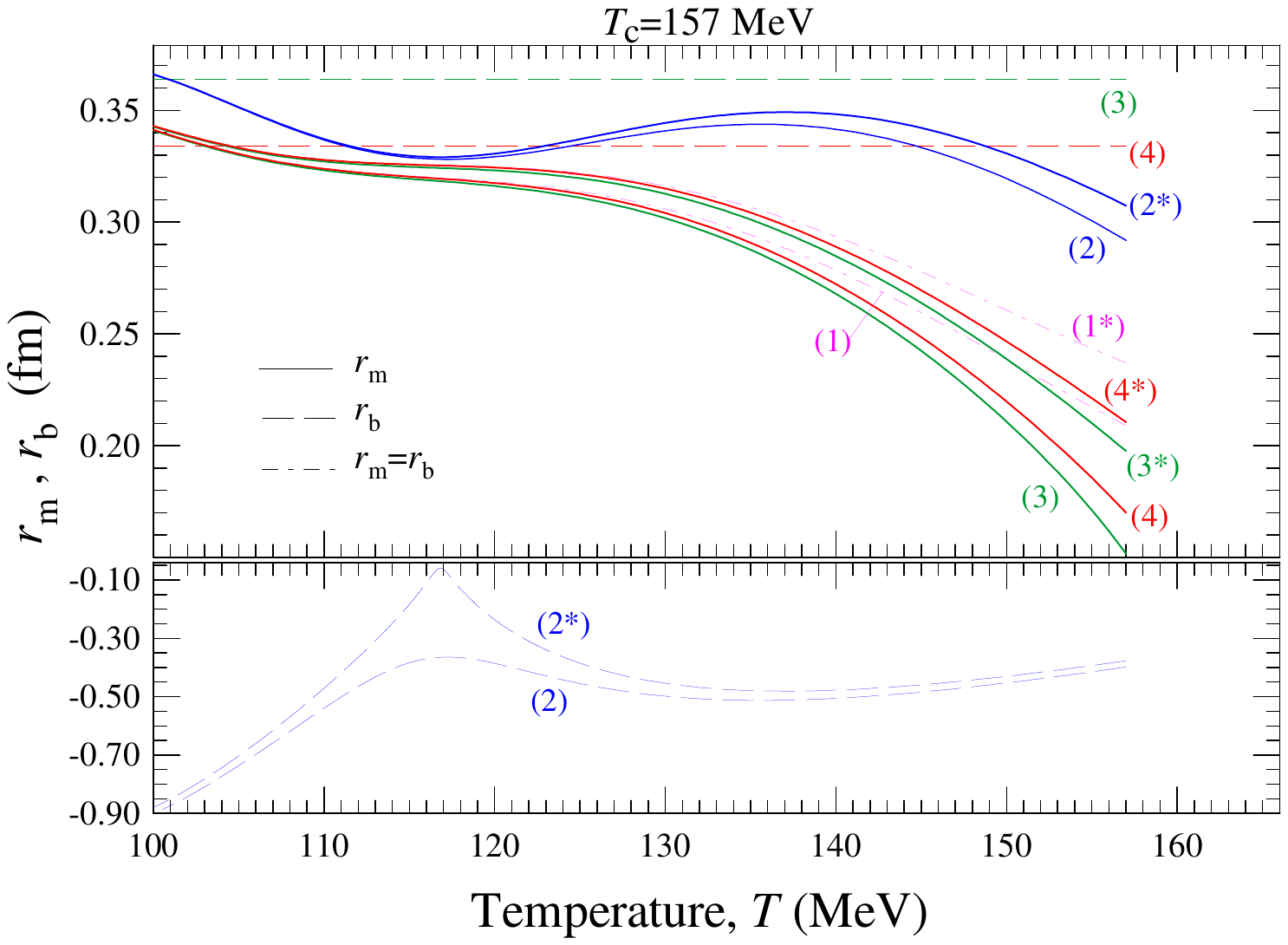}\\
\end{figure}

\begin{figure}[H]
\vspace{-0.5cm}
(iii) \includegraphics[scale=0.6,trim=0.8in 0.8in 0.8in 0.5in,angle=0]{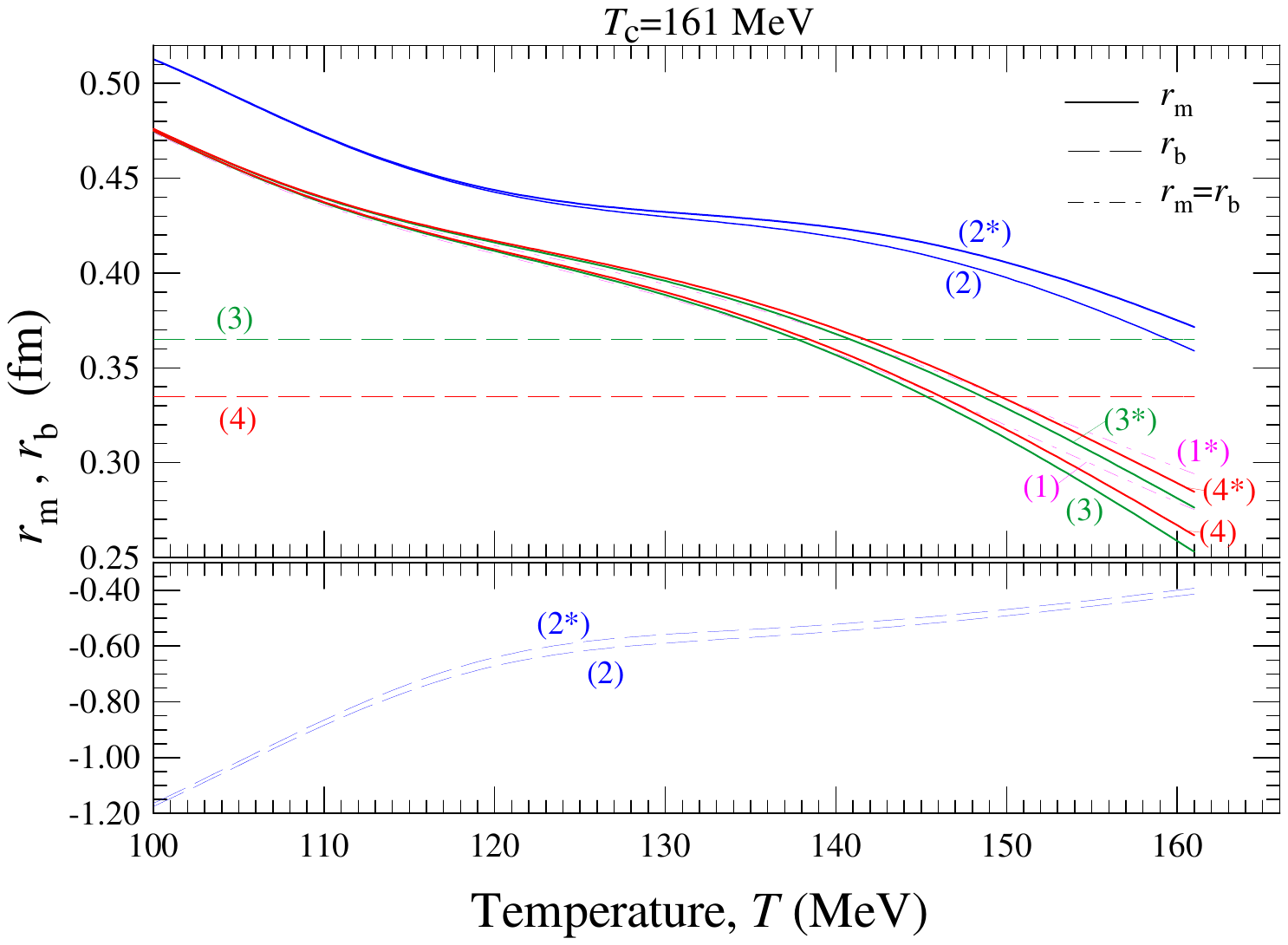}\\
(iv) \includegraphics[scale=0.6,trim=0.8in 0.8in 0.8in 0.in,angle=0]{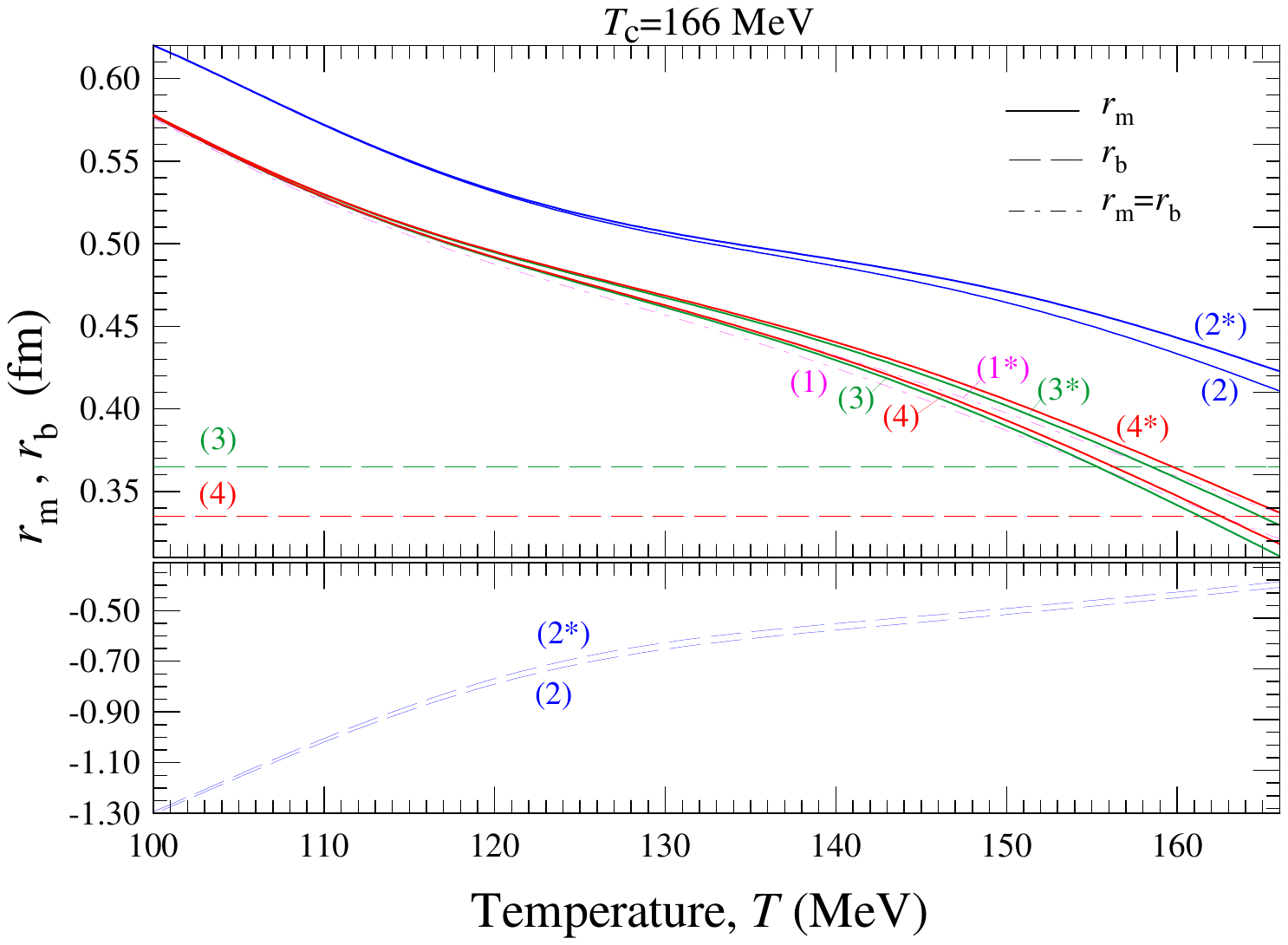}\\
\vspace{-0.3cm}
\caption{\label{fig:R0_Tc=154,157,161,166} {\small 
The solutions $r_m(T)$ and $r_b(T)$ obtained by equating the 
pressures, as well as, the entropy densities, of HG and 
lattice QCD. Dotted-slashed line (1), case A: a common radius for all hadrons.
Lines (2), case B: Continuous line: the radius for all mesons, $r_m(T)$. Slashed line: the 
radius for all baryons, $r_b(T)$. Lines (3), case C: Continuous line: the radius for all 
mesons, $r_m(T)$. Slashed line: the radius for all baryons, held fixed at $r_b=0.365$ fm (as in 
\cite{hadvol}). Lines (4), case C: the same as lines (3), but baryon radius fixed 
at $r_b=0.335$ fm. Each graph corresponds to a specific value for the critical QCD temperature 
$T_c$: (i) $T_c=$154 MeV, (ii) $T_c=$157 MeV, (iii) $T_c=$161 MeV and (iv) $T_c=$166 MeV.
Calculations with different hadron sets are displayed.}}
\end{figure}

\noindent
where $n^v_{HG,m}$ ($n^v_{HG,b}$) is the density of all mesons (baryons) in HG calculated for 
constant eigenvolumes. The last 
equation    
suggests
that when, at certain temperature intervals, $\frac{d v_m}{d T}>0$,
then $\frac{d v_b}{d T}<0$ and vice versa. Also, when $\frac{d v_m}{d T}=0$,
  then
$\frac{d v_b}{d T}=0$, meaning that $v_m(T)$ and $v_b(T)$ should reach simultaneously extremum values. Our results for this case are shown with lines (2) in graphs 
\ref{fig:R0_Tc=154,157,161,166}, where the properties of $v_m(T)$ and $v_b(T)$ which are 
inferred by eq.~(\ref{eq:vmvbders}) can be observed.

Our last effort (case C) is again with two volume parameters, but only with the meson volume 
depending on temperature, e.g.~$v_m(T), v_b$. The baryon eigenvolume is held fixed and the meson 
eigenvolume is determined by solving the equation:  
\begin{equation}
\frac{3P_L(T;T_c)}{T^4}-\frac{3P_{HG}(T,\{\mu\}=0;v_m(T),v_b)}{T^4}=0\;\;.
\end{equation}

\noindent The HG entropy density can be calculated by eq.~(\ref{eq:s_HG}) and equals to the 
corresponding lattice quantity.
The results of this case are shown in graphs \ref{fig:R0_Tc=154,157,161,166} with lines (3), 
where the baryon radius is fixed at $r_b=0.365$ fm (which is the mean baryon radius cited in 
\cite{hadvol}) and with lines (4), where the baryon radius is fixed at $r_b=0.335$ fm (which is 
the lower baryon radius cited in \cite{hadvol}).

Comparing, now, the results from cases A-C, we see that case B leads to baryon volumes which are 
negative in almost all the temperature range and for all $T_c$. 
These results are similar to the case of temperature independent 
baryon volumes determined by the fit.
Again we find that if $v_b$ is left as a free parameter, 
it is easily shifted to negative, unphysical values. So we have to reject this approach. On the 
contrary, fixing the baryon volume at a certain (positive) value, independently of  our choice, allows us to 
solve for the meson volume $v_m(T)$, so that the HG pressure equals the lattice pressure at 
every temperature point. 
The relevant entropy density can be calculated by eq.~(\ref{eq:s_HGder}), where $i$ 
runs
over mesons only. In Fig.~\ref{fig:sL-sHG} it is shown that the Hadron Gas entropy density fits 
exactly the lattice entropy density, as well.

For $T_c=154$ MeV and for cases A and C, the fitted radius partly falls to the unphysical
negative domain for the (vh) set and the set (*). However, the physical
positive values are recovered for the set (**).

\section{Locating the Critical Point through volume models guided by lattice QCD} \label{sec:loccp}

In this section we will utilise HG models, using
hadronic volumes extracted from lattice QCD results,
in order to describe 
the transition of the hadron phase to the chiral phase in the 
entire baryon chemical potential - temperature plane. Our goal is to determine the location of the critical endpoint of the first 
order transition line indicating the entrance into the smooth crossover region. 
We shall present calculations for the location of the critical point for the five hadron sets described in Section \ref{sec:constr}.


\subsection{Critical Point with single Constant Hadron Volume}\label{subsec:v.m.a}

We first consider the model which assumes that all particles have the same eigenvolume $v_0$, 
connected with the hard-core particle radius $r_0$, which is constant with respect to 
temperature and 
chemical potentials. Its value, however, will be determined, for a specific choice of $T_c$, by 
a fit on the lattice 
pressure curve, as in Section \ref{sec:constvary} (through minimisation of 
$\chi^2$ given by eq.~(\ref{eq:chi2})). We will refer to this choice as volume model (a).

The use of a single volume parameter for all hadrons has, also, been used elsewhere. In 
\cite{equalr0_1,equalr0_2,equalr0_3,equalr0_4} the eigenvolume was taken equal for all baryons and mesons. 
In these works the radius for baryons
was given by the hard-core repulsive interaction as 
extracted from nucleon-nucleon scattering \cite{equalr0_nn}, while radius values for other 
baryons were taken similar. For mesons, in the absence of detailed information on
their 


\begin{figure}[H]
\centering
(i)\includegraphics[scale=0.30,trim=0.8in 0.8in 1.in 0.2in,angle=0]{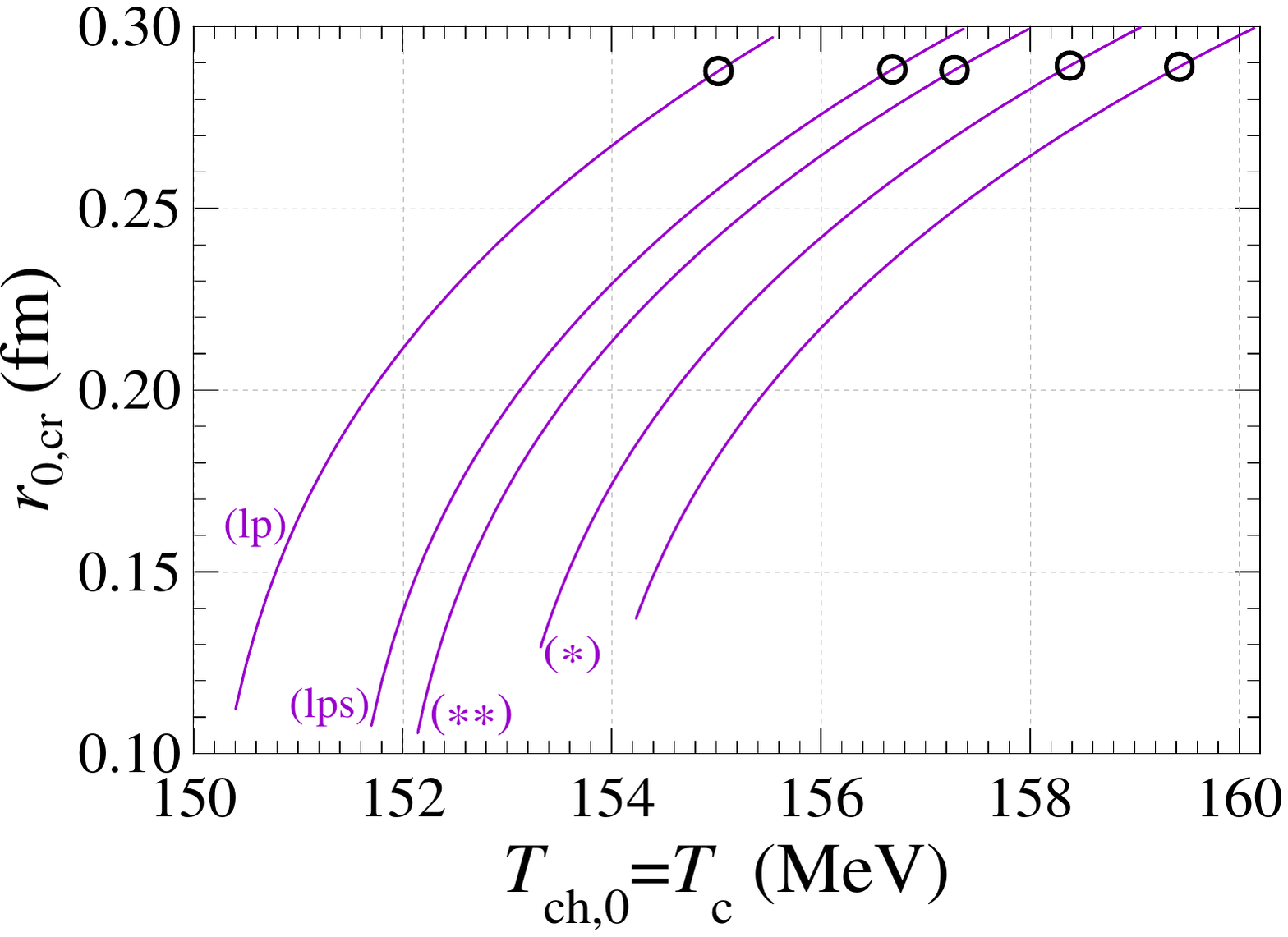}
(ii)\includegraphics[scale=0.30,trim=0.8in 0.8in 1.in 0.2in,angle=0]{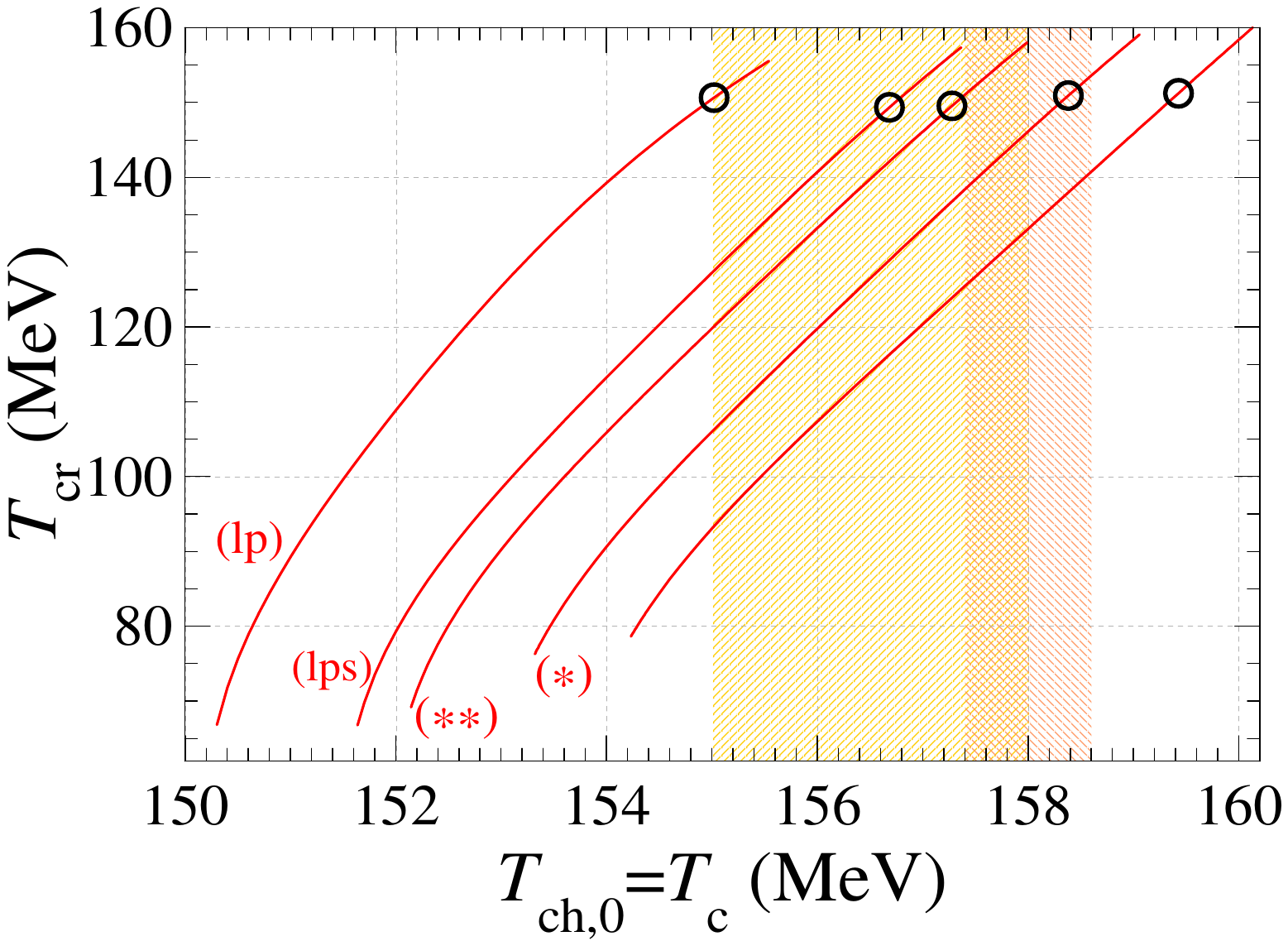}\\
(iii)\includegraphics[scale=0.30,trim=0.8in 0.8in 1.in 0.2in,angle=0]{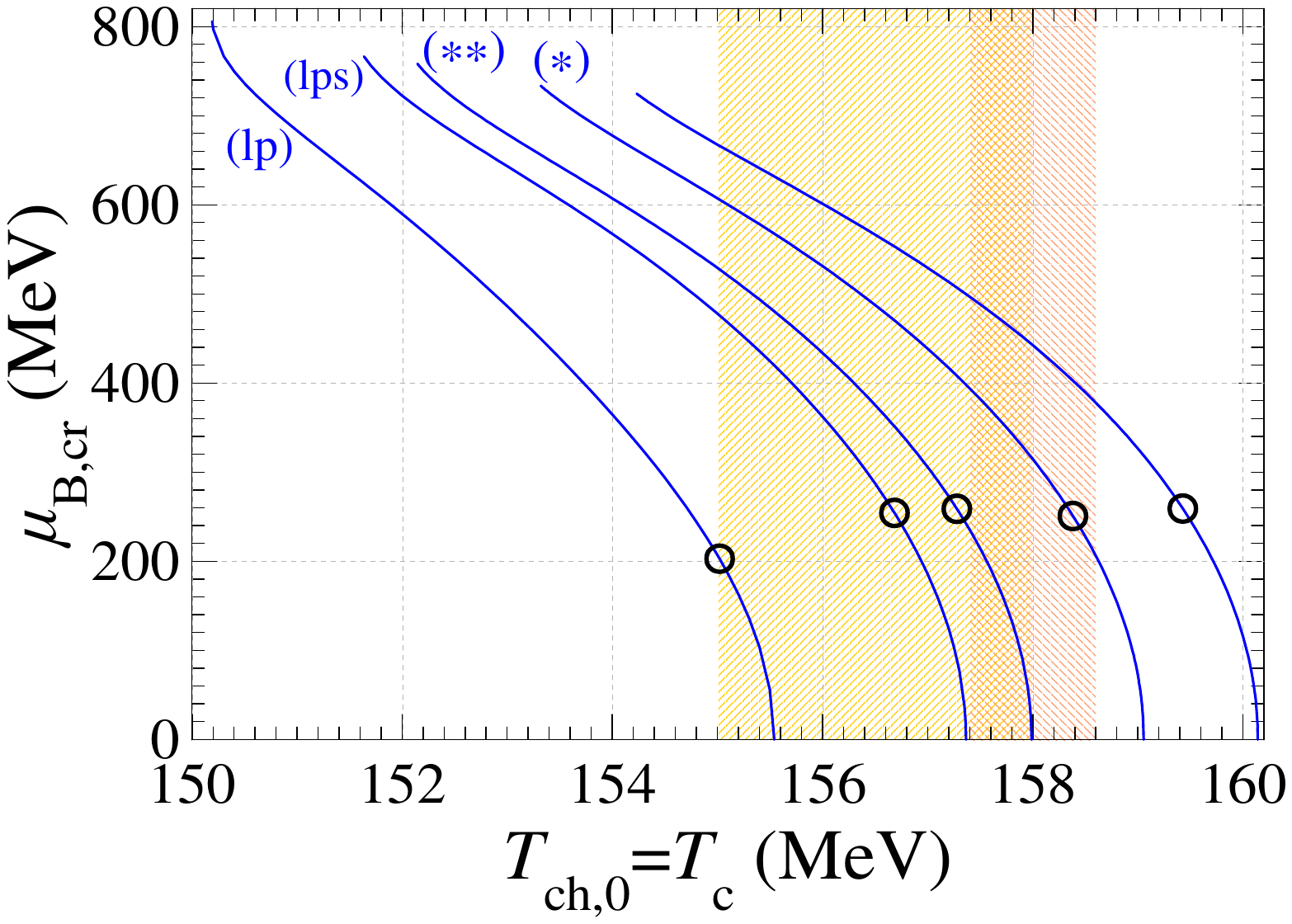}
(iv)\includegraphics[scale=0.30,trim=0.8in 0.8in 1.in 0.2in,angle=0]{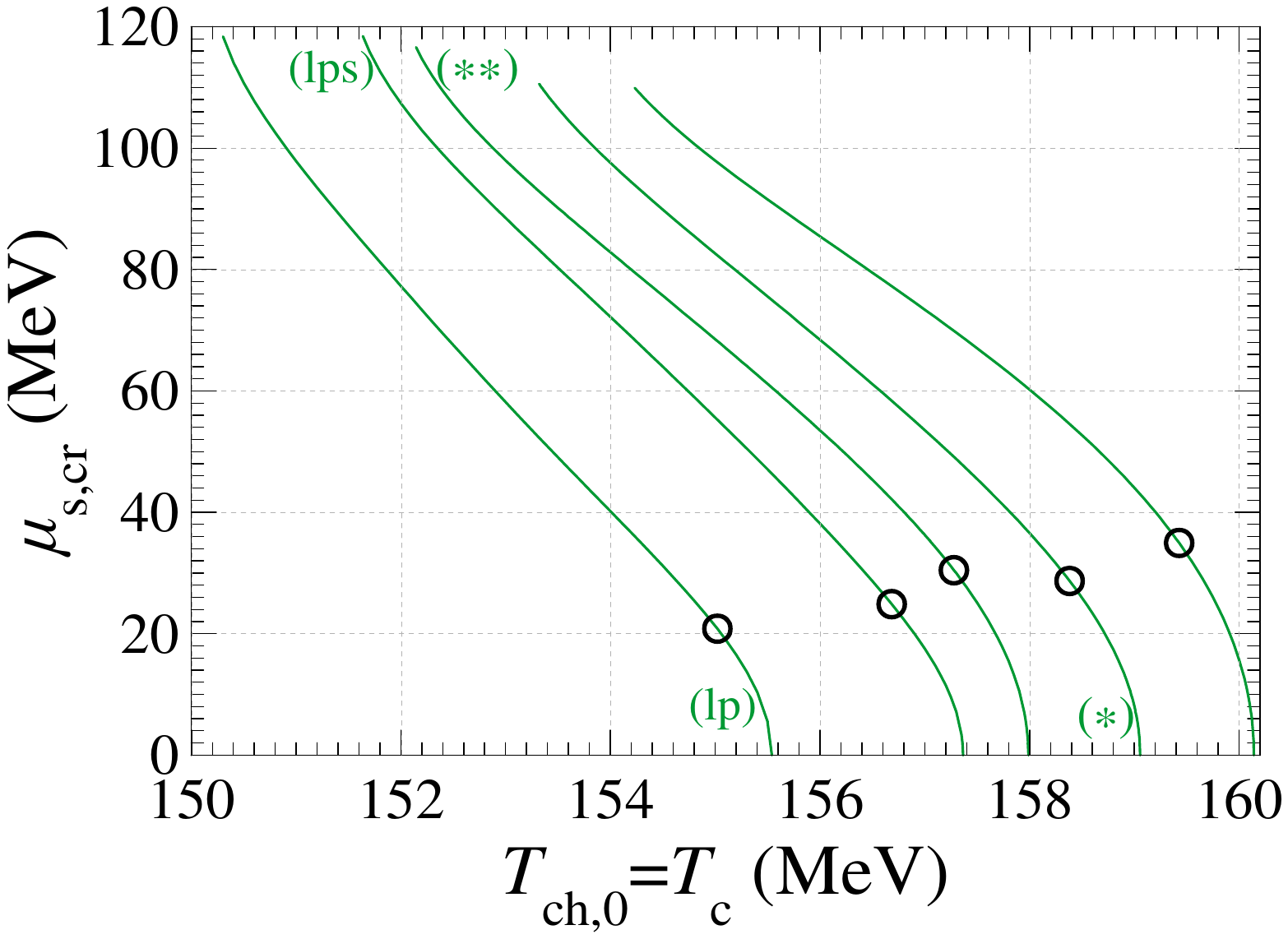}\\
(v)\includegraphics[scale=0.30,trim=0.8in 0.8in 1.in 0.2in,angle=0]{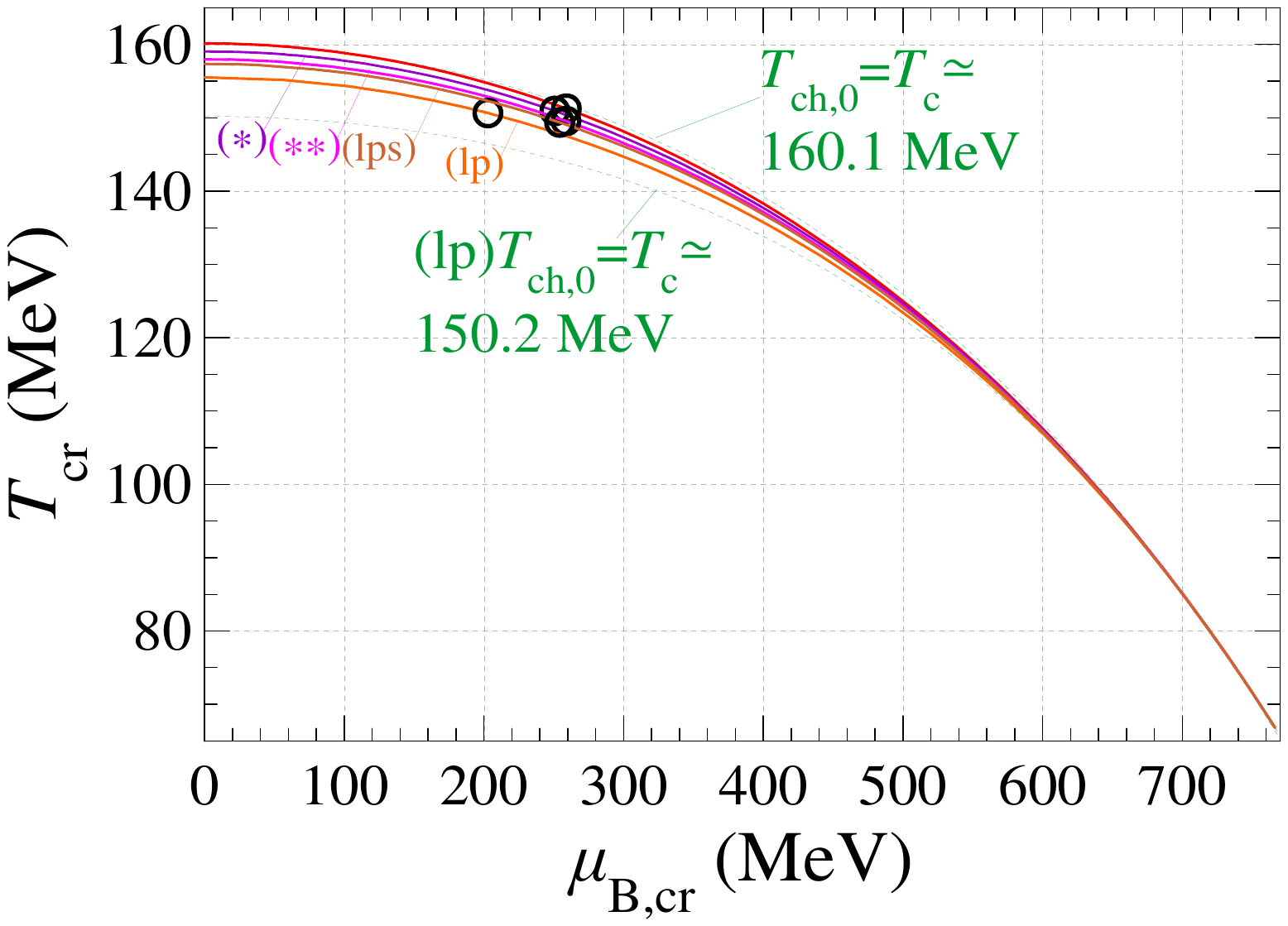} 
(vi)\includegraphics[scale=0.30,trim=0.8in 0.8in 1.in 0.2in,angle=0]{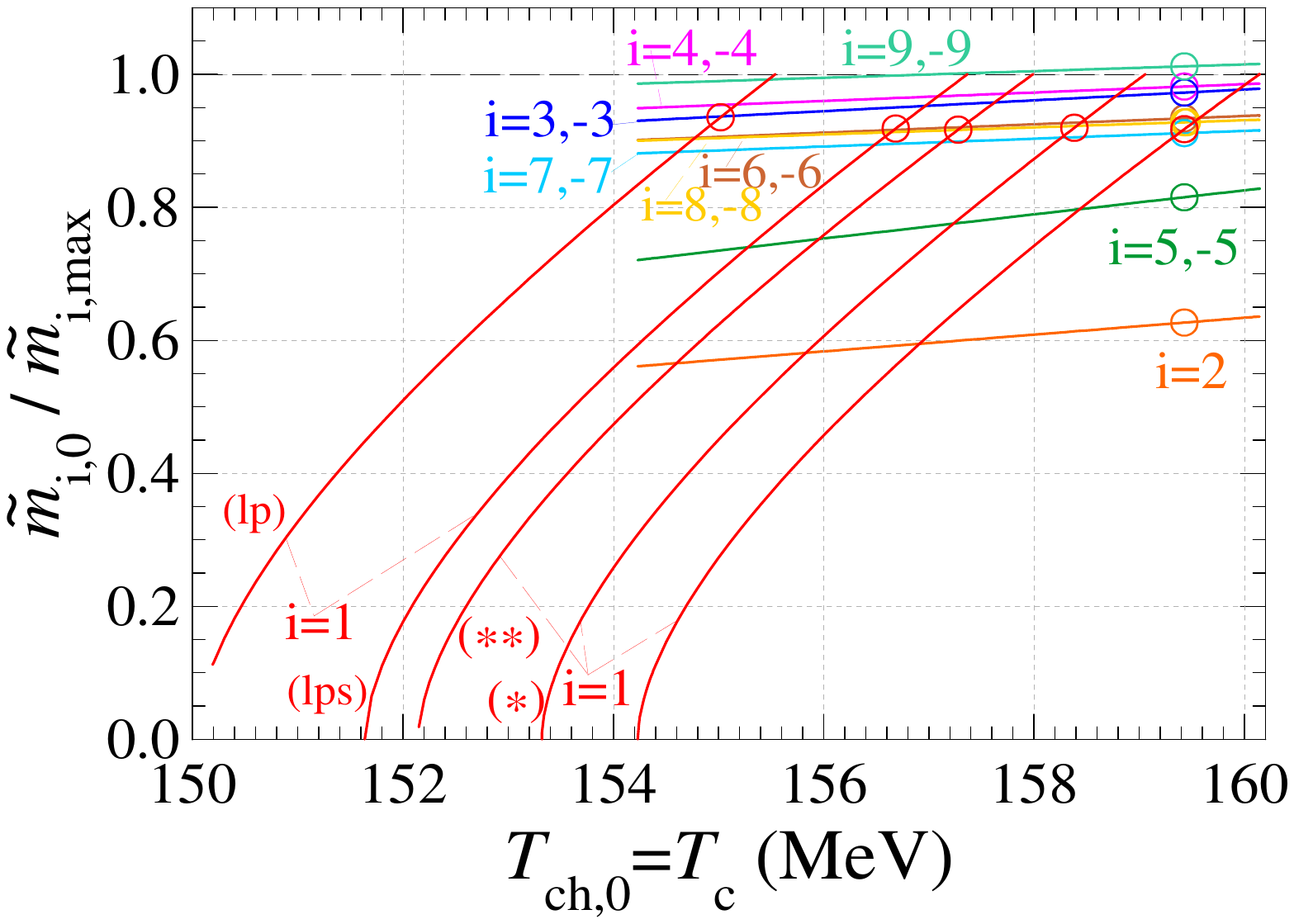}
\vspace{0.5cm}
\caption{\label{fig:cpr0} {\small  The position of the critical point determined for different values of critical lattice 
temperatures, $T_c$, which is always taken equal to the freeze-out
temperature at $\mu_B=0$, $T_{ch,0}$ and for 5 hadron sets. Calculations are carried out in the interval 
$T_c\simeq$(150.2-160.1) MeV. The critical point always resides on the corresponding freeze-out 
curve. Calculations involve constant common hadron radius $r_0$, determined by the lattice 
pressure at $\mu_B=0$ (volume model (a)).
(i) The value of the common hadron radius on the critical point, $r_0$, as function of 
$T_c=T_{ch,0}$. 
(ii) The temperature of the critical point, $T_{cr}$, as function of $T_c=T_{ch,0}$.
(iii) The baryon chemical potential of the critical point, $\mu_{B,cr}$, as function of 
$T_c=T_{ch.0}$.
(iv) The strange quark chemical potential of the critical point, $\mu_{s,cr}$, as function of 
$T_c=T_{ch,0}$.
(v) The location of the critical point, $T_{cr}-\mu_{B,cr}$, for different values of 
$T_c=T_{ch,0}$. The freeze-out curves which correspond to the maximum value of $T_c=$160.1 MeV 
of the (vh) set
and the minimum value of $T_c=$150.2 MeV of the (lp) hadron set are shown.
(vi) The solution for the chiral mass for each family at $\mu_B=$0 MeV as function of 
$T_c=T_{ch,0}$. This chiral mass for the pion family becomes zero at a certain $T_c$ and so, no 
real solutions exist for lower values of $T_c$. In all graphs with circle we present the 
critical point which additionally fulfils the criterion described in 
Section \ref{sec:crit}.
In graphs of $T_{cr}$ and $\mu_{B,cr}$ as function of $T_c$ we depict regions
compatible with $T_c$=156.5$\pm$1.5 MeV reported in \cite{lat_Tc1} (shaded area with
left limit at 155 MeV) and with with $T_c$=158$\pm$0.6 MeV reported in \cite{lat_Tc2} 
(shaded area with right limit at 158.6 MeV).
}}
\end{figure}


\begin{figure}[H]
\centering
(i)\hspace{-2cm}\includegraphics[scale=0.65,trim=0.5in 0.8in 2in 0.5in,angle=0]{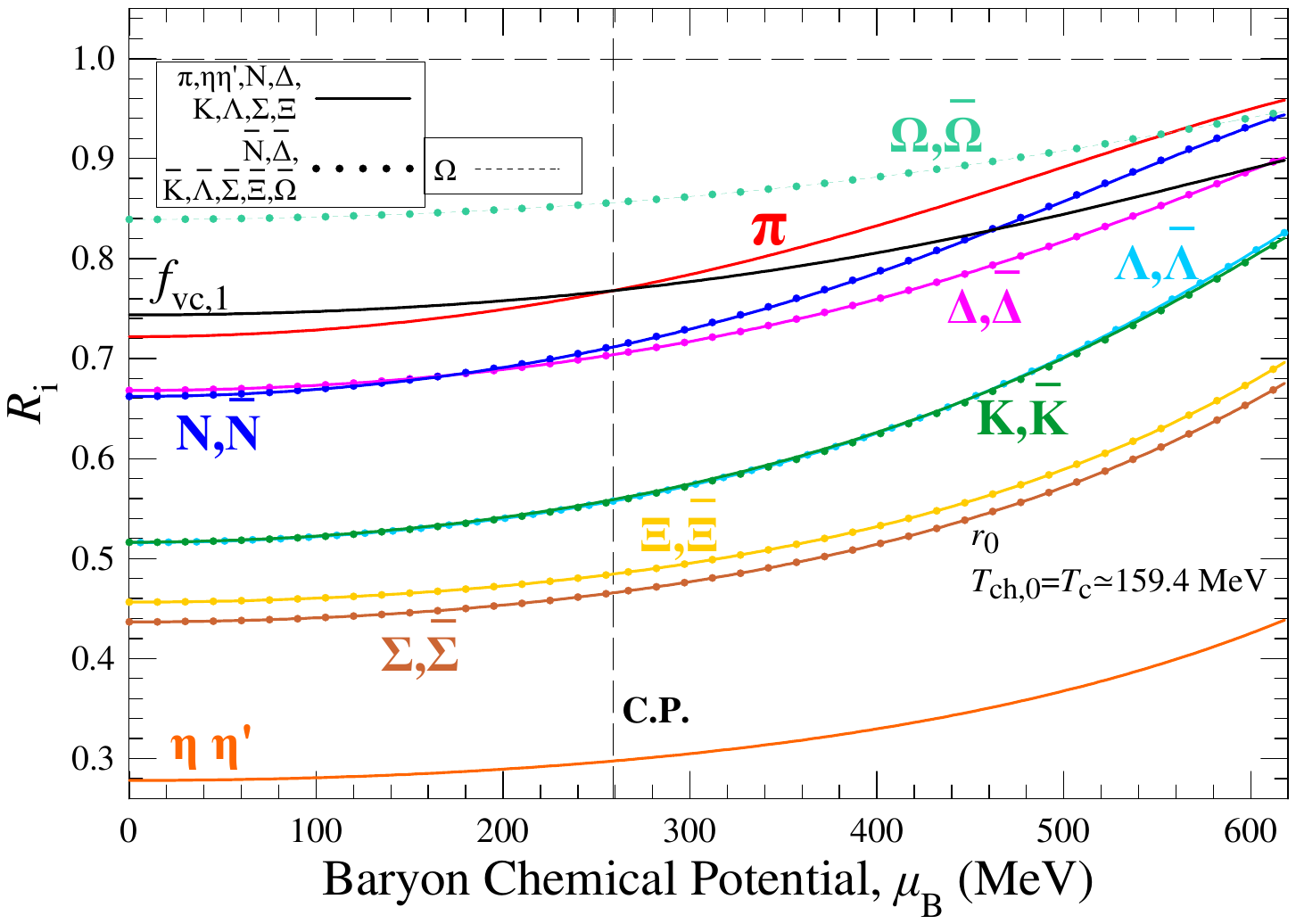}\\
(ii)\hspace{-2cm}\includegraphics[scale=0.65,trim=0.5in 0.8in 2in 0.5in,angle=0]{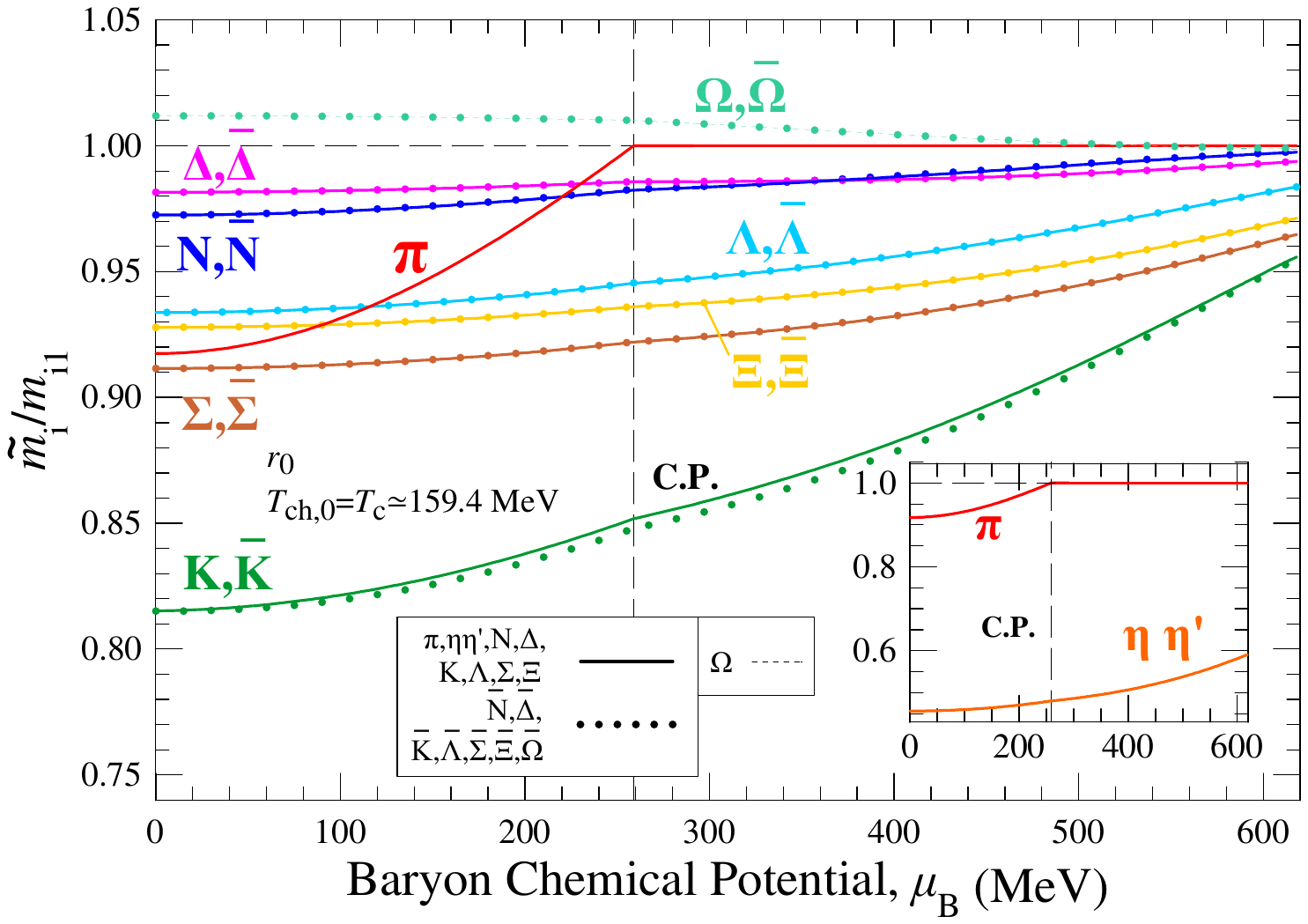}
\end{figure}

\begin{figure}[H]
\centering
(iii)\hspace{-1cm}\includegraphics[scale=0.75, trim=0in 1.3in 0in 1.in]{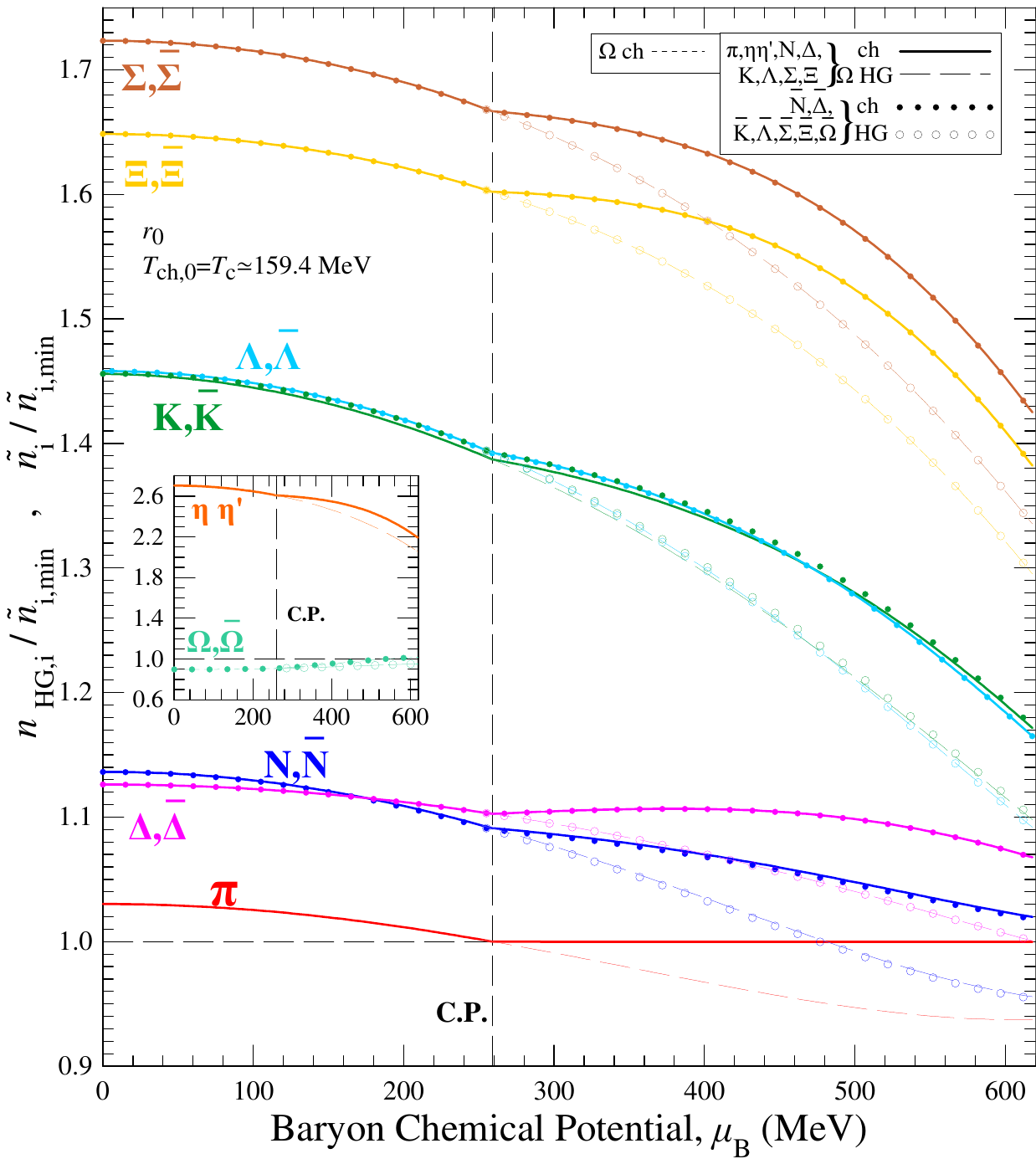}\\
\caption{\label{fig:Tc=159.4} {\small 
Calculations for $T_c=T_{ch,0}\simeq$159.4 MeV and constant common hadron radius, 
$r_0=$0.289 fm, determined by the lattice pressure for $\mu_B=0$, for the (vh) set.
(i) The ratios $R_i$ for the families $i$ as function of $\mu_B$, calculated exactly for these 
specific values of $T_c$ and $r_0$.
Also, the volume correction factor for the pion family,
$f_{vc,1}$, which intersects with $R_1$ at $\mu_B\simeq$258.9 MeV and produces at this 
location the critical point is shown.
(ii) The ratio of the chiral mass for the $i$ family to the respective maximum allowed value, as 
function of $\mu_B$. With continuous lines we depict calculations for particle families, while 
with solid circles we depict calculations for antiparticle families.
(iii) The ratio of the HG or chiral family density to the minimum chiral density for the 
same conditions as function of $\mu_B$. In the crossover region the hadron gas and chiral 
densities are equal, corresponding, thus, to the same curve, while they differ in the 1st order 
transition regime. Densities of particle families are depicted with lines, continuous for the 
chiral case and slashed for the HG case. Densities of antiparticle families are depicted with 
circles, solid for the chiral case and open for the HG case.
The case of the $\Omega$ family is shown with dotted line to demonstrate the
irregular behaviour of the $\Omega,\bar{\Omega}$ families.
}}
\end{figure}
\vspace{0.3cm}

\noindent
interactions at short distance, it was assigned the same radius value, based on the 
similarity of the meson charge radii
compared to baryons and on the energy dependence of the pion-nucleon phase shifts 
\cite{equalr0_pn}.

The choice of eq.~(\ref{eq:freezeout}) as the transition curve, fixes the temperature, $T$, for 
given $T_c,\;\mu_B$. To determine the strange quark chemical potential we apply the zero strangeness 
condition in the HG phase:
\begin{equation} \label{eq:S=0}
\left< S \right>_{HG} (T,\mu_B,\mu_s;T_c) =0 \;\;.
\end{equation}
Last equation can be solved to determine $\mu_s$ for given values of $T_c$, $T$ and $\mu_B$. 
Eqs.~(\ref{eq:freezeout}), (\ref{eq:S=0}) enable us to depict our calculations as function of 
$\mu_B$ alone, for a specific value of $T_c$. The conservation of particle numbers insures the 
strangeness neutrality in the chiral phase, as well.

We proceed by considering the dependence of the position of the critical point on $T_c$, i.e. 
the lattice QCD critical temperature at $\mu_B=0$. In Fig.~\ref{fig:cpr0}(i) we plot the 
calculated radius $r_0$ as function of $T_c$. In Figs.~\ref{fig:cpr0}(ii)-(iv) we show the 
position of the critical point ($T_{cr},\mu_{B,cr},\mu_{s,cr}$, respectively) as a function of 
$T_c$.
In Fig.~\ref{fig:cpr0}(v) we depict the location of the critical point in the ($T,\mu_B$) plane. 
We observe that for the (vh) set and for value $T_c \simeq $160.1 MeV ($r_0 \simeq $0.300 fm) the critical point is 
located at zero baryon chemical potential, while it ceases to exist for higher $T_c$ (or $r_0$) values. 
Decrease of $T_c$ (and the radius $r_0$) shifts the critical point at higher values of chemical 
potential $\mu_B$. 

We have to impose the condition that the solution for the chiral masses is positive for all the 
families. The lowest values for these masses are found for $\mu_B=0$ and these values decrease 
with decreasing $T_c$ (or $r_0$). We find that the chiral mass of the pion family reaches the zero 
value, as $T_c$ decreases, while the rest of the chiral masses are still positive. 
This imposes a constraint on $T_c$, leading to 
the lowest allowed value of $T_c$. However, it may happen that our solutions for the critical point
parameters may lead to a minimum value of $T_c$, where the pion chiral mass at $\mu_B=0$
still attains positive but non-zero values. This condition delimits the solution space 
for the relevant parameters.
The lowest allowed value of $T_c$ for the (vh) set is $\sim$154.2 MeV ($r_0\simeq$0.137 fm) which in turn gives an upper 
value for the position of the critical point at $\mu_{B,cr}\simeq$725 MeV. These findings 
provide a lower limit for the critical temperature $T_{cr}\simeq$78.6 MeV. Thus, our treatment 
excludes scenarios with very low critical temperatures \cite{cplowT}. 
In Fig.~\ref{fig:cpr0}(v) 
we present the solution at zero baryon density for the chiral mass of each family (as the ratio 
to the higher allowed chiral mass of the family) with varying $T_c$.

In summary, for a universal hadronic volume remaining constant with temperature,
after considering all five hadron sets, the critical 
point can be located in the range of $\mu_B\sim$ (805-0) MeV, which corresponds to range of 
$T_c \sim$ (150.2-160.1) MeV and mean hadronic radius in the range of $r_0\sim$ (0.069-0.300) fm. Also 
in Figs.~\ref{fig:cpr0} we depict by open circle the critical point, determined by the criterion 
which will be presented in detail 
in Section \ref{sec:crit}. This critical point for the four first hadron sets 
corresponds to $T_c \simeq$156.7-159.4 MeV ($r_0\simeq$0.288-0.289 MeV) and it is 
located at 
$\mu_{B,cr}\simeq$250.6-258.9 MeV and $T_{cr}\simeq$149.3-151.2 MeV.
For the (lp) hadron set we find that the solution which satisfies this criterion
is located to lower $\mu_{B,cr}\simeq$202.6 MeV. 

We shall finish this subsection by presenting full calculations of the quantities involved for 
the solution for this critical point and for the (vh) set. 
In Fig.~\ref{fig:Tc=159.4}(i) we present the ratios $R_i$. The intersection of the volume 
correction factor of the pion family, $f_{vc,1}$, with the corresponding ratio, $R_1$, determines 
the position of the critical point. In Fig.~\ref{fig:Tc=159.4}(ii) we present the masses that 
solve eq.~(\ref{eq:HGchnBF}) for every family $i$. Since the masses differ considerably among 
the families, we choose to depict the ratio of the chiral mass to the respective mass of the 
HG ground state. As it can be observed, the chiral mass of the pion family starts at its minimum 
value at zero baryon chemical potential, it reaches the maximum value at $\mu_{B,cr}$ and then 
remains constant. The chiral masses of the rest of the families in general increase with the 
increase of $\mu_B$.
A general observation is that all the masses tend, for zero temperature (maximum value of 
$\mu_B$), to the physical mass of the lowest mass hadron in each family. Another observation is 
that the solution for the particle family differs slightly from antiparticle family solution.

In Fig.~\ref{fig:Tc=159.4}(iii) we present the calculations for the densities 
of each family in 
the HG and chiral phase. 
To display values in wide range, the family densities are divided by 
the respective minimum density (maximum value of chiral mass) at the chiral phase for the same 
conditions of temperature and chemical potentials, $\tilde{n}_{i,min}$. For $\mu_B < \mu_{B,cr}$ 
(crossover region) the densities of the HG and the chiral phase are equal (the slashed and 
continuous  lines are identical at this region). For $\mu_B > \mu_{B,cr}$ (1st order transition) 
the density in the HG phase is lower than the density in the chiral phase, since the volume 
expands as the system crosses from chiral to HG phase. This is evident by the fact that the HG 
curves (slashed lines) are lower than chiral curves (continuous lines).
The densities between particle-antiparticle (with the respective normalization) are almost 
equal. All the densities for low temperatures tend to the respective $\tilde{n}_{i,min}$.

Also, Figs.~\ref{fig:Tc=159.4}(ii)-(iii) confirm that the $\Omega,\bar{\Omega}$ families exhibit 
irregular behaviour: the chiral masses of these families attain higher values than the maximum 
allowed ones and the HG
densities for these families at the crossover region are lower than the minimum chiral density 
values. Both of these facts would be remedied if additional mass spectra were present in the 
$\Omega,\bar{\Omega}$ families.

\subsection{Critical Point with single temperature dependent hadron volume} \label{subsec:v.m.b}

The common hadron volume we used in the previous section fails to fit exactly the lattice QCD 
pressure and entropy density,
as it was discussed in Subsection \ref{subsec:const}. In this section we shall use the approach 
of Subsection \ref{subsec:vary} and we shall try to locate the critical point using the same 
radius for all hadrons, which now depends on temperature, $r_0(T)$. This radius is fixed by the 
fit on the lattice pressure at vanishing baryon density. However, the assumption that $r_0(T)$ 
is 
independent of the chemical potentials enable us to transfer the value we have determined from 
the lattice results at $\mu_B=0$
to every value of baryon chemical potential. This will be referred to as volume model (b). 
For the transition curve we use the parametrisation of eq.~(\ref{eq:freezeout}), leaving $T_c$ as a 
free parameter.

We calculate the position of the critical point,
 i.e.~the value of the critical parameters 
($T_{cr}$, $\mu_{B,cr}$, $\mu_{s,cr}$) by solving the following set of equations:
\begin{equation} \label{eq:PLPHG_r0}
P_L(T;T_c)=P_{HG}(T,\mu_B=0,\mu_s=0;r_0(T))\;,
\end{equation}
\begin{equation} \label{eq:den1_1}
\sum_j n_{HG,1j}(T,\mu_{B},\mu_s ; r_0(T)) =
n^{pt}_{HG,11}(T,\mu_{B},\mu_s ; \tilde{m}_1=m_{\pi}) \;,
\end{equation}
\begin{equation} \label{eq:freezeout_1}
T=f_{fr} (\mu_B;T_c) \;,
\end{equation}
\begin{equation} \label{eq:S=0_1}
<S>(T,\mu_{B},\mu_s)=0 \;.
\end{equation}

\begin{figure}[H]
\centering
(i)\includegraphics[scale=0.30,trim=1.3in 0.8in 1.in 0.2in,angle=0]{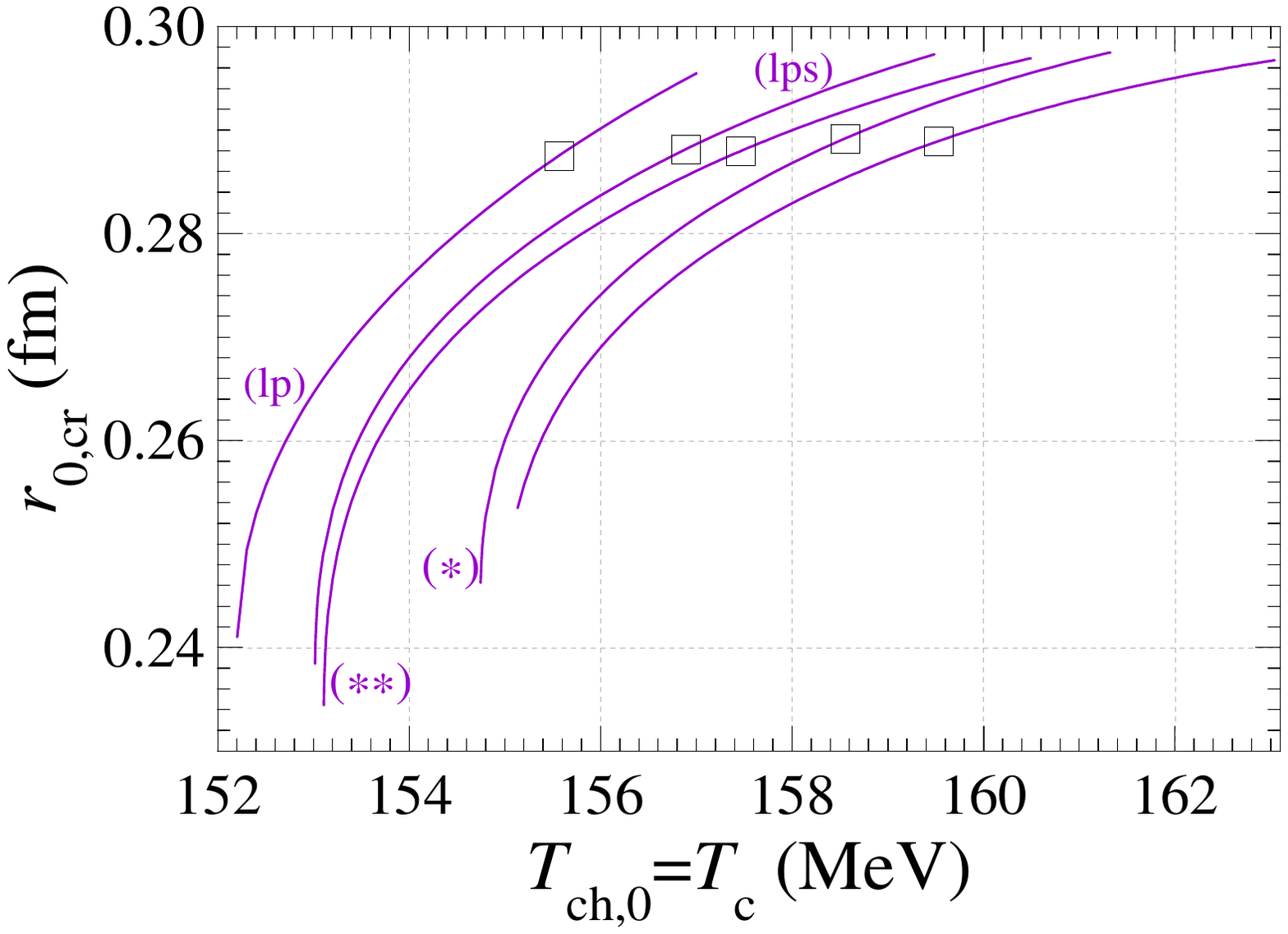} \hspace{0.15cm}
(ii)\includegraphics[scale=0.30,trim=0.8in 0.8in 1.in 0.2in,angle=0]{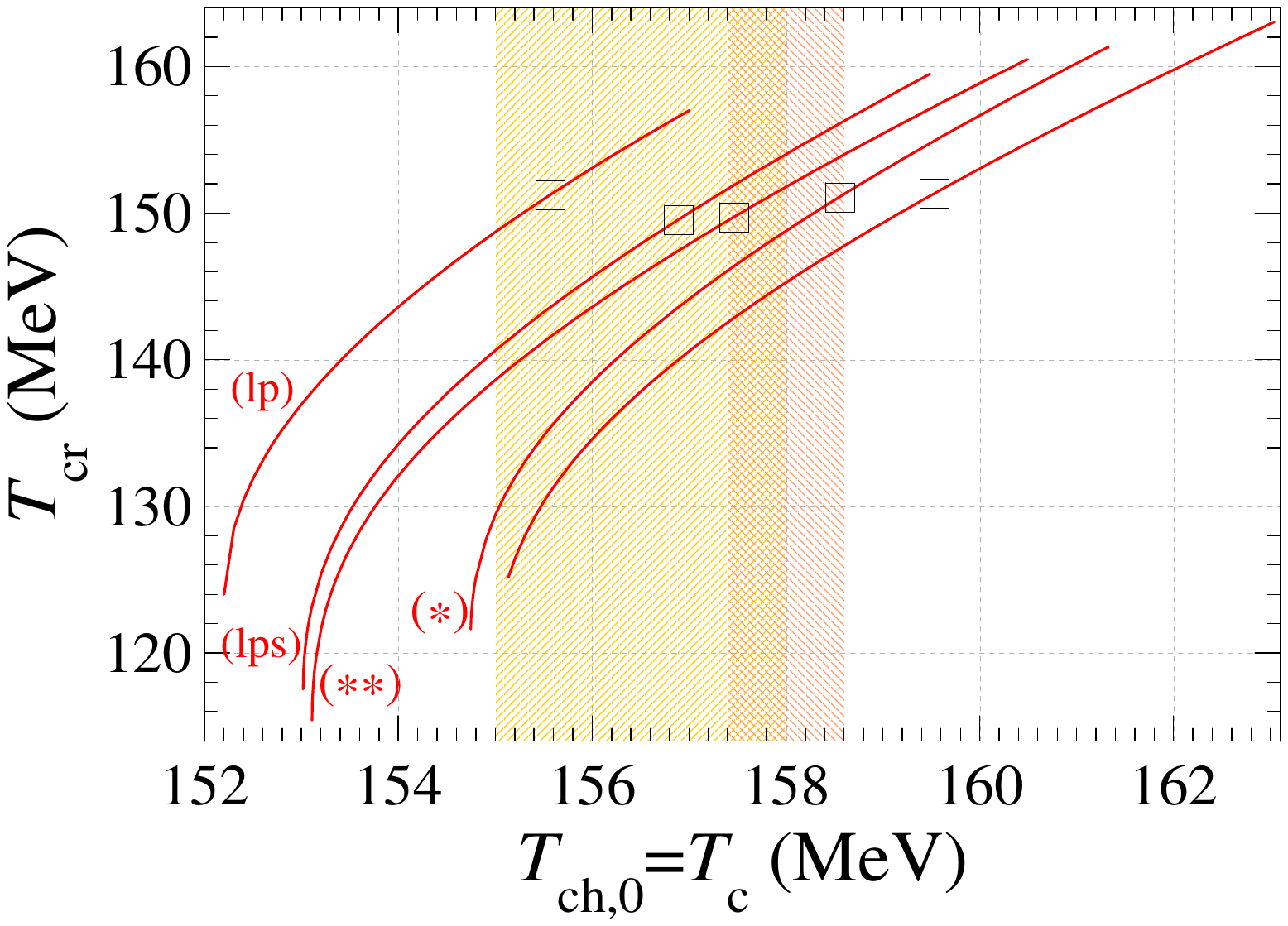}\\
(iii)\includegraphics[scale=0.30,trim=0.8in 0.8in 1.in 0.2in,angle=0]{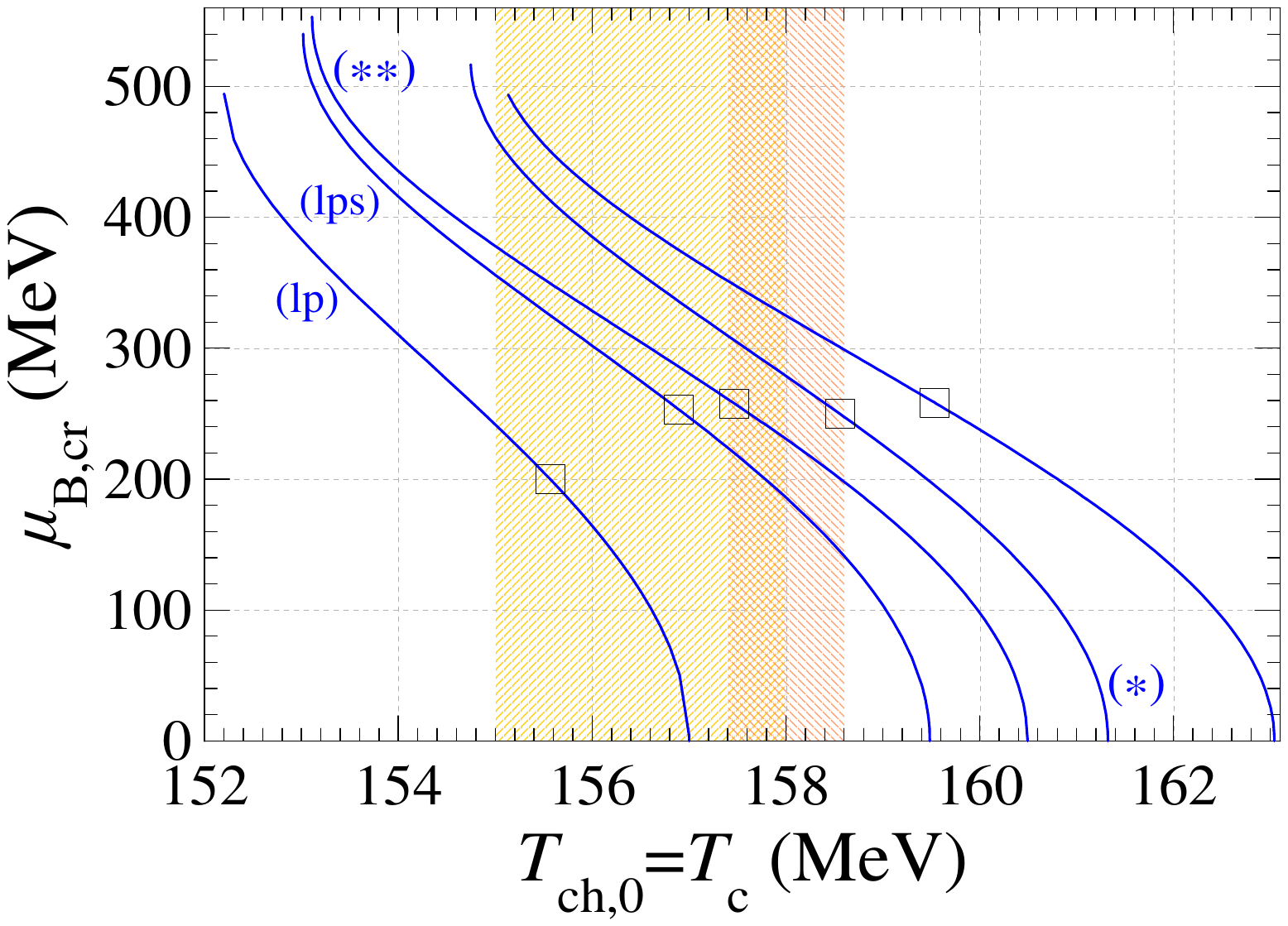}
(iv)\includegraphics[scale=0.30,trim=0.8in 0.8in 1.in 0.2in,angle=0]{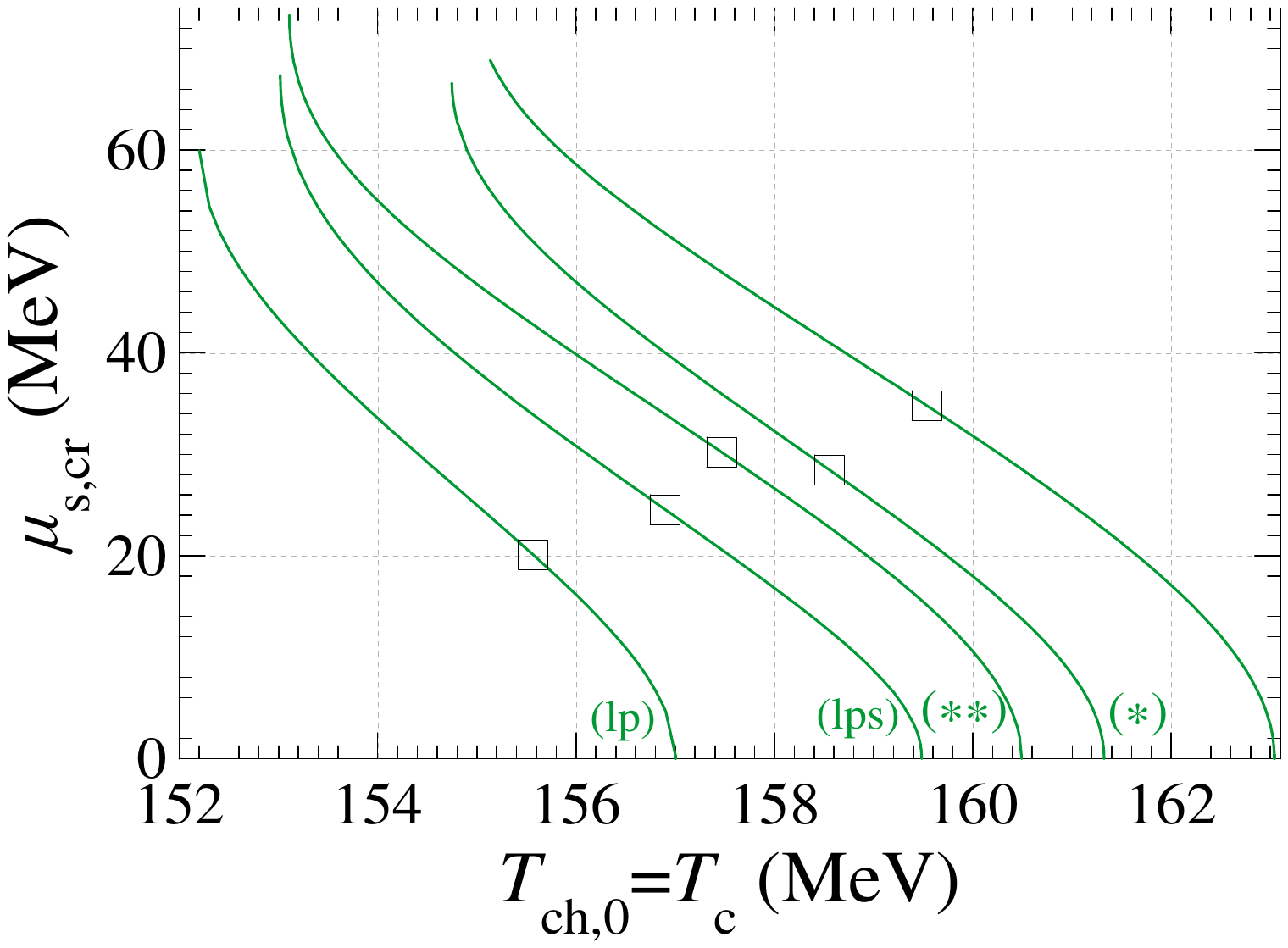}\\
(v)\includegraphics[scale=0.30,trim=0.8in 0.8in 1.in 0.2in,angle=0]{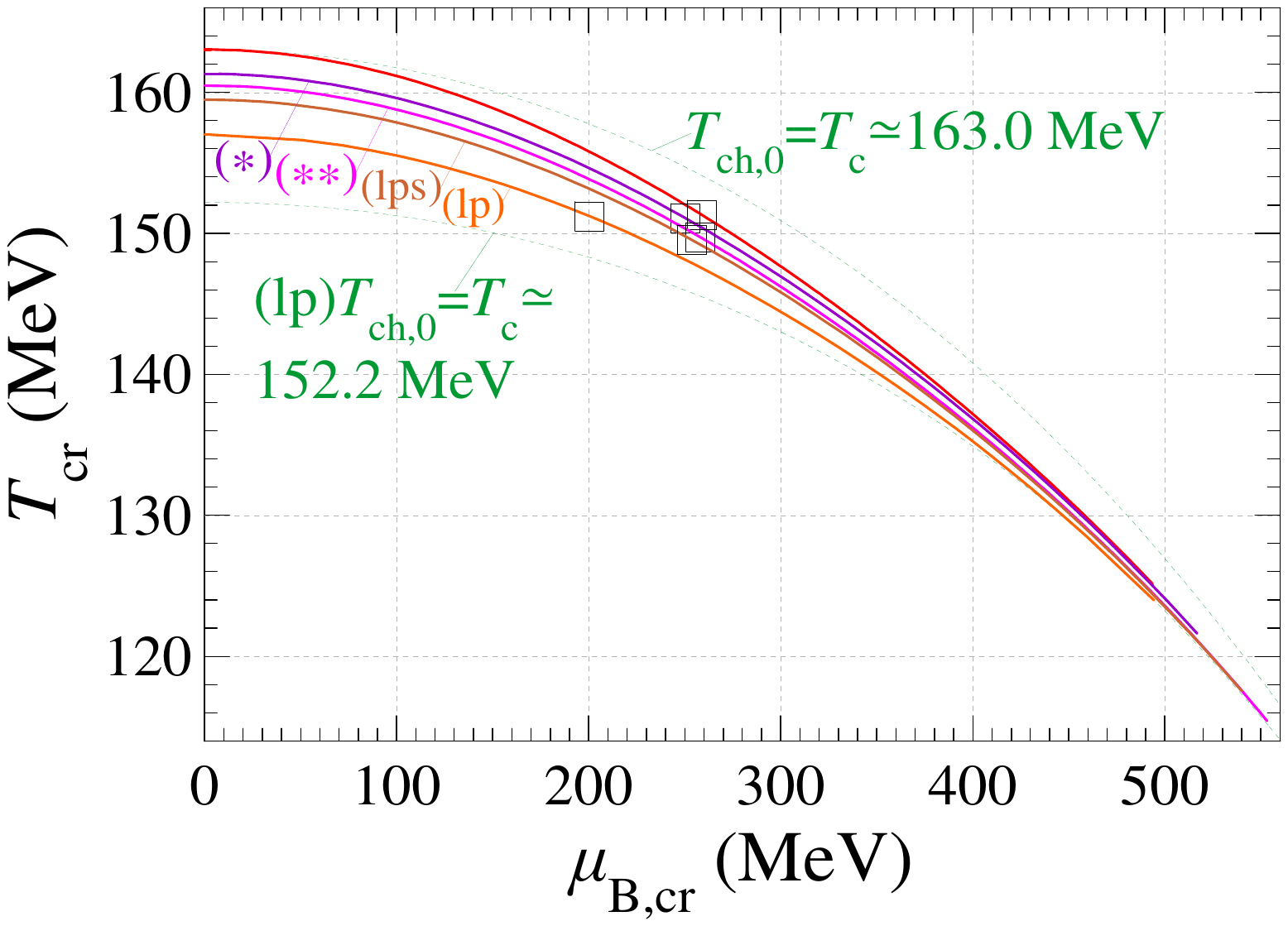} 
(vi)\includegraphics[scale=0.30,trim=0.8in 0.8in 1.in 0.2in,angle=0]{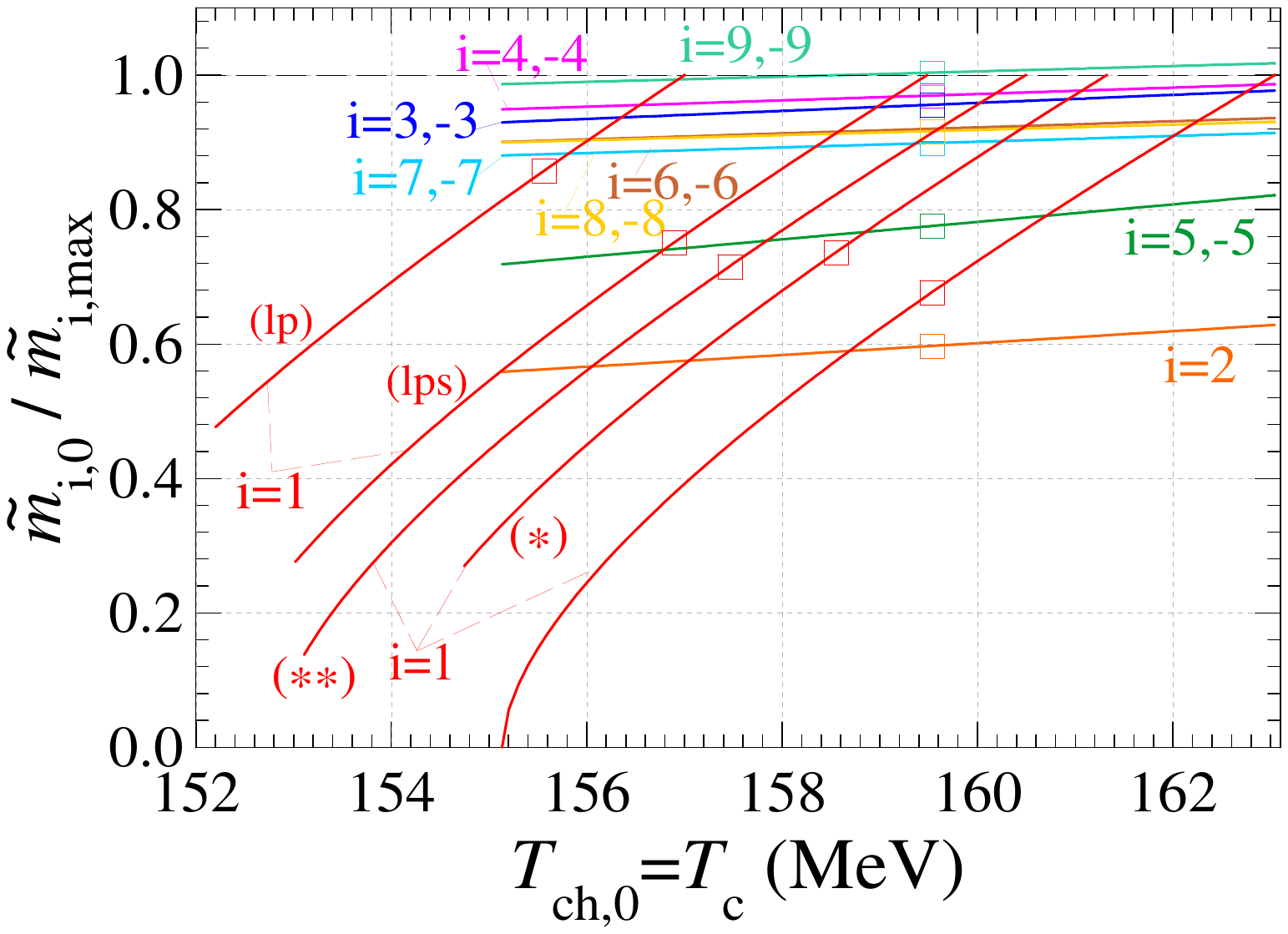}
\vspace{0.5cm}
\caption{\label{fig:cpr0(T)} {\small  
Graph similar to Fig.~\ref{fig:cpr0}.
Calculations are carried out in the interval $T_c\simeq$(152.2-163.0) MeV 
and involve common hadron radius which depends on temperature, $r_0(T)$, determined by the 
lattice pressure at $\mu_B=0$ (volume model (b)) for 5 hadron sets.
In (v) the freeze-out curves which correspond to the maximum value of $T_c=$163.0 MeV of
the (vh) set and 
the minimum value of $T_c=$152.2 MeV of the (lp) hadron set are shown. 
In all graphs with open squares we present the 
critical point which additionally fulfils the criterion of Section \ref{sec:crit}.
}}
\end{figure}

However, we find that the lower value of $T_c$ that results in non-negative 
solution for the chiral pion mass at $\mu_B=0$ is $T_c \simeq $152.2-155.1 MeV
(depending on the hadron set we use). Also, for $T_c \simeq $157.0-163.0 MeV 
the critical point is located at zero baryon density.  
In Fig.~\ref{fig:cpr0(T)} we show our 
results for the position of the QCD critical point for all the accepted values of $T_c
$=152.2-163.0 MeV. In Fig.~\ref{fig:cpr0(T)}(i) we plot the calculated
radius $r_0(T)$ which 
corresponds to the temperature of the critical point as a function of $T_c$. In 
Figs.~\ref{fig:cpr0(T)}(ii)-(iv) we depict the 
position of the point 
($T_{cr},\mu_{B,cr},\mu_{s,cr}$, respectively) with varying $T_c$.

In Fig.~\ref{fig:cpr0(T)}(v) 
we show the solution at zero baryon density for the chiral mass of each family (as the ratio to 
the higher allowed chiral mass of the family) for different values of $T_c$. Our findings are 
similar to the results of the hadronic radius which is independent of the temperature. 
However, the higher value of $\mu_B$ that allows for a critical point decreases.
Thus, the critical point can be located in the range of $\mu_B \sim$ (553-0) MeV, which 
corresponds to mean hadronic radius at the 
critical point 
$r_{0}(T)$ in the range of (0.234-0.298) fm.
Also in Figs.~\ref{fig:cpr0(T)} we depict by open square the critical point which is determined 
by the criterion discussed in the 
next Section \ref{sec:crit}. This corresponds to $T_c \simeq$155.6-159.5 MeV and it is located at 
$\mu_{B,cr}\simeq$250.0-258.5 MeV for the 4 first hadron sets or at $\mu_{B,cr}\simeq$200.2 MeV 
for the (lp) hadron set and $T_{cr}\simeq$149.6-151.3 MeV.

\begin{figure}[H]
\centering
(i)\hspace{-2cm}\includegraphics[scale=0.65,trim=0.5in 0.8in 2in 0.5in,angle=0]{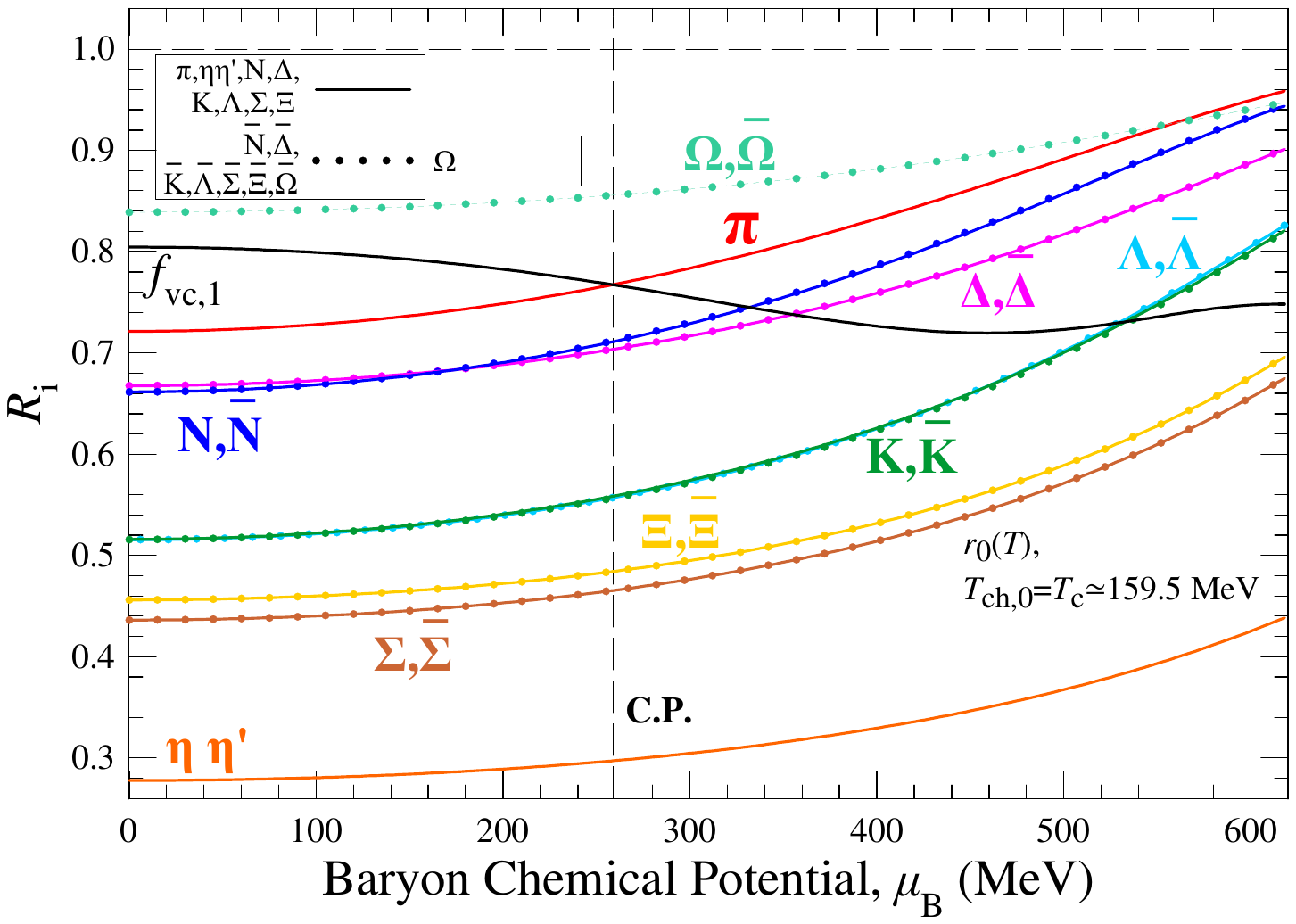}\\
(ii)\hspace{-2cm}\includegraphics[scale=0.65,trim=0.5in 0.8in 2in 0.5in,angle=0]{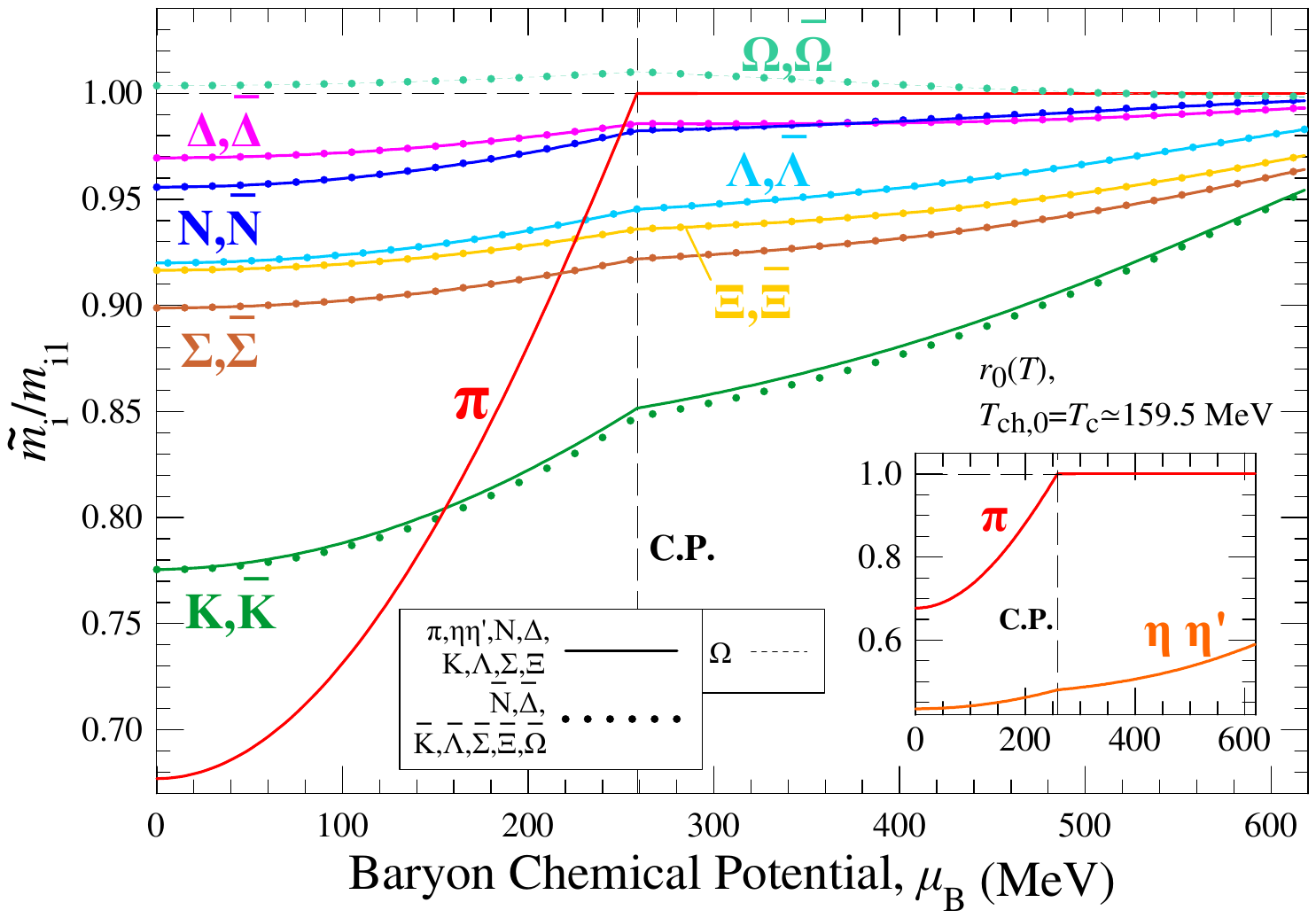}
\end{figure}

\begin{figure}[H]
\centering
(iii)\hspace{-1cm}\includegraphics[scale=0.75, trim=0in 1.2in 0in 1.in]{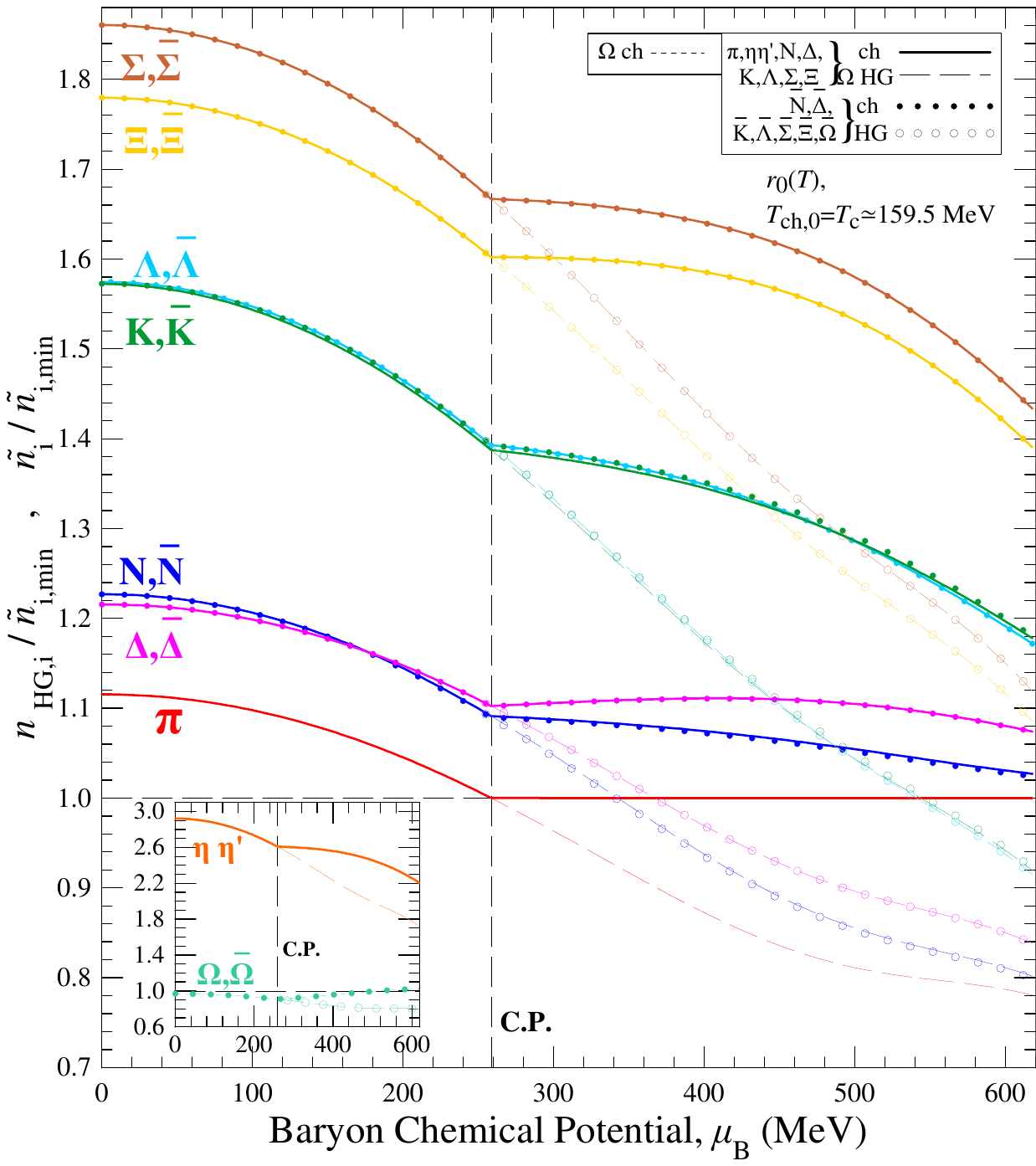}\\
\vspace{0.5cm}
\caption{\label{fig:Tc=159.5} {\small 
Graph similar to Fig.~\ref{fig:Tc=159.4}.
Calculations are carried out for $T_c=T_{ch,0}\simeq$159.5 MeV and temperature dependent common 
hadron radius, $r_0(T)$, determined by the lattice pressure for $\mu_B=0$,
for the (vh) set.
In (i) the shown volume correction factor for the pion family, $f_{vc,1}$, intersects with 
$R_1$ at $\mu_B\simeq$258.5 MeV to produce at this location the critical point.
}}
\end{figure}



In Figs.~\ref{fig:Tc=159.5} we show results for this value of $T_c$, which is determined
with the use of the (vh) set. In (i) we show the ratios 
$R_i$ for the different families. The 
intersection of the pion family with the volume correction factor determines the position of 
the critical point. In (ii) we show the corresponding solutions for the chiral masses for 
each family and in (iii) we present 
the densities in the HG and chiral phases for all families.

\subsection{Critical Point with meson and baryon radii} \label{subsec:v.m.c}

Turning to a more realistic scenario for the hadron volumes, we introduce separate volume radii 
for the mesons, $r_m$ and for the 
baryons, $r_b$. 
Different hadron volumes have, also, been used through the bag-model 
where the volume varies with the hadron mass 
\cite{matching,hadvol2} or energy \cite{matching}. 
In \cite{hadvol} different eigenvolumes have been used for several groups of hadrons.
Here we assume that the mesonic radius only depends on the temperature, $r_m(T)$ and the 
bayonic radius only depends on the
baryon chemical potential, $r_b(\mu_B)$. We, also, assume that $r_b$ remains constant at the 
crossover region with respect to 
$\mu_B$. So, the baryon chemical potential begins to affect the baryonic 
volume at the first 
order transition region, where the 
baryonic density has increased above a certain level. Having $r_b$ fixed, we determine the radius 
$r_m(T)$ from a fit on the lattice pressure for a 
specific value $T_c$. Then, these two values are passed on to the whole crossover region. 
Moreover, the fact that the radii are 
independent on the baryon chemical potential 
($\frac{d v_b}{d \mu_B}=0$) allows us to 
calculate densities through eq.~(\ref{eq:nv_HG}), instead of eq.~(\ref{eq:n_HG}), 
for values of baryon chemical potential up to the critical point.
This will be referred to as volume model (c).

In the 1st order transition curve the baryon volume 
{\it does} depend on $\mu_B$. 
Applying eq.~(\ref{eq:n_HG}), we have for mesonic families
$\frac{\partial v_j}{\partial \mu _j} = 0$, so:
\begin{equation} \label{eq:n_mes}
n_{HG,j}=n_{HG,j}^v ,\;\;j = 1,2,5,-5\;\;.
\end{equation}

For the baryonic and antibaryonic families eq.~(\ref{eq:n_HG})
can be rewritten as
\begin{equation} \label{eq:n_bar-abar1}
n_{HG,j} = n_{HG,j}^v 
\left( 1 - 
P\frac{\partial v_j}{\partial \mu_B}
\frac{\partial \mu_B}{\partial \mu_j}
\right)\;\;.
\end{equation}
For baryons we have
$\mu_j=\mu_B+n\mu_S$ with $n=0,-1,-2,-3$, so 
$\frac{\partial \mu_B}{\partial \mu_j}=1$. 
Similarly, for the antibaryons we have
$\frac{\partial \mu_B}{\partial \mu_j}=-1$.
Thus, eq.~(\ref{eq:n_bar-abar1}) is transformed to:
\begin{equation} \label{eq:n_bar-abar2}
n_{HG,j} = n_{HG,j}^v 
\left( 1 - 
cP\frac{\partial v}{\partial \mu_B}\right),
\;\; c=1 (j = 3,4,6,7,8,9),\;\;
c=-1 (j =-3,-4,-6,-7,-8,-9).
\end{equation}
In the last equation for this model (c) the volume is the
common volume to all baryons and antibaryons
$v=v_b(\mu_B)$, which is connected to the respective radius $r_b$.

The extra parameter for the baryon radius, $r_b$, allows us to impose the constraint that the 
chiral masses of both the pion and 
nucleon families acquire their maximum values at the critical point. For a specific choice of 
$T_c=T_{ch,0}$ we have a freeze out 
curve which links the freeze out temperature with the baryon chemical potential. So, in order to 
calculate the position of the 
critical point we have to evaluate the value of the critical parameters ($T_{cr}$, $\mu_{Bcr}$, 
$\mu_{scr}$), as well as, the values 
of the meson radius, $r_m$ and baryon radius, $r_b$. This is accomplished by solving, for given 
$T_c$, the following set of five equations:
\begin{equation} \label{eq:PLPHG_rmrb}
P_L(T;T_c)=P_{HG}(T,\mu_B=0,\mu_s=0;r_m(T),r_b)\;,
\end{equation}
\begin{equation} \label{eq:den1_2}
 n^v_{HG,1}(T,\mu_{B},\mu_s ; r_m(T), r_b) =
 \tilde{n}_1(T,\mu_{B},\mu_s ; \tilde{m}_1=m_{\pi})\;,
\end{equation}
\begin{equation} \label{eq:den3_2}
 n^v_{HG,3}(T,\mu_{B},\mu_s ; r_m(T), r_b) = 
\tilde{n}_3(T,\mu_{B},\mu_s ; \tilde{m}_3=m_{n})\;,
\end{equation}
\begin{equation} \label{eq:freezeout_2}
T=f_{fr} (\mu_B;T_c)\;,
\end{equation}
\begin{equation} \label{eq:S=0_2}
<S>_{HG} (T,\mu_{B},\mu_s ; r_m(T), r_b)=0\;.
\end{equation}

\begin{figure}[H]
\vspace{0.cm}
\centering
(i)\includegraphics[scale=0.31,trim=1.2in 0.8in 1.in 0.2in,angle=0]{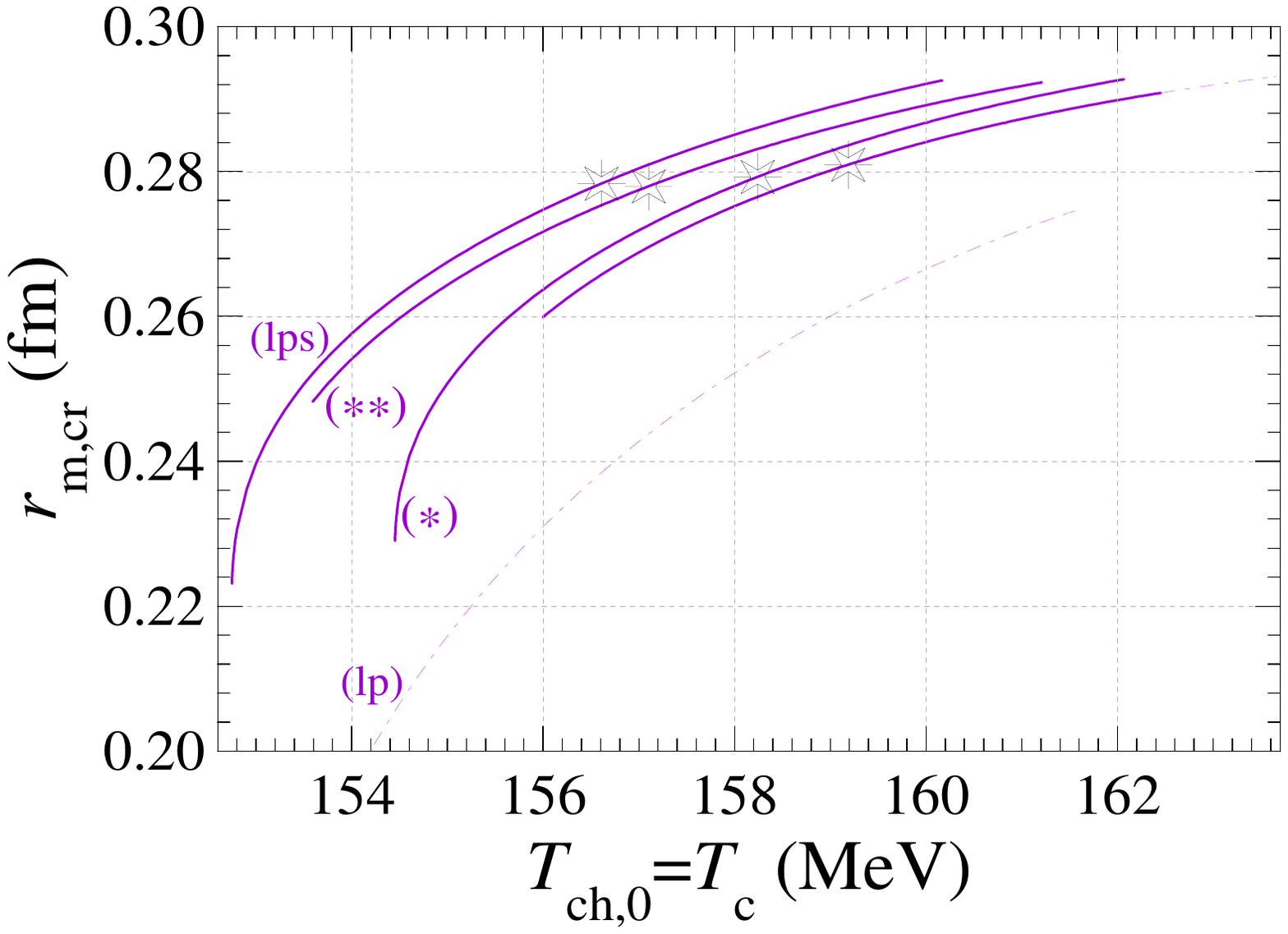} 
(ii)\includegraphics[scale=0.31,trim=0.8in 0.8in 1.in 0.2in,angle=0]{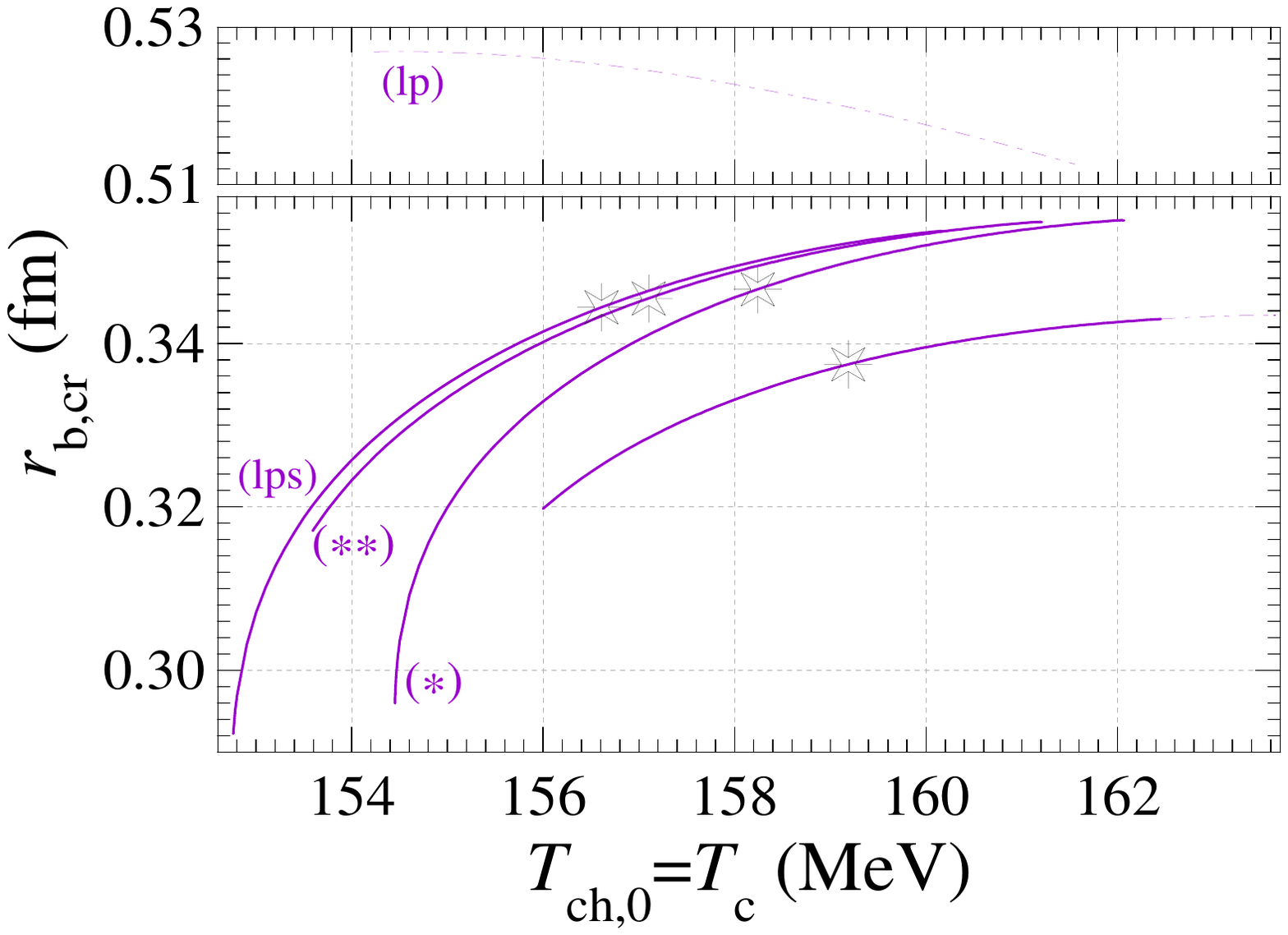}\\
\hspace{0.5cm}(iii)\includegraphics[scale=0.31,trim=1.2in 0.8in 1.in 0.2in,angle=0]{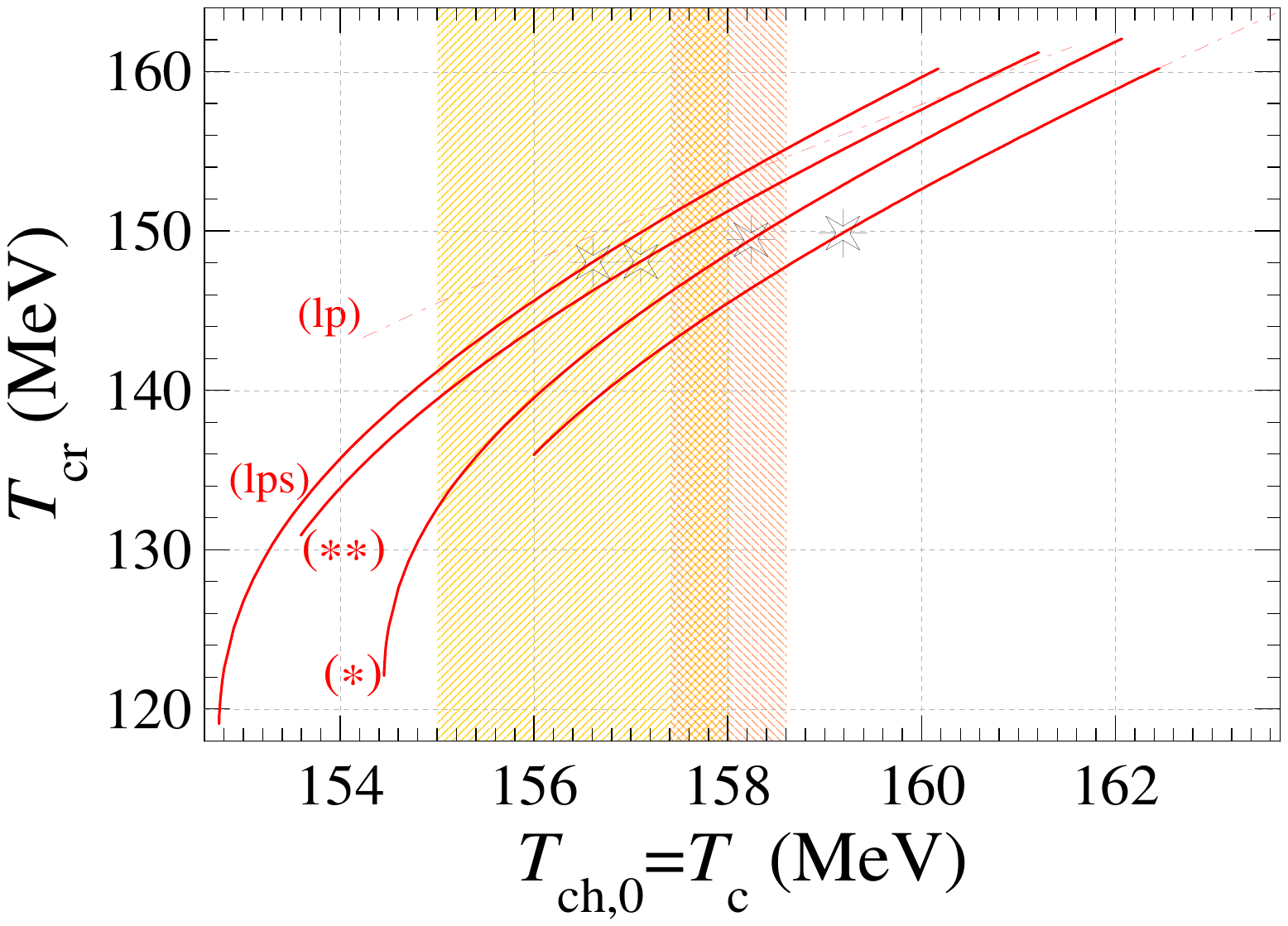}
(iv)\includegraphics[scale=0.31,trim=0.8in 0.8in 1.in 0.2in,angle=0]{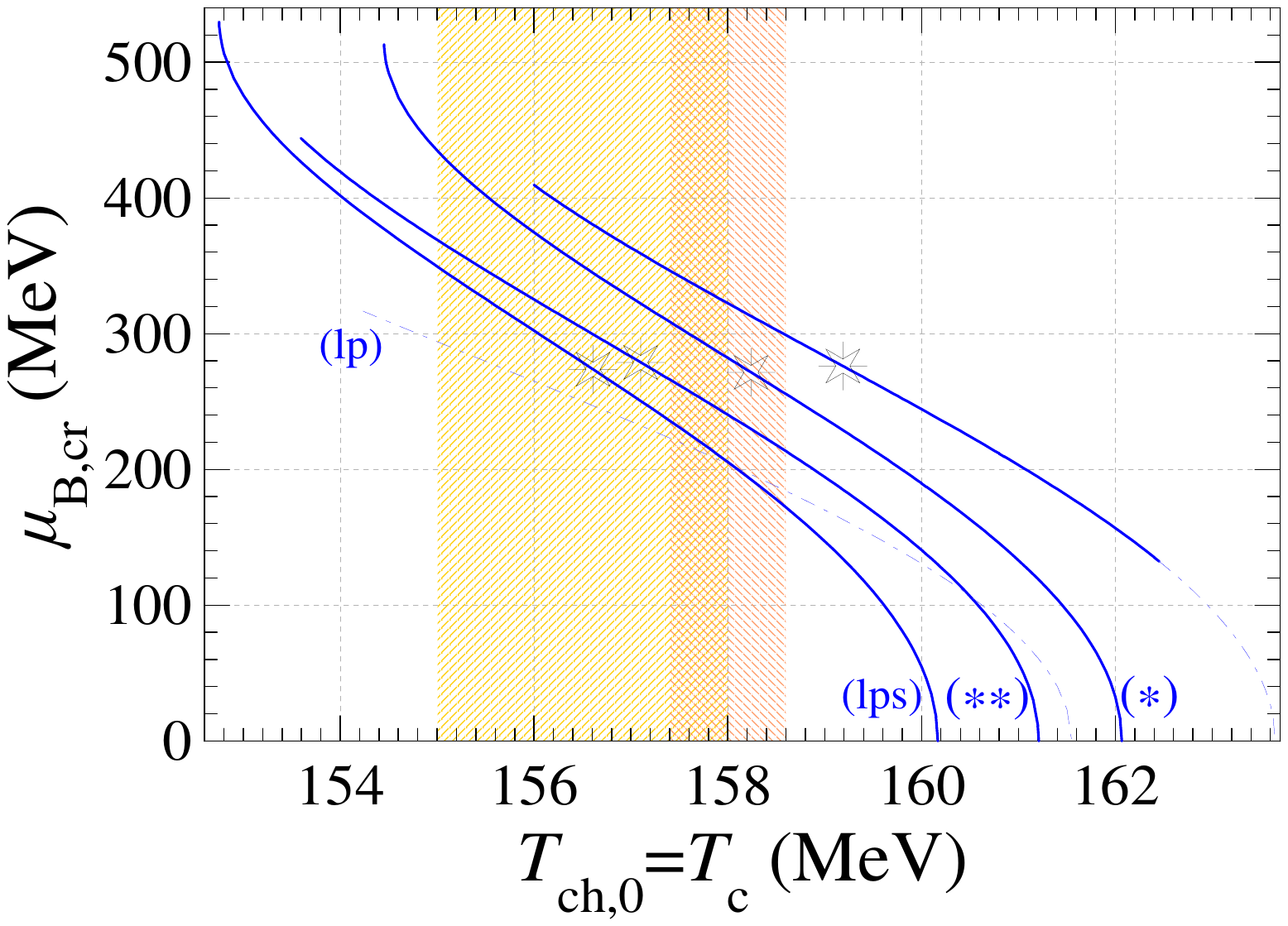}\\
\hspace{0.9cm}(v)\includegraphics[scale=0.31,trim=1.2in 0.8in 1.in 0.2in,angle=0]{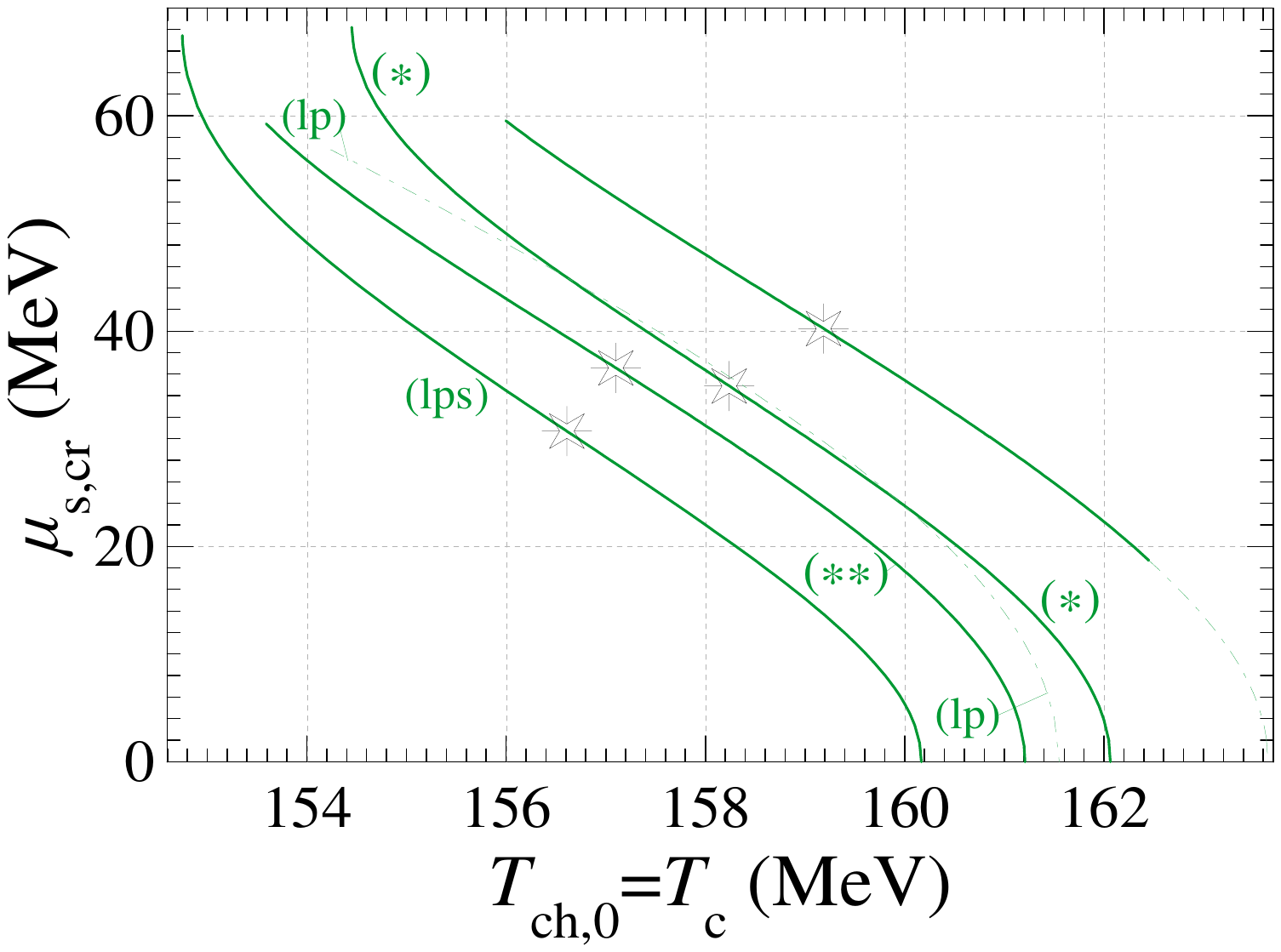}
\hspace{-0.3cm}(vi)\includegraphics[scale=0.31,trim=0.8in 0.8in 1.in 0.2in,angle=0]{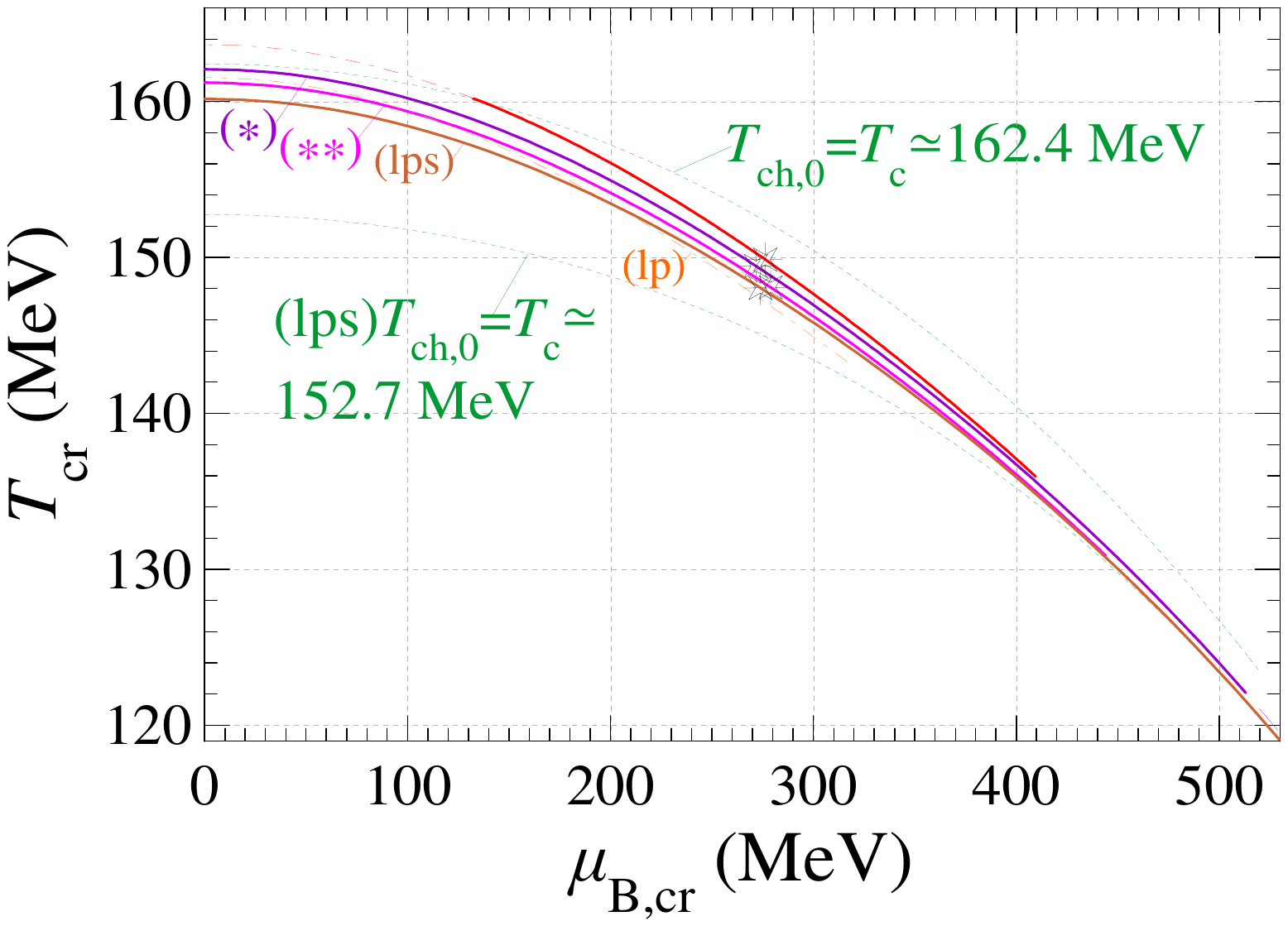} \\
(vii)\includegraphics[scale=0.31,trim=1.2in 0.8in 1.in 0.2in,angle=0]{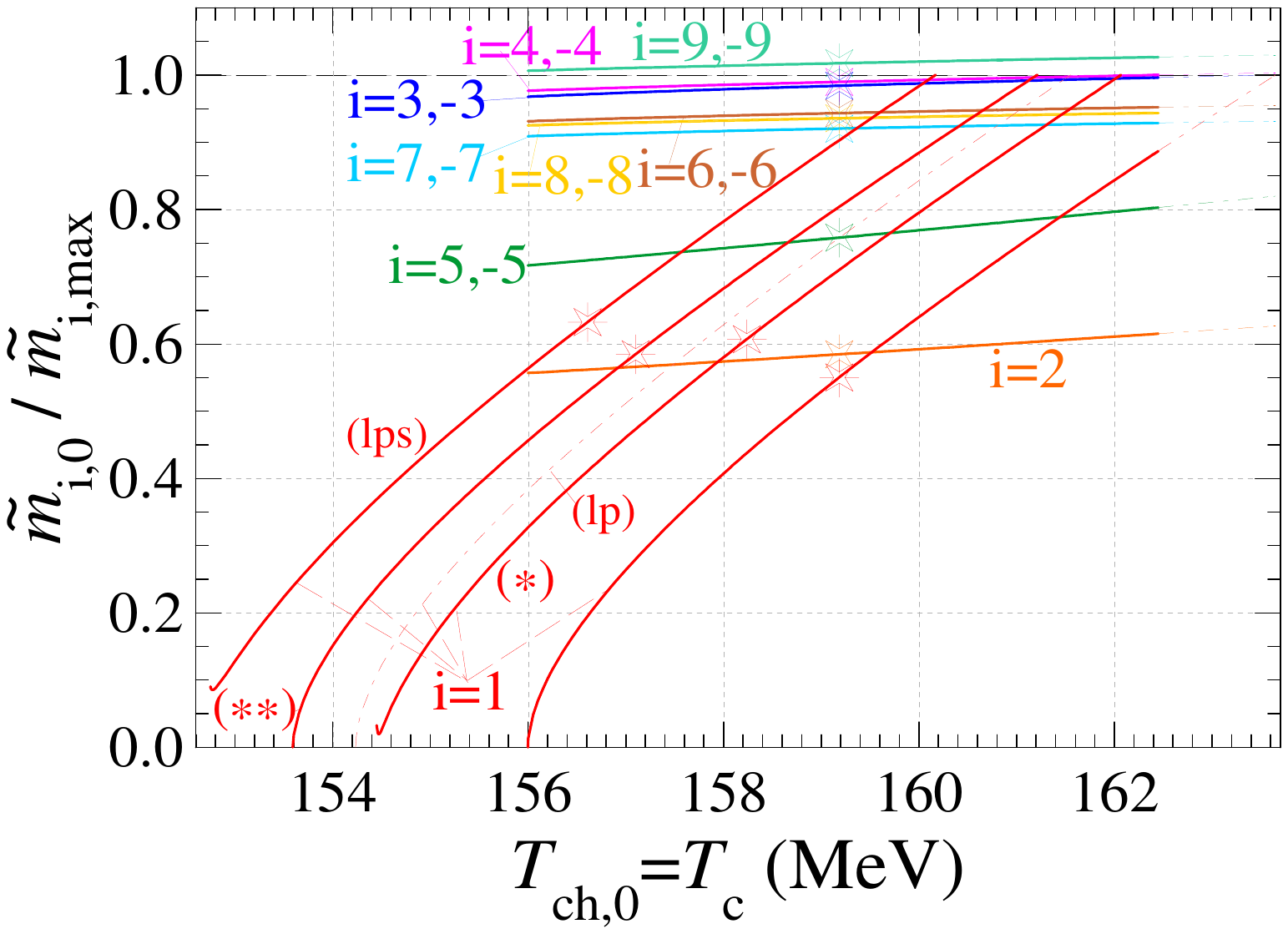}
\end{figure}

\begin{figure}[H]
\caption{\label{fig:cprmrb} {\small  
Graph similar to Fig.~\ref{fig:cpr0}.
Calculations are carried out in the interval $T_c\simeq$(152.7-163.6) MeV
and involve a 
meson radius which depends on temperature, $r_m(T)$ and a baryon radius which depends on 
baryon chemical potential, $r_b(\mu_B)$ for 5 hadron sets. The radius $r_m(T)$ is determined by the lattice pressure at 
$\mu_B=0$ for the value of $r_b$ at the critical point (volume model (c)).
In (i), (ii) we depict the value of the meson, baryon radius on the critical point, 
$r_{m,cr}$, 
$r_{b,cr}$, as a function of $T_c=T_{ch,0}$, respectively.
Graphs (iii)-(vii) correspond to Graphs \ref{fig:cpr0}(ii)-(v), respectively.
In (vi) we show the freeze-out curves which 
correspond to the maximum value of $T_c=$162.4 MeV of the (vh) set and the minimum value of $T_c=$152.7 MeV for the (lps) hadron set. 
In all graphs with stars 
we present the critical points which additionally fulfil the criterion described in Section 
\ref{sec:crit}. Also, we include regions (slashed-dotted lines) where the $\Delta$ family acquires higher chiral mass than the 
maximum allowed value, in cases of the (vh) and the (lp) sets.
}}
\end{figure}


Our solutions are limited by 
the values of $T_c$ which produce zero pion chiral mass at $\mu_B=0$ or by the minimum value of $T_c$ 
that produces a solution and the values of $T_c$ which produce a critical
point at $\mu_B=$0.
Also, for the (vh) set  
 the $\Delta$ family develops a chiral mass 
which is slightly above the maximum allowed value.
To stay on the safe side we exclude
these cases for the solution of a critical point.
For the (lp) set this situation appears in all the solutions. For
this reason we exclude
these solutions, but for completeness we depict them on the graph.
After applying all these constraints and for the rest of hadron sets we find that the critical 
point is located in the interval $\mu_B \simeq $513-0 MeV and the critical region 
in $\mu_B$ is further suppressed compared to the case of the common hadronic radius.
In Fig.~\ref{fig:cprmrb} we show 
with solid lines our results for the 4 first hadron sets for the position of the 
QCD critical point for the values for $T_c$=152.7-163.6 MeV.
With slashed-dotted lines we depict the solutions which are excluded
due to the high chiral mass of the $\Delta$ family for the (lp) hadron set and the (vh) set
with $T_c$=162.4-163.6 MeV. 
In Figs.~\ref{fig:cprmrb}(i)-(ii) we plot the calculated 
mesonic radius $r_m$ and
baryonic radius $r_b$ which correspond to the conditions of the 
critical point as function of $T_c$. In 
Figs.~\ref{fig:cprmrb}(iii)-(v) we depict the position of the critical point ($T_{cr},\mu_{B,cr},
\mu_{s,cr}$, respectively) as a function of 
$T_c$. In Fig.~\ref{fig:cprmrb}(vi) we illustrate the solution at zero baryon density for the 
chiral mass of each family 
(as the ratio to the higher allowed chiral mass of the family) with varying $T_c$. Also, in 
Figs.~\ref{fig:cprmrb} we show by stars 
 the critical points which further fulfil the criterion described in the next Section 
\ref{sec:crit}. This corresponds to 
$T_c \simeq$156.6-159.2 MeV and it is located at $\mu_{B,cr}\simeq$271.6-278.9 MeV and 
$T_{cr}\simeq$148.1-149.9 MeV.

Next we can evaluate the thermodynamic variables on the crossover curve, for a given value of 
$T_c$. The value of $r_b$ remains fixed 
to the value which was found from the equations that determined the critical point for this 
value of $T_c$. So we obtain 
($T,\mu_s;r_m$) for every value of 
$\mu_B$ by solving equations 
(\ref{eq:PLPHG_rmrb}), (\ref{eq:freezeout_2}) and 
(\ref{eq:S=0_2}), while the chiral masses for all families are calculated by 
eq.~(\ref{eq:HGchnBF}) with $V_{HG}=\tilde{V}$.

Turning, then, to the first order transition line, we have to evaluate for specific $T_c$ 
the set of variables 
($T,\mu_s, v_{er}; r_m,r_b$), since now, we allow $r_b$ to change. Given a value of $r_b$, we 
determine $r_m(T)$ from the lattice 
pressure curve for zero baryon density. The chiral masses of the pion and the nucleon family 
remain fixed at their corresponding maximum values.
Since we allow the baryon radius to depend on $\mu_B$ and the meson radius does not depend on 
$\mu_B$, we have to replace eqs.~(\ref{eq:den1_2}) and (\ref{eq:den3_2}) by the equations:
\begin{equation} \label{eq:den1_4}
 n^v_{HG,1} = v^{-1}_{er} \tilde{n}_{1,min} \Rightarrow
v^{-1}_{er}=\frac{n^v_{HG,1}}{\tilde{n}_{1,min}}\;,
\end{equation}
\[
n_{HG,3} = v^{-1}_{er} \tilde{n}_{3,min} \Rightarrow 
n^v_{HG,3}\left(1- 
P_{HG} \frac{\partial v}{\partial \mu_B} \right)=
v^{-1}_{er} \tilde{n}_{3,min} \Rightarrow
P_{HG} \frac{\partial v}{\partial \mu_B} =
1-v^{-1}_{er} \frac{\tilde{n}_{3,min}}{n^v_{HG,3}} \Rightarrow
\]
\begin{equation} \label{eq:den3_4}
\Rightarrow \frac{\partial v}{\partial \mu_B}=
\frac{1}{P_{HG}}
\left( 1-\frac{n^v_{HG,1}}{n^v_{HG,3}} \frac{\tilde{n}_{3,min}}{\tilde{n}_{1,min}} \right)\;,
\end{equation}
where $v=v_b(\mu_B)$. The zero strangeness condition has to be 
imposed, as well, which in view of 
eqs.~(\ref{eq:n_mes}) and (\ref{eq:n_bar-abar2}), acquires the form:

\begin{figure}[H]
\centering
(i)\hspace{-2cm}\includegraphics[scale=0.65,trim=0.5in 0.8in 2in 0.5in,angle=0]{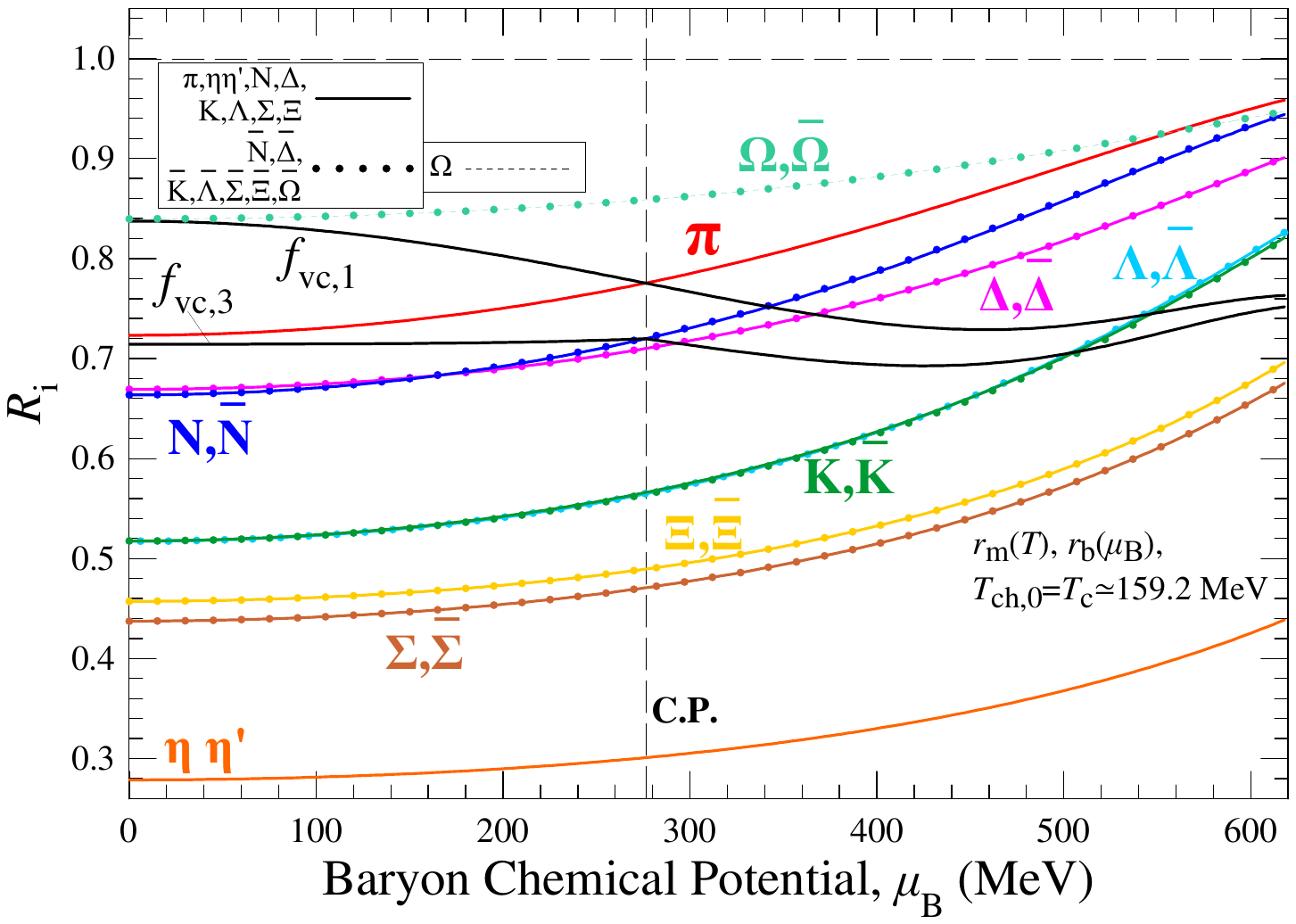}\\
(ii)\hspace{-2cm}\includegraphics[scale=0.65,trim=0.5in 0.8in 2in 0.5in,angle=0]{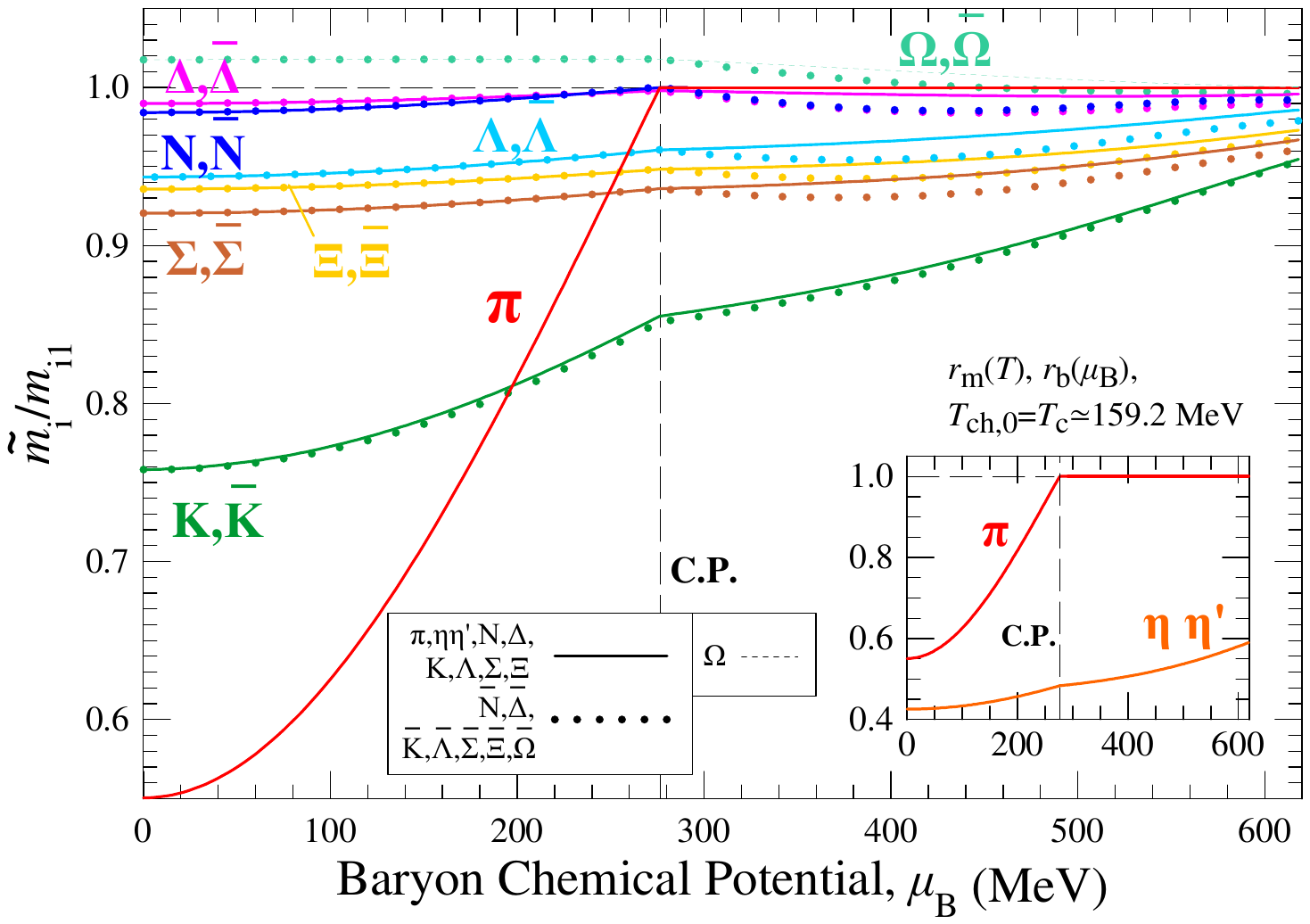}
\end{figure}

\begin{figure}[H]
\centering
(iii)\hspace{-1cm}\includegraphics[scale=0.75, trim=0in 1.2in 0in 1.in]{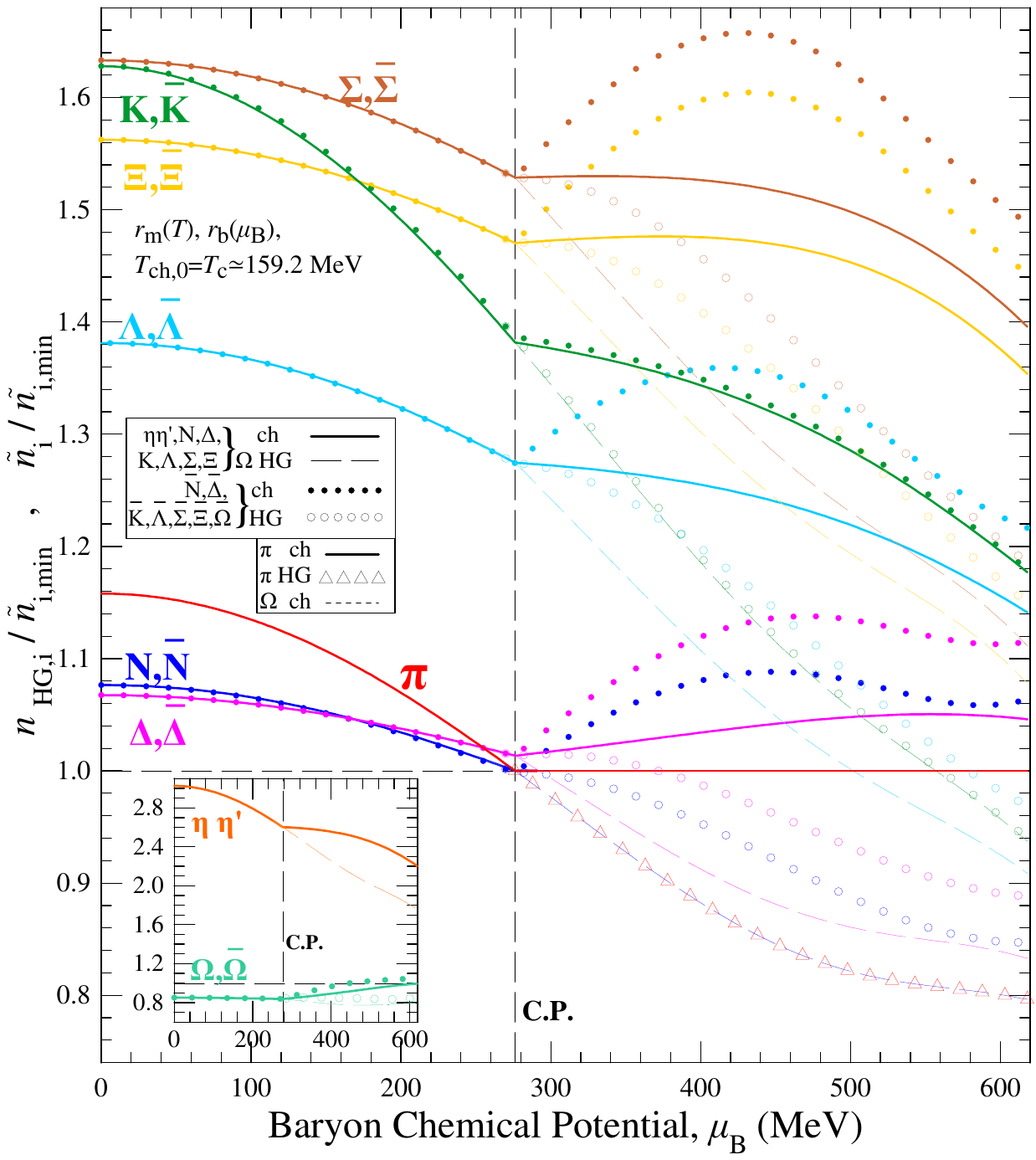}
\caption{\label{fig:Tc=159.2} {\small 
Graph similar to Fig.~\ref{fig:Tc=159.4}.
Calculations are carried out for $T_c=T_{ch,0}\simeq$159.2 MeV and different meson $r_m(T)$ and 
baryon $r_b(\mu_B)$ radii, determined by the lattice pressure for $\mu_B=0$, for the
(vh) set.
In (i) the shown volume correction factors for the pion family, $f_{vc,1}$, and the 
nucleon family, $f_{vc,3}$,
simultaneously intersect with $R_1$ and $R_3$, respectively, 
at $\mu_B\simeq$276.2 MeV and 
produce at this location the critical point.
}}
\end{figure}


\begin{equation} \label{eq:S=0_4}
<S>_{HG} =0 \Rightarrow n_{HG,S}=0 \Rightarrow
n_{HG,S}^v + P\frac{\partial v}
{\partial \mu_B} n_{HG,|S|,b+\bar{b}}^v  = 0\;,
\end{equation}
\noindent
where the subscript $|S|,\;b+\bar{b}$ denotes the
absolute strangeness enclosed in the baryonic and antibaryonic
families, i.e.
\begin{equation} \label{eq:|S|}
n_{HG,|S|,b+\bar{b}}^v=
n_{HG,6}^v + n_{HG,-6}^v + n_{HG,7}^v + n_{HG,-7}^v+
2\left(n_{HG,8}^v + n_{HG,-8}^v \right)+
3\left(n_{HG,9}^v + n_{HG,-9}^v \right)\;.
\end{equation}
Eqs.~(\ref{eq:PLPHG_rmrb}), (\ref{eq:freezeout_2}),
(\ref{eq:den1_4})-(\ref{eq:S=0_4}) form a system of five equations which allows us to 
determine the set of variables 
($T,\mu_s, v_{er};$ $r_m, r_b$), for a given value of $\mu_B$. The situation is simplified since 
eqs.~(\ref{eq:freezeout_2}), (\ref{eq:den1_4})
allow us to solve for $T$ and $v_{er}$ directly. 
So, eqs.(\ref{eq:PLPHG_rmrb}), (\ref{eq:den3_4}) 
and (\ref{eq:S=0_4}) form a subset 
of three equations (one of which is a differential equation, though), which allows to evaluate 
$r_m$, $r_b$ and $\mu_s$. The chiral 
masses for all families except the pion ($i=1$) and the nucleon family ($i=3$) are calculated 
by eq.~(\ref{eq:HGchnBF}) with $V_{HG}=v_{er} \tilde{V}$.

In Fig.~\ref{fig:Tc=159.2} we depict all the calculations (as a function of $\mu_B$) for a 
specific value of $T_c\simeq$159.2 MeV 
which, also, fulfils the criterion described in Section \ref{sec:crit}, for the 
(vh) set.
In (i) we plot the ratios $R_i$, where the simultaneous intersections of the volume correction 
factors for the pion and the nucleon
family with the corresponding ratios $R_i$ determine the position of the critical point. 
In (ii) we plot
the chiral masses of the 
hadronic families, where, at 
the first order transition line, the chiral masses of the pion and the nucleon families
remain constant, retaining their maximum allowed values. 
In (iii) we display the densities of the families, with suitable 
normalisation, at the hadronic and chiral phases.
We observe that in the 1st order transition curve, where
the baryon volume depends on $\mu_B$, the densities of the
baryon families exhibit different behaviour than the 
corresponding antibaryon families due to the 
different sign that enters eq.~(\ref{eq:n_bar-abar2}).

\subsection{Critical Point from lattice QCD at finite baryon chemical potential}\label{subsec:v.m.d}

Lattice calculation at finite values of baryon chemical potential
have been carried out in 
\cite{lat_muB1,lat_muB2,lat_muB3,lat_muB4,lat_muB5,lat_muB6}.
We use the results of the pressure $p/T^4$ as function of temperature 
$T$ for fixed values of $\hat{\mu}_B \equiv \mu_B/T$ from $0$ to 
$3.5$ with $0.5$ MeV step from \cite{lat_muB1}. These values are determined for vanishing strangeness chemical potential, $\mu_S$. The temperature values of interest are in the range $125-170$ 
MeV. The lower value is given by the presented calculations and the 
upper is an approximate upper bound for the critical 
lattice temperature for $\mu_B=0$, which sets an upper temperature bound in the existence of the Hadron Gas. The values from the graph are then fitted with a 4th degree polynomial (as function of $T$) to produce a curve that fits the pressure points for each value of fixed $\hat{\mu}_B$. Totally, we have 8 such curves, which
give us the pressure as function of $T$ at specific values of
$\hat{\mu}_B$. Then, each time we need the pressure at
a given point ($T_1$, $\hat{\mu}_{B,1}$), we evaluate from these 8 curves the pressure
values for temperature $T_1$ and 8 values of $\hat{\mu}_B$. On these 8 values we perform
a fit with a 6th degree polynomial, but only with the even powers present (since the
pressure has to be an even function of $\hat{\mu}_B$). This procedure gives us the pressure as function of chemical potential for the specific temperature $T_1$ and we can just
read the value for the specific $\hat{\mu}_{B,1}$.

We, also, need to allow the critical temperature at $\mu_B=0$, 
$T_{0,c} \equiv T_c(0)$ to vary.
In such case we define the dimensionless temperature
$t \equiv \frac{T}{T_{0,c}}$. If the critical temperature is shifted to $T_{0,c}'$ the temperature $T'$ for which $t$ stays unchanged is
\begin{equation}\label{eq:t}
t = t' \Rightarrow \frac{T}{T_{0,c}} = \frac{T'}{T'_{0,c}} \Rightarrow 
T' = T\frac{T'_{0,c}}{T_{0,c}}
\Rightarrow T' = T a \;,
\end{equation}
where we have defined
\begin{equation}\label{eq:a}
a \equiv \frac{T'_{0,c}}{T_{0,c}} = \frac{T'}{T} \;.
\end{equation}

For vanishing baryon chemical potential
we demand that the dimensionless pressure 
$\hat{P} \equiv \frac{P}{T^4}$ stays the same at the same $t$, so
\begin{equation}\label{eq:P_t}
\hat{P}'(t') = \hat{P} (t) \Rightarrow 
\hat{P}'\left( T' \right) = 
\hat{P}\left( \frac{T'}{a} \right) \Rightarrow
P'\left( T' \right) = a^4
P\left( \frac{T'}{a} \right) \;.
\end{equation}
This was the case for volume models (a), (b) and (c).

In \cite{lat_susc,lat_Tc2} the critical temperature is taken to be $T_c=$158 MeV.  
For finite baryon chemical potential the critical temperature
changes as (\cite{lat_susc,lat_Tc2}):
\begin{equation}\label{eq:Tc_muB}
T_c \left( \mu _B \right) = T_{0,c}
\left[ 1-k_2 \left( \frac{\mu _B}{T_{0,c}} \right)^2 \right]
\;.
\end{equation}

Demanding eq.~(\ref{eq:Tc_muB}) to stay unchanged, if we
change the critical temperature from $T_{0,c}$ to $T'_{0,c}$,
amounts to
\begin{equation}\label{eq:T'c_muB}
T'_c \left( \mu'_B \right) = T'_{0,c}
\left[ 1-k_2 \left( \frac{\mu'_B}{T'_{0,c}} \right)^2 \right] \;.
\end{equation}

Since $T'_c \left( \mu'_B \right)$ is the temperature that
corresponds to the lattice with ${T'_{0,c}}$, it should be related
to $T_c \left( \mu_B \right)$ through eq. (\ref{eq:t}).
Combing eqs.~(\ref{eq:t})-(\ref{eq:T'c_muB}) we get
\begin{equation}
T'_c \left( \mu'_B \right) = 
T_c \left( \mu_B \right)\frac{T'_{0,c}}{T_{0,c}}
\Rightarrow
\frac{\mu'_B}{T'_{0,c}}=\frac{\mu_B}{T_{0,c}}
\Rightarrow
\frac{\mu'_B}{T'}=\frac{\mu_B}{T} \Rightarrow
\hat{\mu}'_B=\hat{\mu}_B\;.
\end{equation}
Thus, for eq.~(\ref{eq:Tc_muB})
to remain unchanged the baryon chemical 
potential has to be shifted to
\begin{equation}\label{eq:hat_muB}
\hat{\mu}'_B=\hat{\mu}_B
\Rightarrow
{\mu}'_B=\mu_B \frac{T'_{0,c}}{T_{0,c}}
\Rightarrow
{\mu}'_B=\mu_B a \;.
\end{equation}

We assume the same behaviour for any other chemical potential 
describing the system:
\begin{equation}\label{eq:hat_muU}
\hat{\mu}'_U=\hat{\mu}_U \;.
\end{equation}
Generalising eq.~(\ref{eq:P_t}) for the pressure as function of
any set of chemical potentials, as well, we get
\begin{equation}\label{eq:P_t,mu_U}
P'(T',\ldots,\mu'_B,\ldots) = 
a^4 
P\left( \frac{T'}{a},\ldots,\frac{\mu'_U}{a},\ldots \right)
\Rightarrow
\hat{P}'(T',\ldots,\hat{\mu}'_U,\ldots) = 
\hat{P}\left( T,\ldots,\hat{\mu}_U,\ldots \right) \;.
\end{equation}
Assuming that for a quantity $\hat{f}$ the following relation holds:
\begin{equation}\label{eq:f_t,mu_U}
\hat{f}'(T',\ldots,\hat{\mu}'_U,\ldots) = 
\hat{f}\left( T,\ldots,\hat{\mu}_U,\ldots \right)\;,
\end{equation}
then we have
\begin{equation}\label{eq:df}
\left. \frac{\partial \hat{f}'(T',...,\hat{\mu}'_U,...)}
{\partial \hat{\mu}'_U} \right|_{T',...}= 
\left. \frac{\partial \hat{f}(T'/a,...,\hat{\mu}'_U,...)}
{\partial \hat{\mu}'_U} \right|_{T'/a,...}=
\left. \frac{\partial \hat{f}(T,...,\hat{\mu}_U,...)}
{\partial \hat{\mu}_U} \right|_{T,...}
\;.
\end{equation}
Applying the above procedure to any kind of chemical potentials and any
order of derivatives, we get that for any susceptibility:
\vspace{-0cm}
\begin{equation}\label{eq:susc}
\chi_{ijk}^{'BSQ}(T',{\hat \mu '_B},{\hat \mu '_S}) = 
\chi_{ijk}^{BSQ}(T,{\hat \mu _B},{\hat \mu _S}) \;.
\end{equation}
Using equations (\ref{eq:f_t,mu_U})-(\ref{eq:df}) for the dimensionless pressure,
we find for the dimensionless baryon density 
$\hat{n}_B \equiv \frac{n_B}{T^3}$:
\begin{equation}\label{eq:nB}
{{{\hat n'}_B}(T',{{\hat \mu '}_B}) = 
{{\hat n}_B}(T,{{\hat \mu}_B})} \;.
\end{equation}

After dealing with derivatives of chemical potentials, we turn to temperature derivatives. Assuming that we have defined a new dimensionless quantity:
\begin{equation}\label{eq:h}
\hat h(T,...,{\hat \mu _U},...) \equiv 
{T^k}{\left. {\frac{{{\partial ^k}
\hat f(T,...,{{\hat \mu }_U},...)}}
{{\partial {T^k}}}} \right|_{...,{{\hat \mu }_U},...}}\;,
\end{equation}
then (\ref{eq:f_t,mu_U}) leads to:
\begin{equation}\label{eq:der_h_T}
{T'^k}{\left. {\frac{{{\partial ^k}\hat f'(T',...,{{\hat \mu '}_U},...)}}{{\partial {{T'}^k}}}} \right|_{...,{{\hat \mu '}_U},...}} = 
{T'^k}\frac{1}{{{a^k}}}{\left. {\frac{{\partial \hat f(T'/a,...,{{\hat \mu '}_U},...)}}{{\partial {{\left( {T'/a} \right)}^k}}}} \right|_{...,{{\hat \mu '}_U},...}} \;.
\end{equation}

\begin{figure}[H]
\vspace{-1.cm}
\centering
(i)\includegraphics[scale=0.51,trim=1.2in 0.1in 0.7in 0.2in,angle=0]{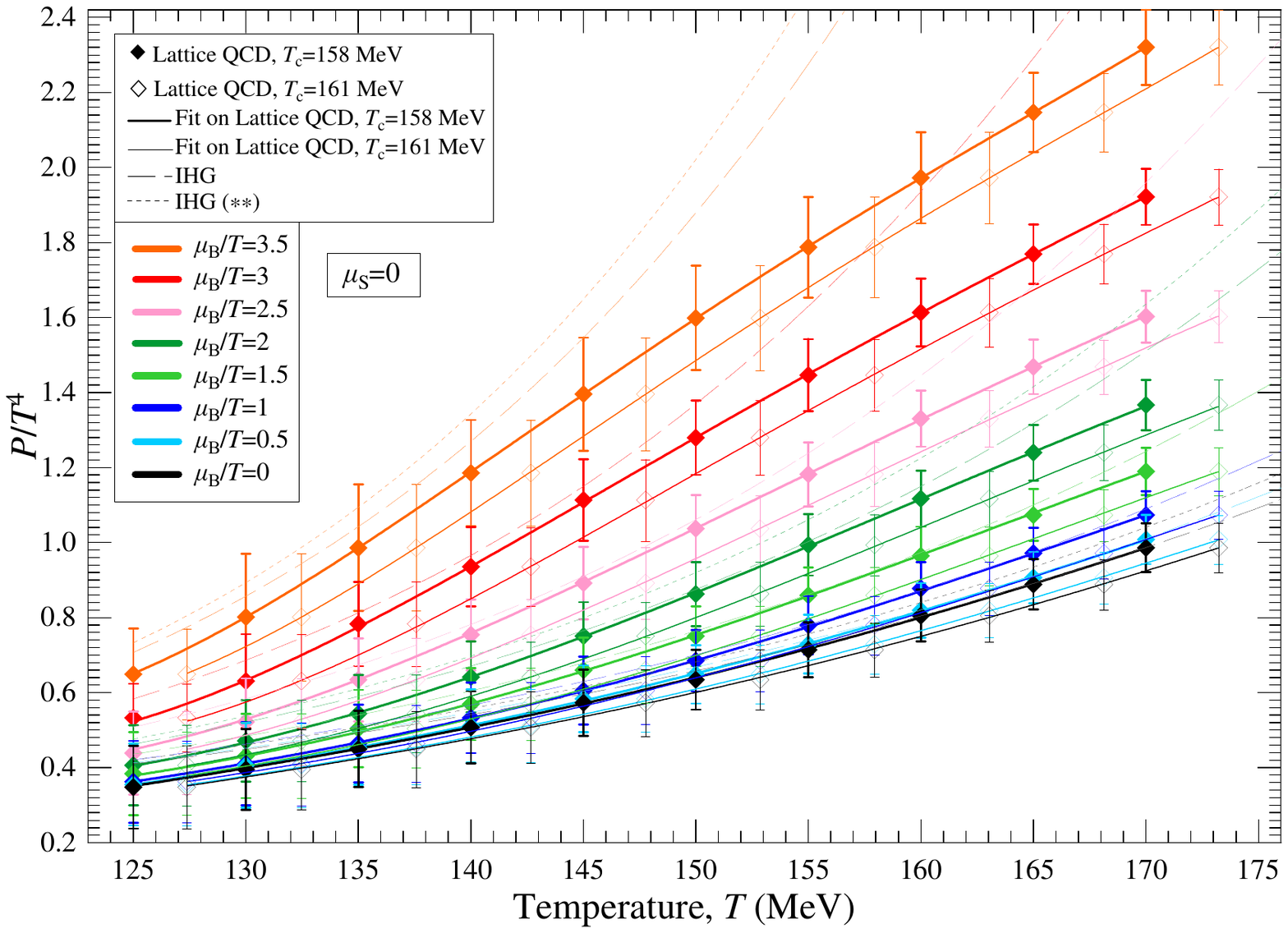}\\
(ii)\includegraphics[scale=0.51,trim=1.2in 0.1in 0.7in 0.2in,angle=0]{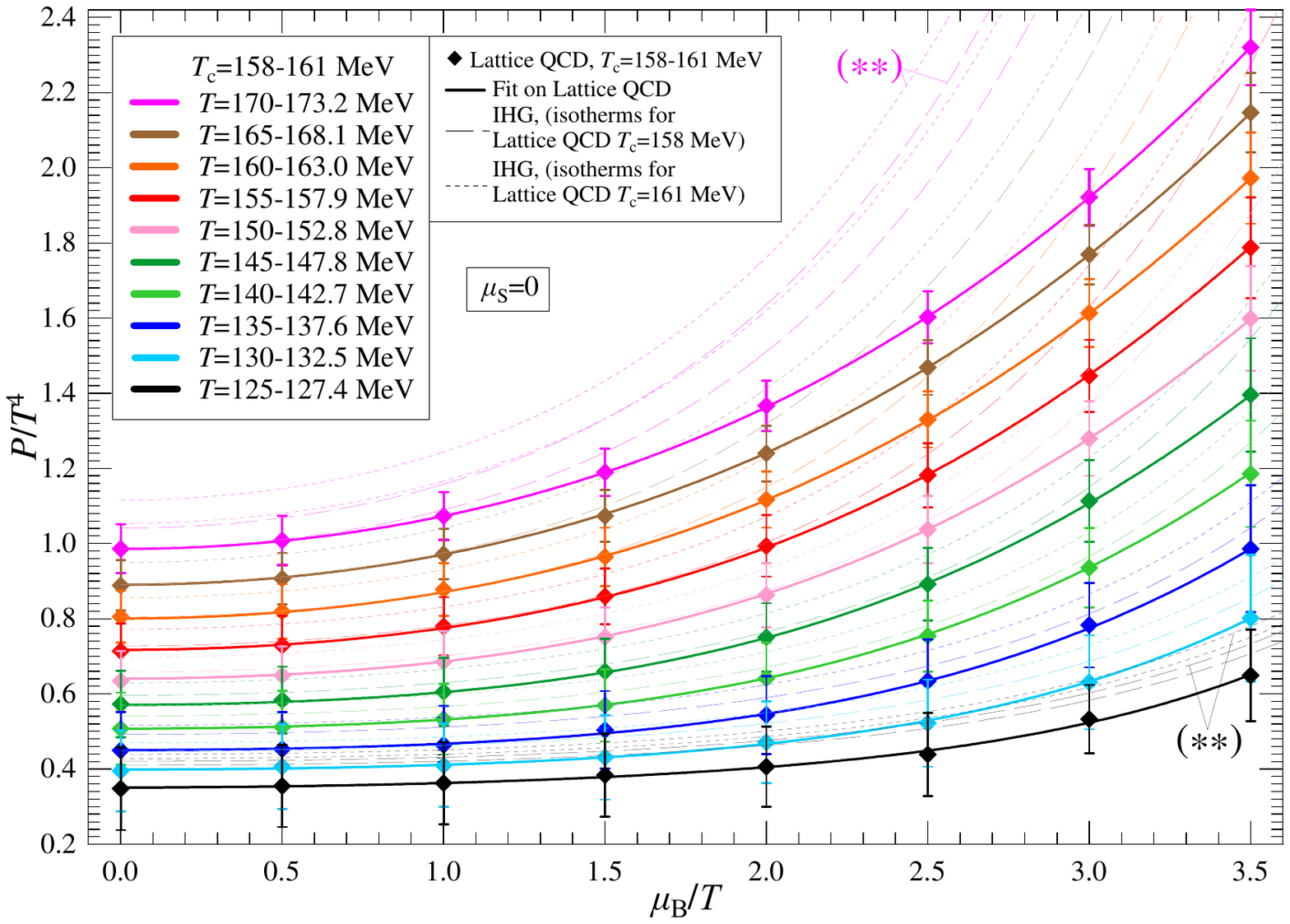}\\
\vspace{-0.2cm}
\centering
\caption{\label{fig:PL-AB} {\small  
(i) The pressure $P/T^4$ for constant $\mu_B/T$ as
function of temperature. Lattice QCD calculations extracted 
from Fig.~4 of \cite{lat_muB1} at
$T_c=$158(161) MeV appear with closed (open) rhomboid points. 
Our fits on these points are depicted with thick (thin)
continuous lines. Corresponding Ideal Hadron Gas (IHG) 
calculations appear with slashed lines (one set of curves
since IHG formalism does not depend on $T_c$).
(ii) The pressure $P/T^4$ for constant temperature as
function of $\mu_B/T$. Since the quantity $P/T^4$ remains the
same as function of $\hat{\mu}_B$ under the shift of $T_c$
(eq.~(\ref{eq:P_t,mu_U})), the lattice QCD calculations
and the corresponding fits are unchanged for the two values of 
$T_c$ which are considered, however, they correspond to 
different isotherms, according to the displayed legend.
The IHG curves which correspond to lattice curves of
$T_c=$158(161) MeV are displayed with slashed (dotted) lines.
All calculations in (i) and (ii) are carried out at vanishing strangeness
chemical potential ($\mu_S=0$).
Each set of curves from lower to higher values of 
pressure correspond to gradually increasing values
of $\mu_B/T$ in (i) and $T$ in (ii).}}
\end{figure}
\vspace{-0.cm}



\begin{figure}[H]
\centering
(i)\includegraphics[scale=0.30,trim=0.8in 0.0in 0.5in 0.1in,angle=0]{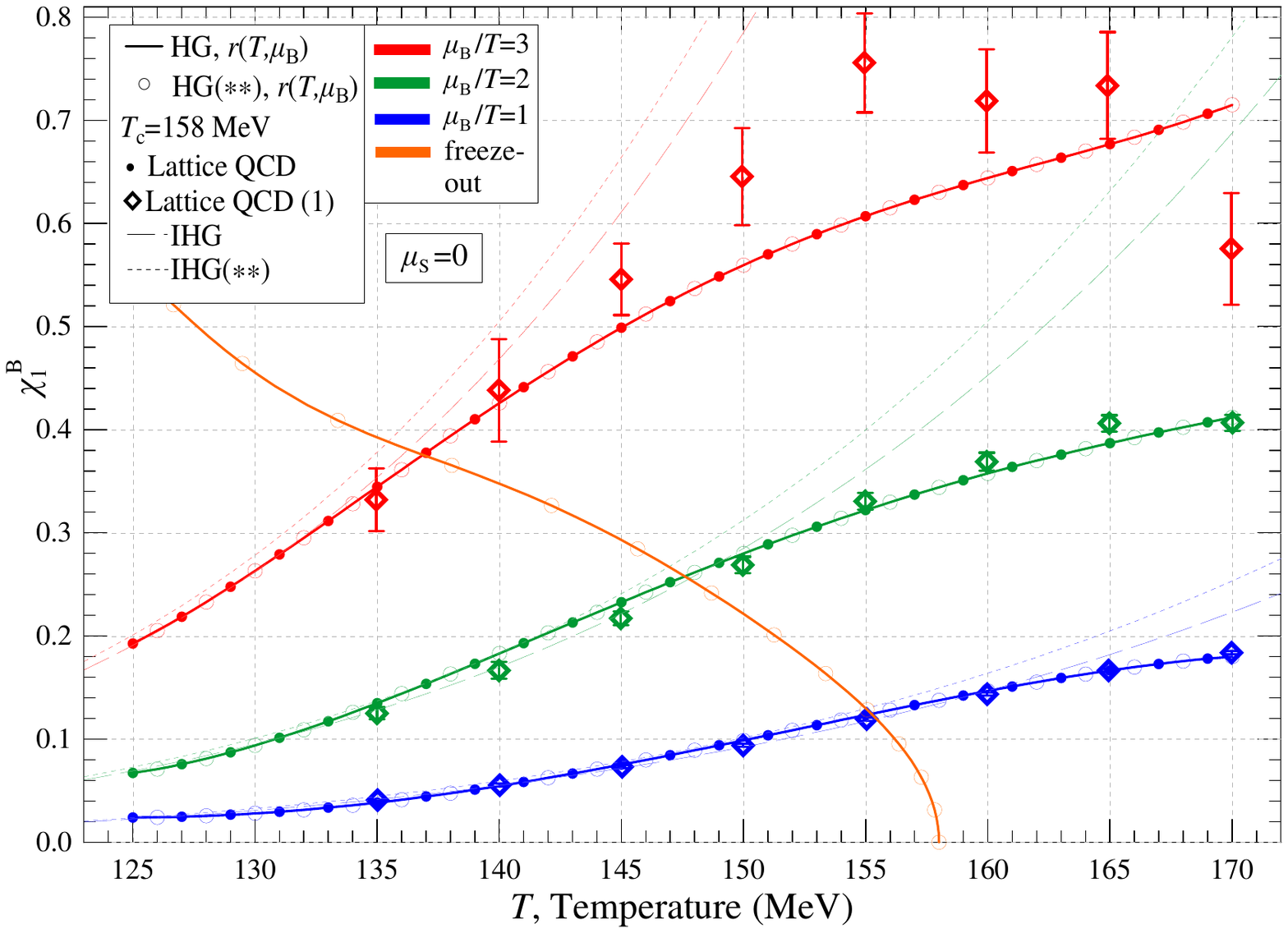} 
(ii)\includegraphics[scale=0.30,trim=0.8in 0.0in 0.5in 0.1in,angle=0]{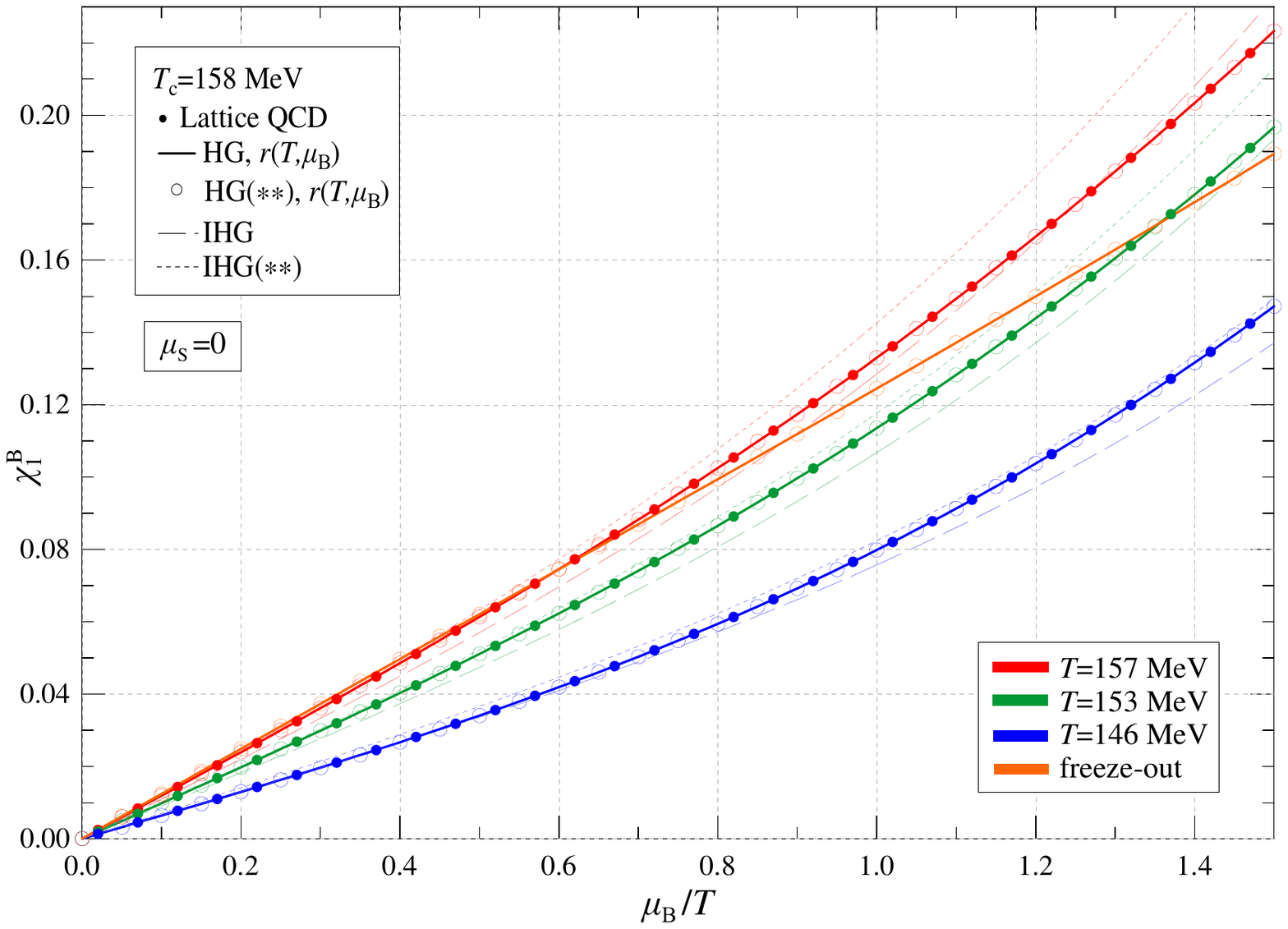}\\
(iii)\includegraphics[scale=0.30,trim=0.8in 0.0in 0.5in 0.1in,angle=0]{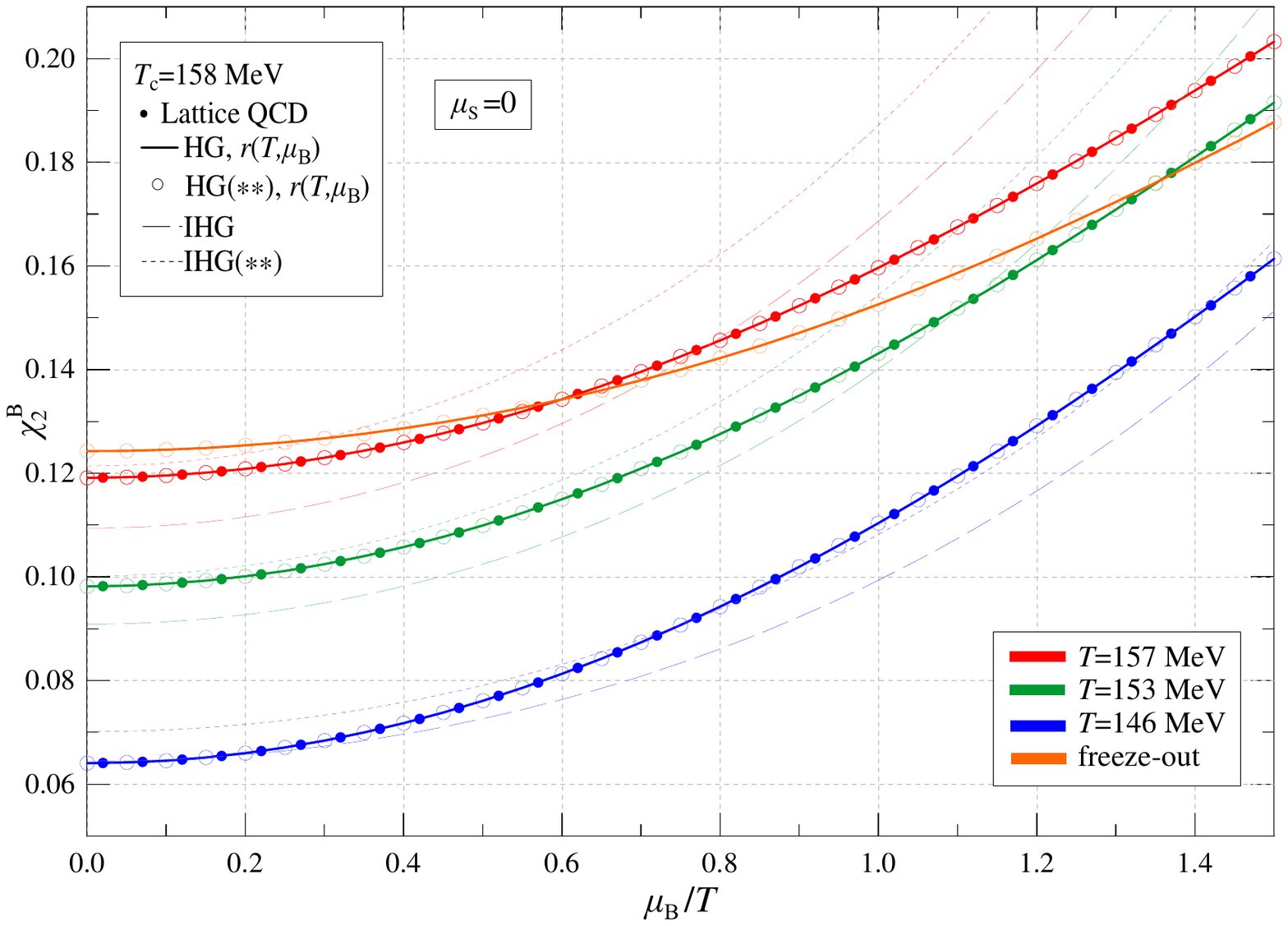}
(iv)\includegraphics[scale=0.30,trim=0.8in 0.0in 0.5in 0.1in,angle=0]{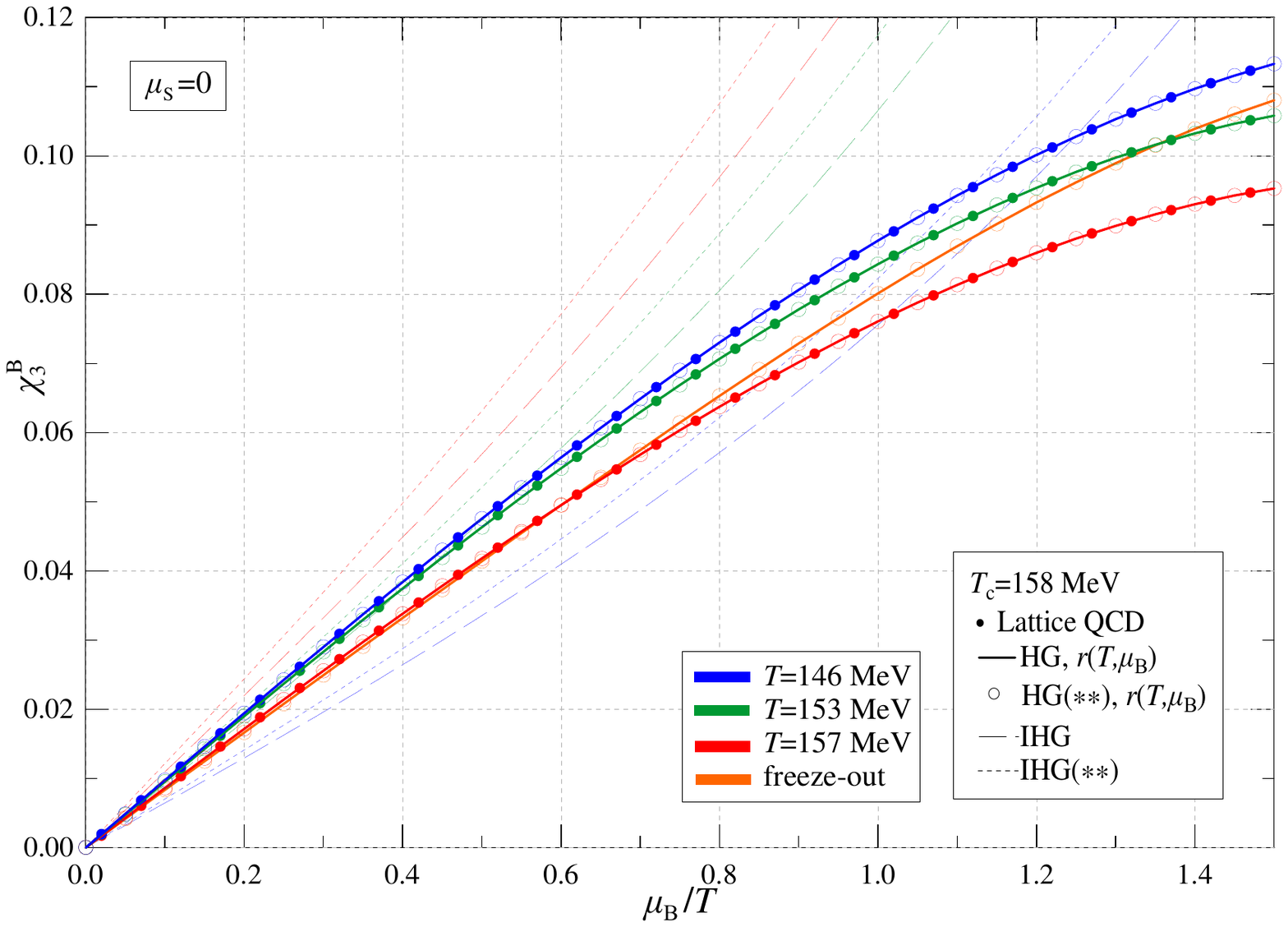}\\
(v)\includegraphics[scale=0.30,trim=0.8in 0.0in 0.5in 0.1in,angle=0]{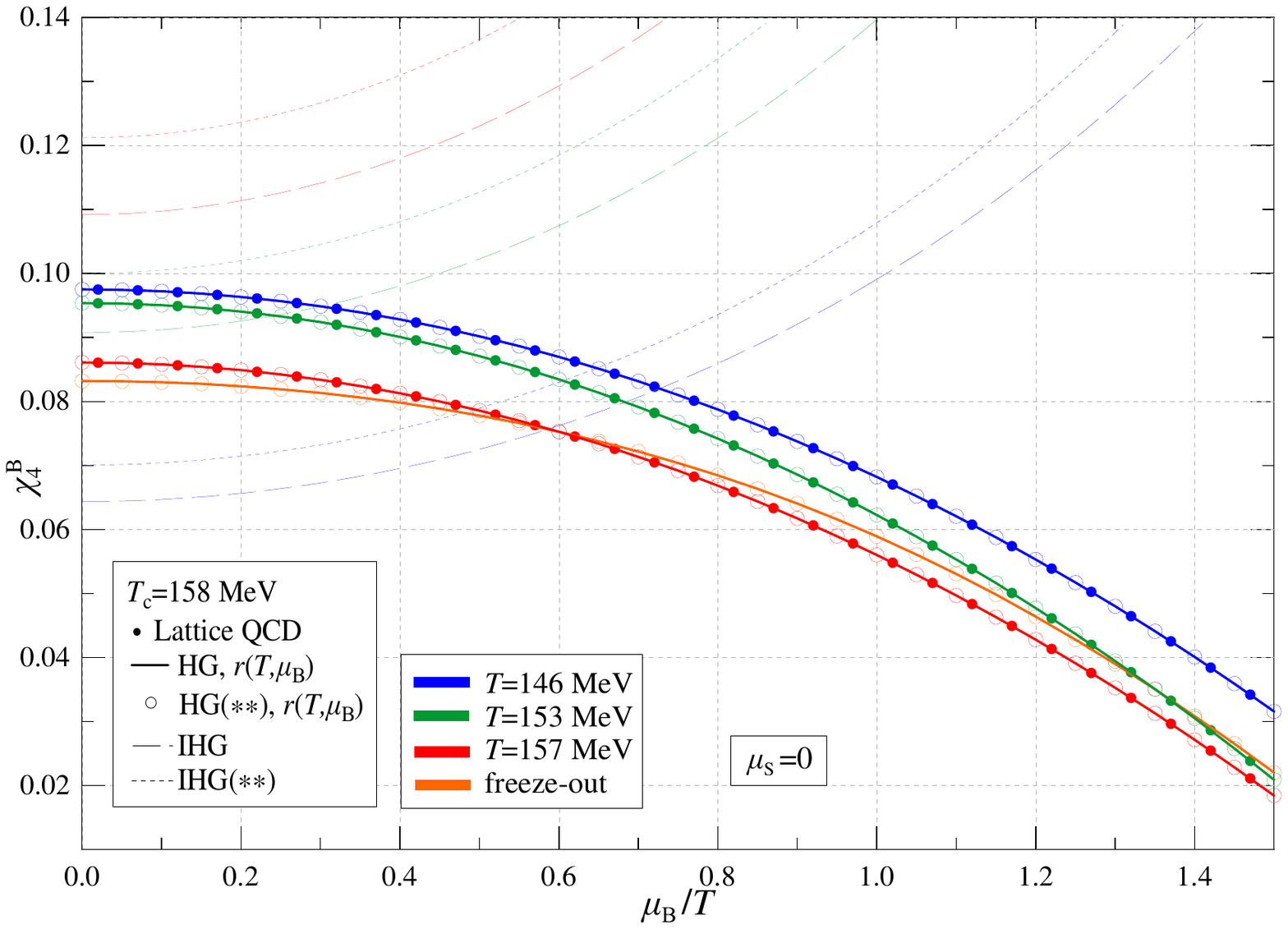}
(vi)\includegraphics[scale=0.30,trim=0.8in 0.0in 0.5in 0.1in,angle=0]{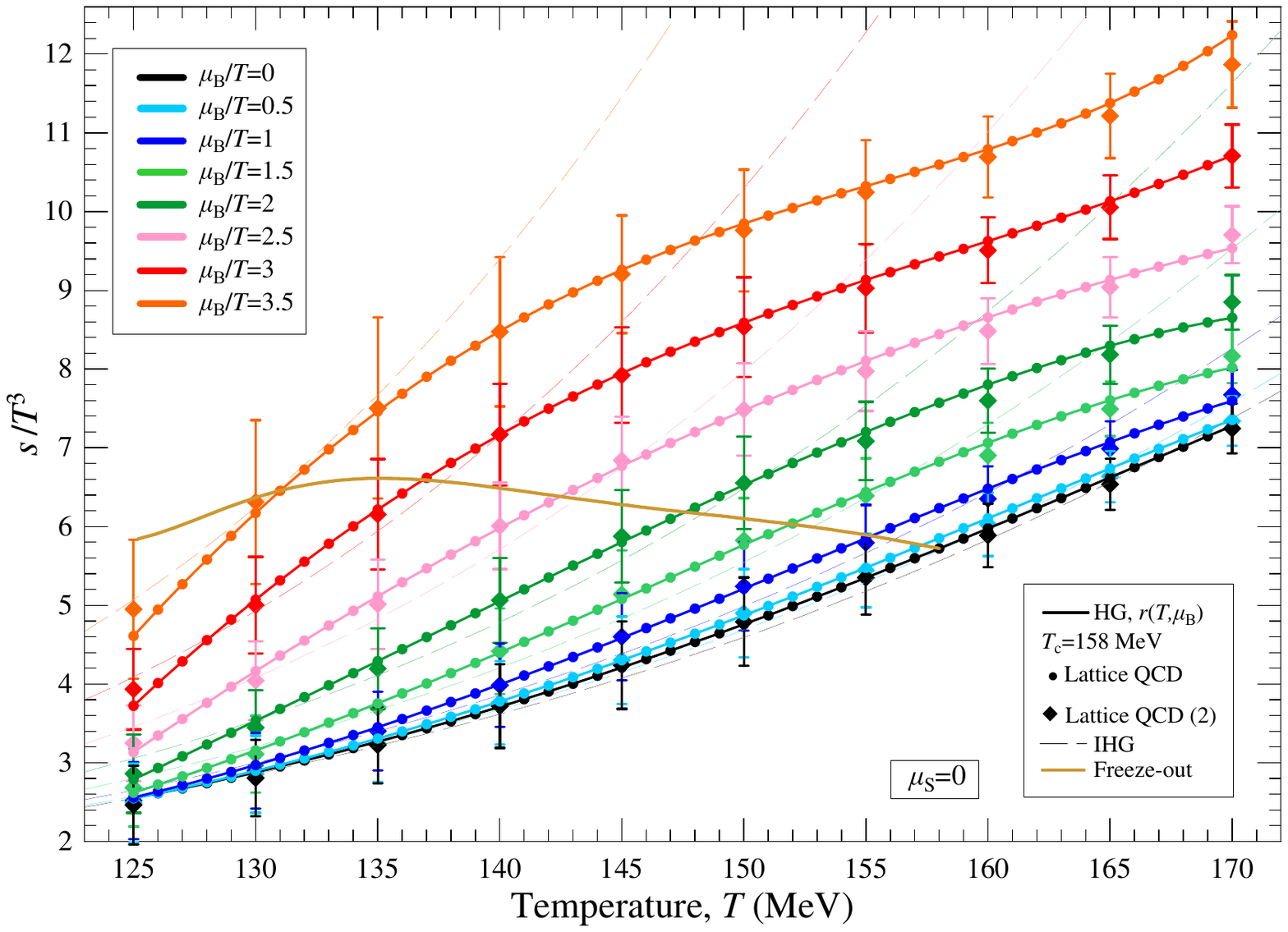}\\
\centering
\caption{\label{fig:xB} {\small  
(i)-(v) Baryon susceptibilities 
$\chi^B_j=\left(\frac{\partial}
{\partial \hat{\mu_B}}\right)^j \frac{P(T,\hat{\mu}_B)}{T^4}$,
$j=1,2,3,4$
for $T_c=$158 MeV.
(vi) Entropy density $s/T^3$ for $T_c=$158 MeV.
Solid circles are evaluations
from the fits on the lattice QCD pressure from Fig.~4 of 
\cite{lat_muB1}, while continuous thick lines are the 
susceptibilities from our HG model which uses  
hadron volumes with radius $r(T,\mu_B)$ evaluated from 
eq.~(\ref{eq:PLPHG_r0TmuB}). We plot with thin continuous lines
the boundaries set by the freeze-out curve within which the HG 
needs to hold. We, also, show the respective IHG calculations.
In (i) we include as open rhomboid points direct calculation
of $\chi^B_1$ from Fig.~1 of \cite{lat_muB1} (published and arXiv version) (N$^2$LO points)
(lattice QCD (1)).
In (vi) we include as full rhomboid points direct calculation
of $s/T^3$ from of \cite{lat_muB1} (Fig.~4 of the published and Fig.~9 of the arXiv version) 
(lattice QCD (2)).
}}
\vspace{0cm}
\end{figure}


So:
\vspace{-0.cm}
\begin{equation}\label{eq:h'}
\hat{h}'(T',...,{\hat{\mu}'_U},...) = 
\hat{h}(T,...,{\hat{\mu}_U},...) \;.
\end{equation}

Also, for a dimensionless entropy density $\hat{s} \equiv \frac{s}{T^3}$, we have:
\begin{equation}\label{eq:s} 
T\frac{\partial \hat{P}}{\partial T} = - 4\hat P + \hat s \;.
\end{equation}
Identifying in (\ref{eq:h}) $\hat{f}=\hat{P}$ and $k=1$ and using (\ref{eq:h'}), (\ref{eq:s}) and (\ref{eq:P_t,mu_U}) we arrive at:
\begin{equation}\label{eq:s'} 
{\hat s'(T',...,{{\hat \mu '}_U},...) = \hat s(T,...,{{\hat \mu }_U},...)} \;.
\end{equation}

In Fig.~\ref{fig:PL-AB}~(i)-(ii) we have plotted the lattice QCD results for the 
normalised pressure $\hat{P}$ for
vanishing value of strangeness chemical potential ($\mu_S=0$)
 extracted from Fig.~4 in \cite{lat_muB1} 
depicted as solid rhomboid points. These
correspond to the value of $T_c=$158 MeV. On the same graphs
we present our reconstruction which is carried through the fit
on these points. Also, we plot the corresponding results of the Ideal Hadron Gas (IHG), i.e.
Hadron Gas with zero hadron volumes. 
We find that the
lattice QCD pressure for $T_c=$158 MeV is lower than the IHG pressure for the
(vh) set at the area ($T,\mu_B$) we are examining. This IHG pressure, in turn, 
is lower at the same conditions than the pressure of the (*), (**), (lps) and (lp) sets.
Since the lattice QCD pressure for $T_c=$161 MeV is lower than the one for $T_c=$158 MeV
at the same conditions, the above argument is valid for this case, also. 
This means that a positive solution for the hadron volumes, making the HG and lattice QCD pressures identical for all hadron sets and $T_c\ge$158 MeV, exists.

We proceed by evaluating a volume parameter equal for all hadrons
which depends on temperature and baryon chemical potential so that
the lattice QCD and Hadron Gas pressure be equal at any point of
the $\mu_S=0$ plane. This amounts
to solving, for specific $T_c$, the equation:
\vspace{0.cm}
\begin{equation} \label{eq:PLPHG_r0TmuB}
P_L(T,\mu_B,\mu_S=0;T_c)=
P_{HG}(T,\mu_B,\mu_S=0;r(T,\mu_B))\;,
\end{equation}
where $r(T,\mu_B)$ is the radius of the common hadron volume.
At this point we make the assumption that the common hadron
volume does not depend on the strangeness chemical potential,
so we use the radii, calculated from eq.~(\ref{eq:PLPHG_r0TmuB}), as input to evaluate
our Hadron Gas model (called model (d)) for any point
($T,\mu_B,\mu_S$).

In Fig.~\ref{fig:xB} we depict the baryon susceptibilities. For the lattice
QCD we can only evaluate these susceptibilities for $\mu_S=0$
(solid lines on the graphs).
The corresponding susceptibilities from the HG (circular points)
provide a consistency test for our model.
In Fig.~\ref{fig:xB}(i) and (vi) we also include direct evaluation
of $x^B_1$ and $s/T^3$, respectively, from \cite{lat_muB1}.
We note that the HG model needs to hold only for temperatures lower than the freeze-out curve
described by eq.~(\ref{eq:freezeout}).
For comparison we include the calculations of the Ideal HG.

To locate the critical point for specific $T_c=T_{ch,0}$, apart
from eqs.~(\ref{eq:PLPHG_r0TmuB}) and (\ref{eq:freezeout_2}), we have to solve the set of equations:
\vspace{0.cm}
\begin{equation} \label{eq:den1_3}
 n_{HG,1}(T,\mu_{B},\mu_s ; r(T,\mu_B)) =
 \tilde{n}_1(T,\mu_{B},\mu_s ; \tilde{m}_1=m_{\pi})\;,
\end{equation}
\vspace{-0.6cm}
\begin{equation} \label{eq:S=0_3}
<S>_{HG} (T,\mu_{B},\mu_s ; r(T,\mu_B))=0 \;.
\end{equation}
Eq.~(\ref{eq:S=0_3}) acquires the same form as eq.~(\ref{eq:S=0_4}), where we now identify
$v=v(T,\mu_B)$. The derivative 
$\frac{\partial v}{\partial \mu_B}$ can be provided
numerically from the values of $r$ at adjacent points
evaluated through eq.~(\ref{eq:PLPHG_r0TmuB}).

\begin{figure}[H]
\centering
(i)\includegraphics[scale=0.30,trim=1.3in 0.8in 0.5in 0.2in,angle=0]{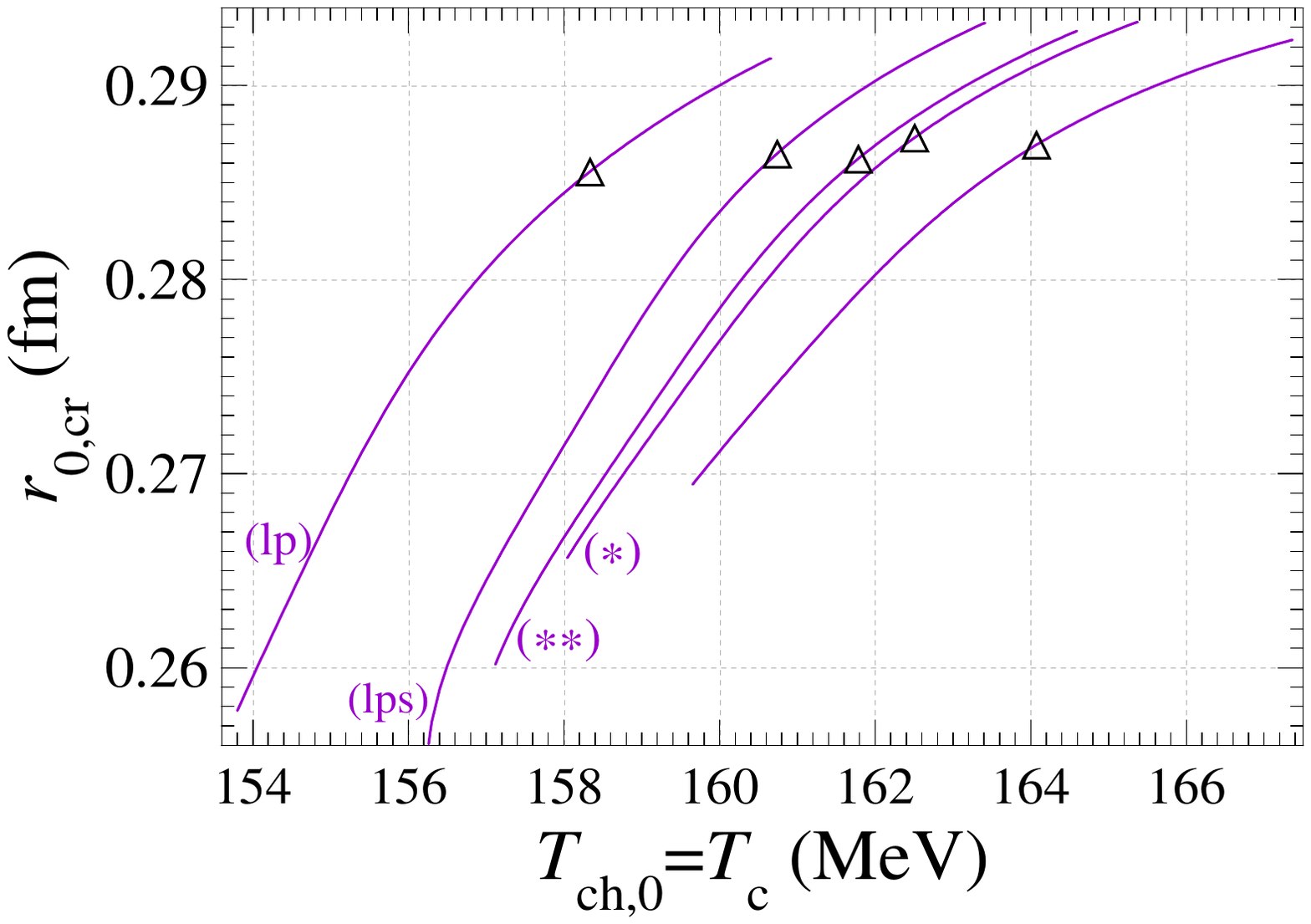} 
(ii)\includegraphics[scale=0.30,trim=0.8in 0.8in 1.in 0.2in,angle=0]{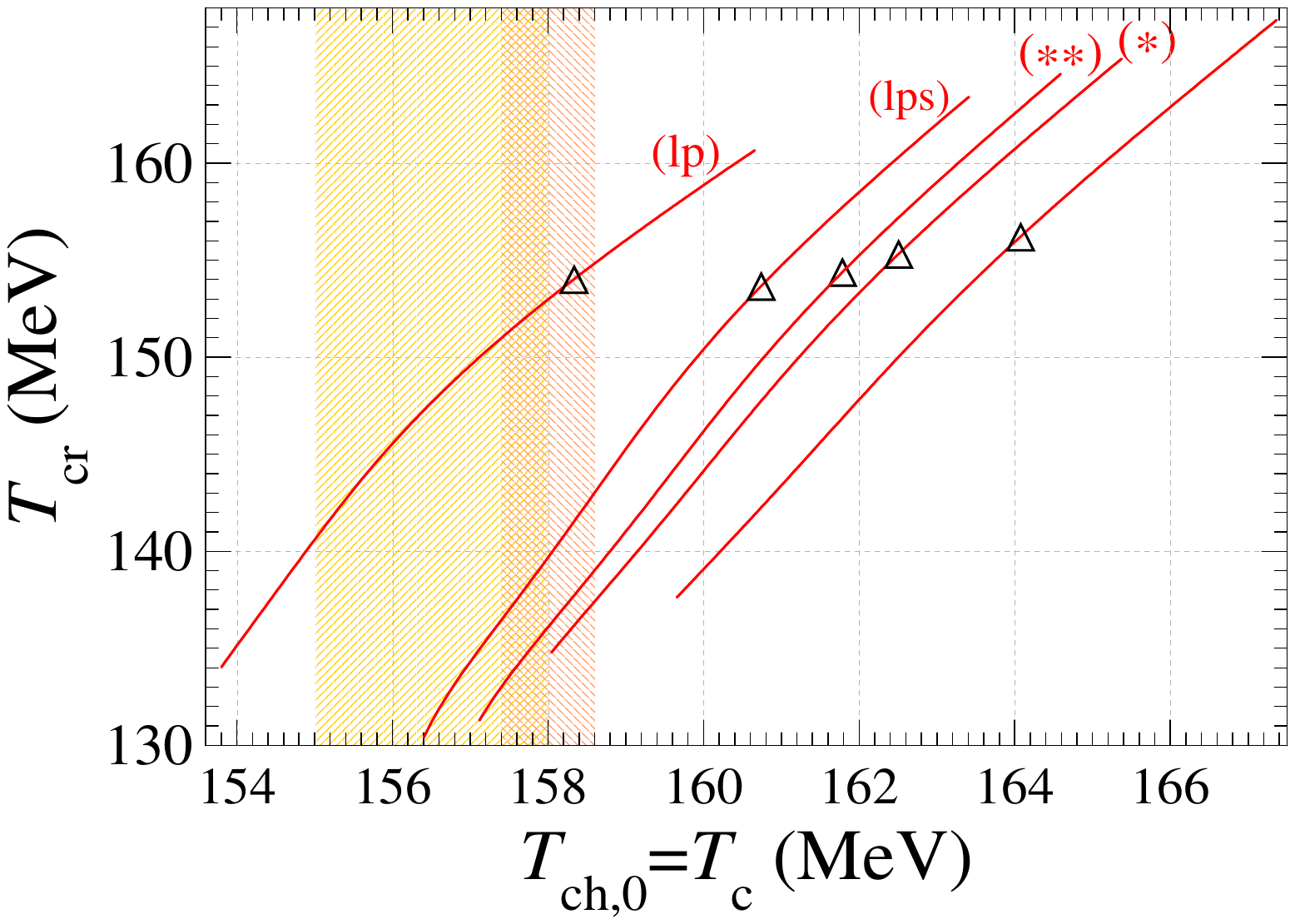}\\
(iii)\includegraphics[scale=0.30,trim=0.8in 0.8in 1.in 0.2in,angle=0]{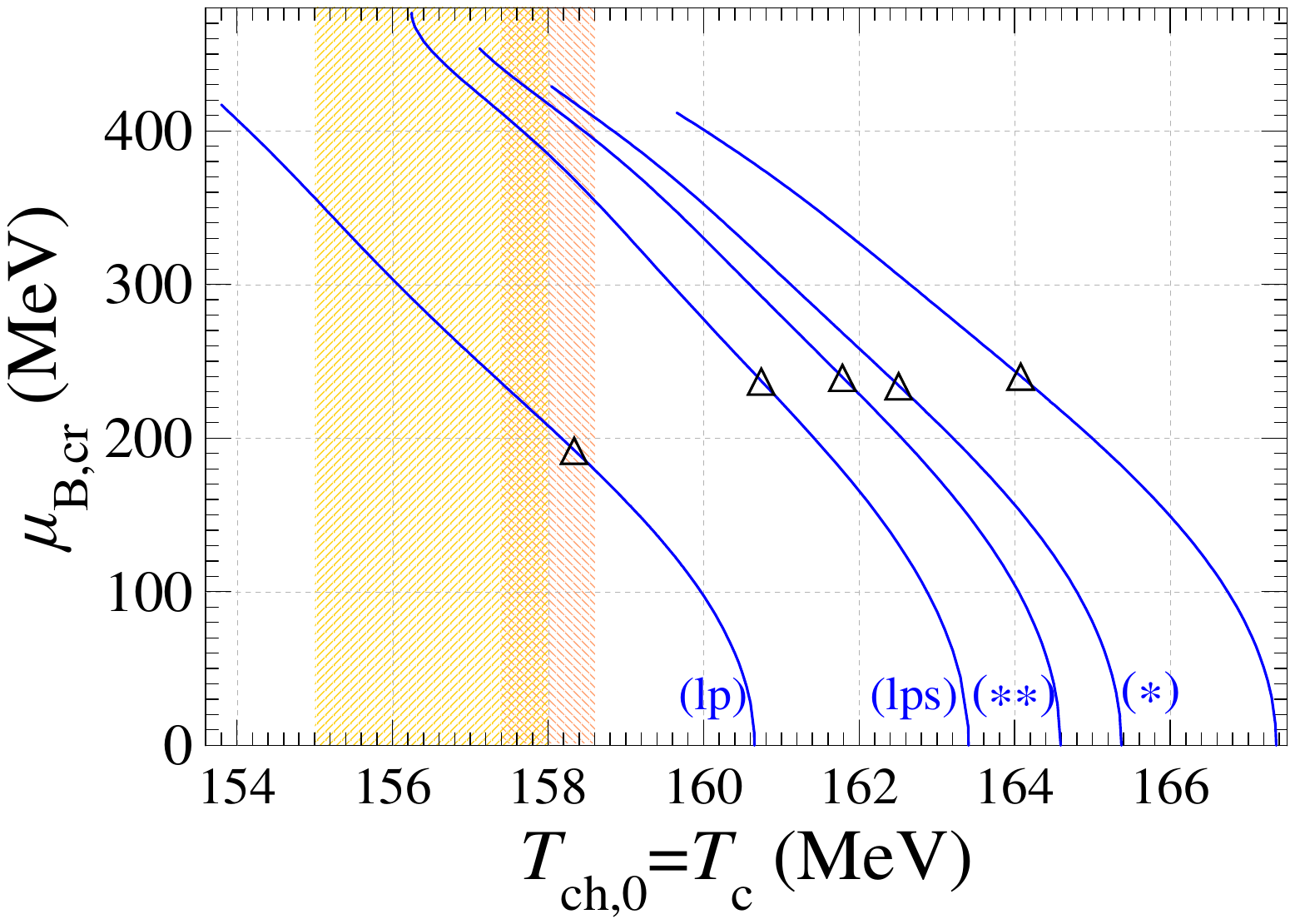}
(iv)\includegraphics[scale=0.30,trim=0.8in 0.8in 1.in 0.2in,angle=0]{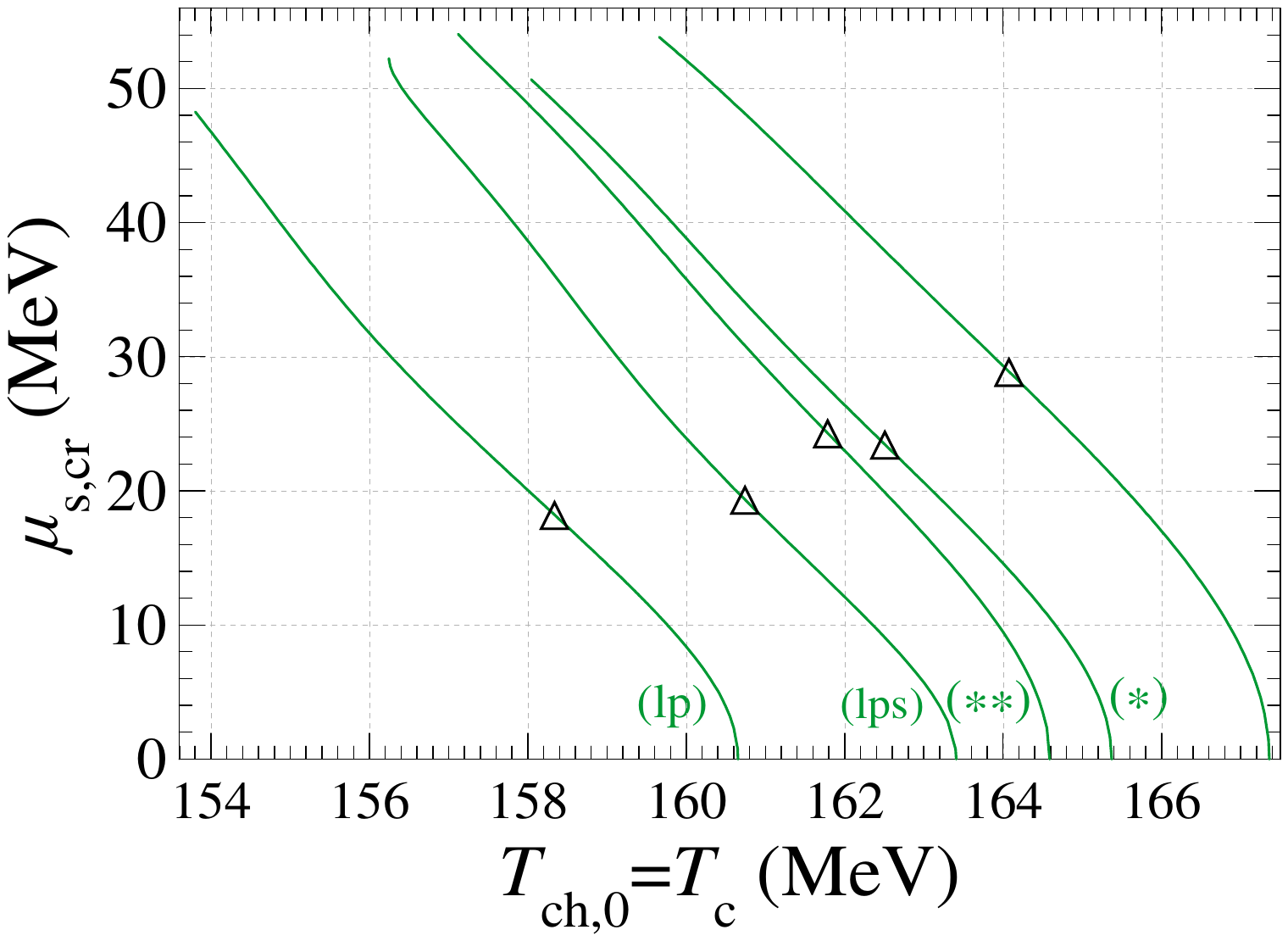}\\
(v)\includegraphics[scale=0.30,trim=0.8in 0.8in 1.in 0.2in,angle=0]{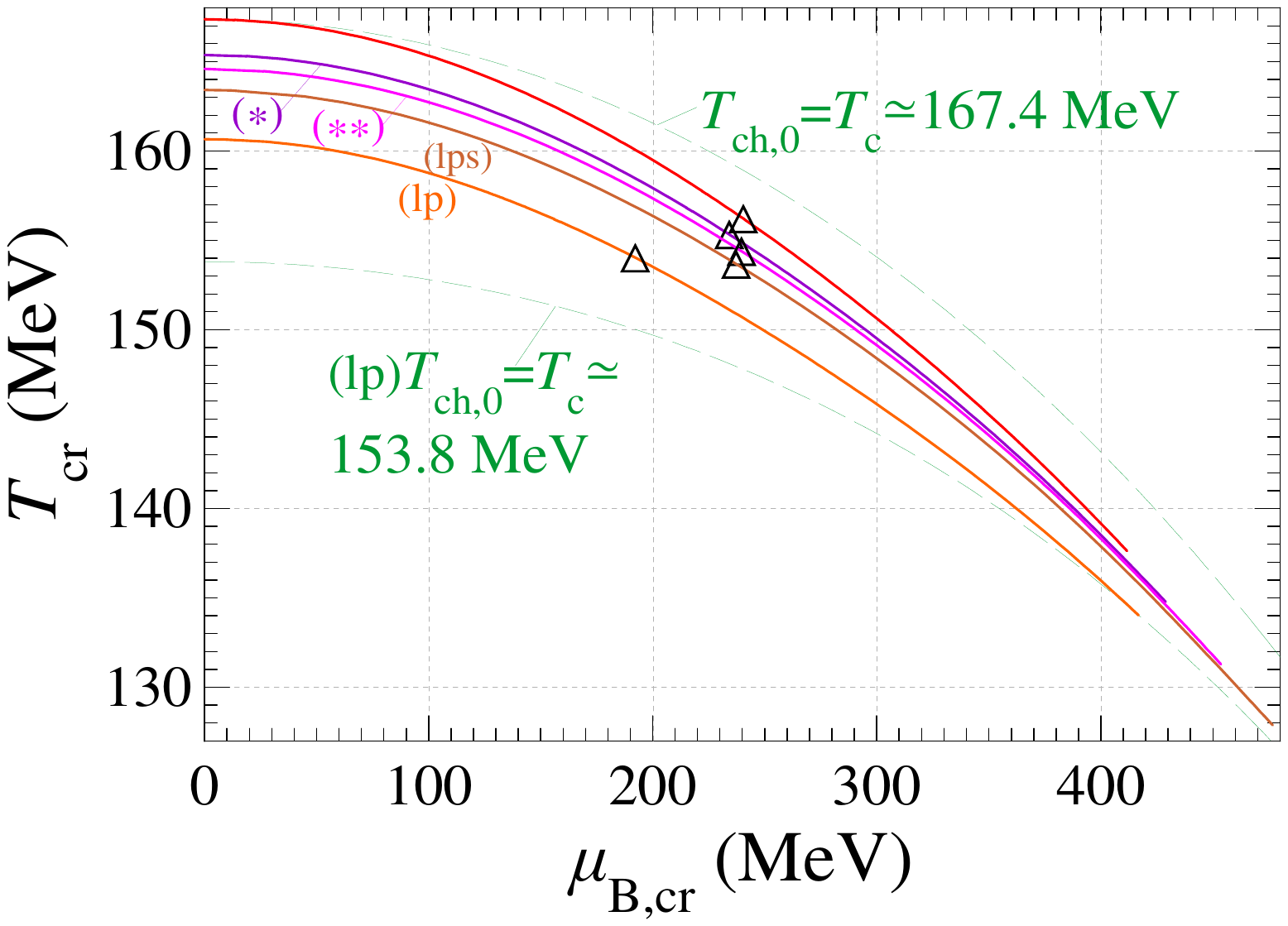} 
(vi)\includegraphics[scale=0.30,trim=0.8in 0.8in 1.in 0.2in,angle=0]{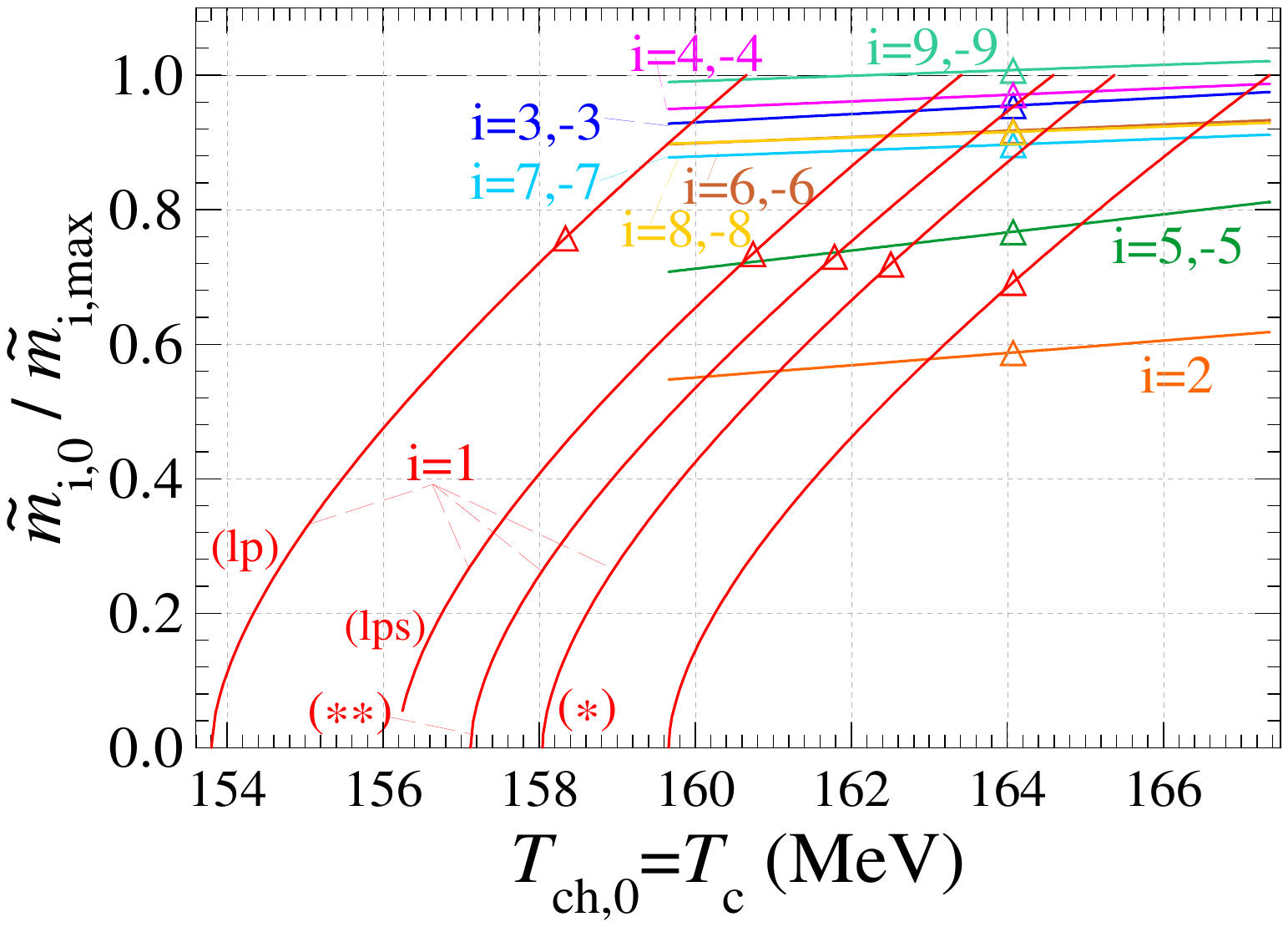}\\
\vspace{0.1cm}
\caption{\label{fig:cpr_TmuB} {\small  
Graph similar to Fig.~\ref{fig:cpr0}.
Calculations are carried out in the interval $T_c\simeq$(153.8-167.4) MeV 
and involve common hadron radius which depends on temperature and 
baryon chemical potential, $r(T,\mu_B)$, determined by the 
lattice pressure at $\mu_S=0$ (volume model (d)), for 5 hadron sets.
In (v) the freeze-out curves which correspond to the maximum value of $T_c=$167.4 MeV for
the (vh) set and  
the minimum value of $T_c=$153.8 MeV for the (lp) hadron set are shown. 
In all graphs with triangles we present the 
critical points which additionally fulfil the criterion of Section \ref{sec:crit}.}}
\end{figure}

We present solutions for the critical point in the same manner as
in the previous Subsections (\ref{subsec:v.m.a}-\ref{subsec:v.m.c})
in Figs.~\ref{fig:cpr_TmuB}(i)-(vi).

We find that for $T_c= T_{ch,0} \gtrsim$153.8-159.7 MeV there is positive solution for the chiral pion 
mass at $\mu_B=0$ and that for $T_c\simeq$160.7-167.4 MeV 
the position of the critical point is located at zero baryon chemical potential, thus, the values of $T_c$ 
between these values
produce acceptable solutions for the critical point.
We note that the rather high values of $T_c$ in the
interval (166-167.4) MeV stay within the limit
of the freeze-out temperature at low baryon chemical
potential $T_f=166 \pm 3$ MeV recorded in
\cite{fr_muB=0}.
Also, in 
Figs.~\ref{fig:cpr_TmuB}(i)-(vi) we show by triangle 
the critical points which further fulfil the criterion that the densities of $q$ and $s$ quarks, which are contained in mesons
and baryons, are equal (described in the next Section 
\ref{sec:crit}). This corresponds to 
$T_c \simeq$158.3-164.1 MeV and it is located at $\mu_{B,cr}\simeq$234.2-240.3 MeV for the 4
first hadron sets or $\mu_{B,cr}\simeq$192.1 MeV for the (lp) set and 
$T_{cr}\simeq$153.7-156.2 MeV.



\vspace{-0.5cm}



Next we can evaluate the thermodynamic variables on the crossover and the 1st order transition curve, for a specific
value of 
$T_c$ which we choose to be the aforementioned one for the
(vh) set, $T_c=$164.1 MeV. The results are shown in 
Figs.~\ref{fig:Tc=164.1}(i)-(iii).
We observe in Fig.~\ref{fig:Tc=164.1}(ii) that for some families
the antibaryon chiral masses, at high values of $\mu_B$ rise
above their maximum allowed values.
On this effect we have 
to comment that:
1) The calculations are carried out at high $\mu_B\sim 480$ MeV and relative low $T\sim 130$ MeV

\begin{figure}[H]
\centering
(i)\hspace{-2cm}\includegraphics[scale=0.65,trim=0.5in 0.8in 2in 0.5in,angle=0]{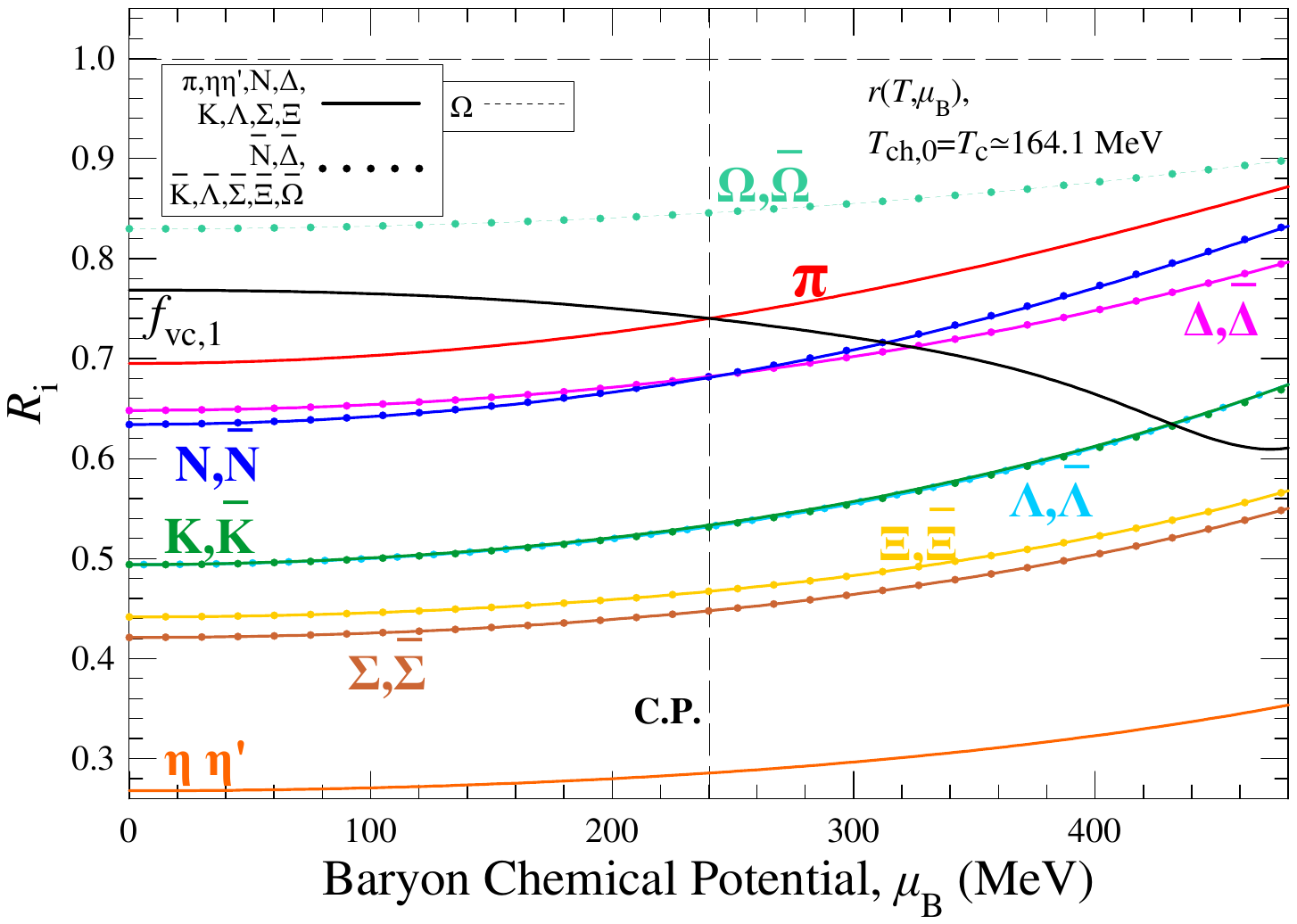} \\
(ii)\hspace{-2cm}\includegraphics[scale=0.65,trim=0.5in 0.8in 2in 0.5in,angle=0]{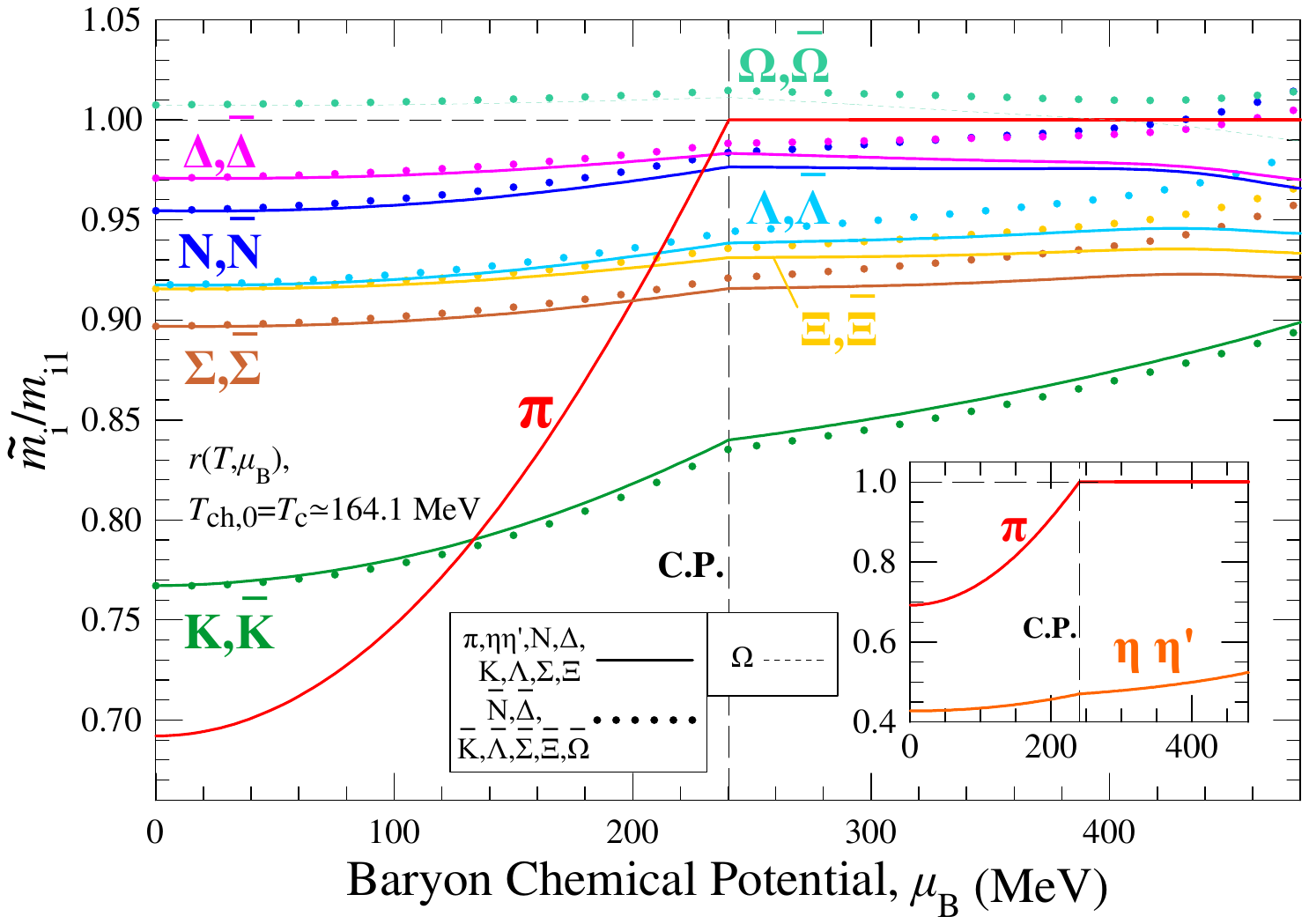}
\end{figure}
\vspace{2cm}
\begin{figure}[H]
\centering
(iii)\hspace{-1cm}\includegraphics[scale=0.75, trim=0in 1.3in 0in 1.in]{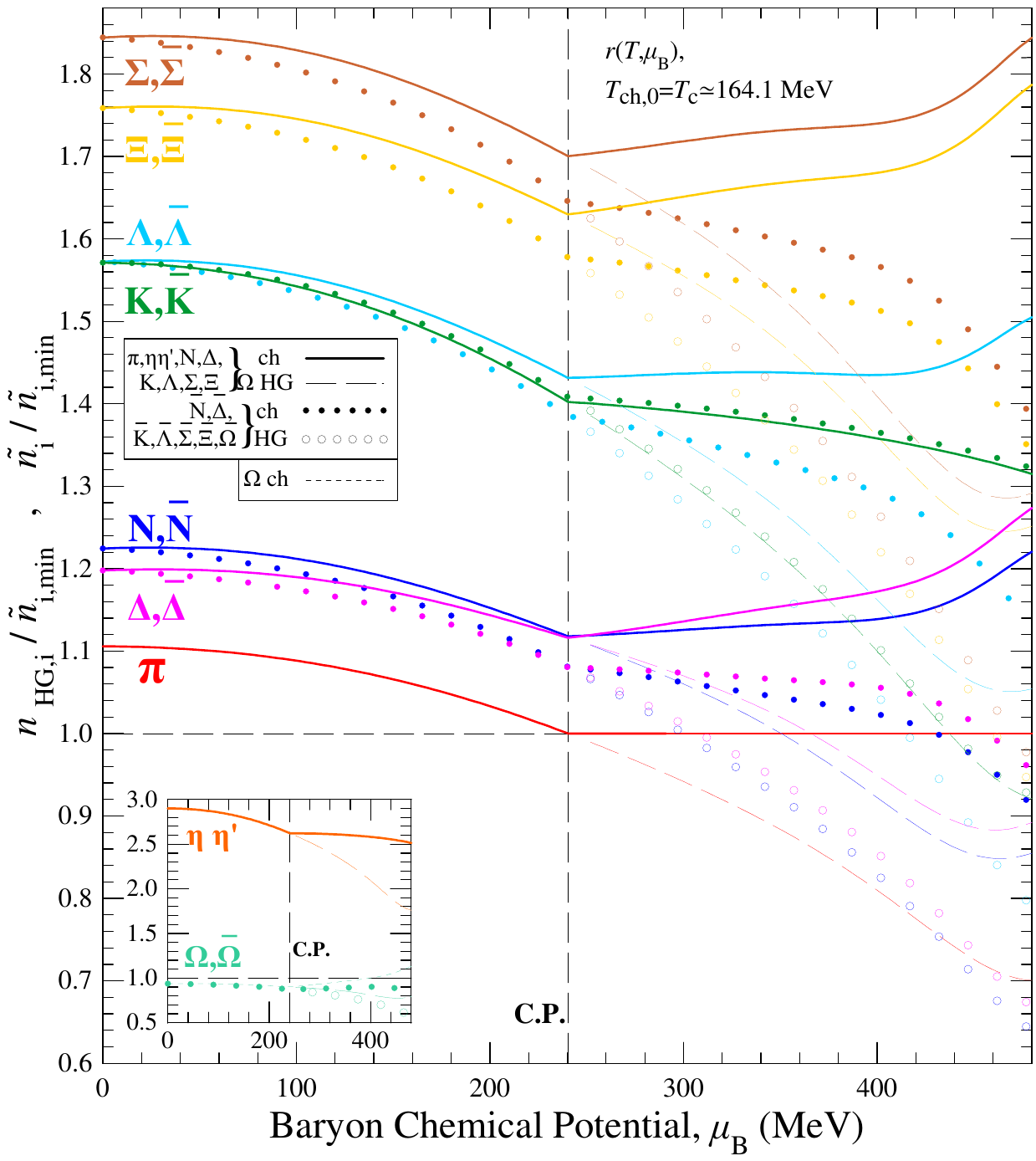}
\caption{\label{fig:Tc=164.1} {\small 
Graph similar to Fig.~\ref{fig:Tc=159.4}.
Calculations are carried out for $T_c=T_{ch,0}\simeq$164.07 MeV and 
common hadronic radius $r_m(T,\mu_B)$, determined by the lattice pressure 
for $\mu_S=0$ for the (vh) set.
In (i) the shown volume correction factor for the pion family, $f_{vc,1}$, intersects with 
$R_1$ at $\mu_B\simeq$240.32 MeV to produce at this location the critical point.
}}
\end{figure}


\noindent
where the lattice
QCD calculations may not be so precise.
2) Only one volume parameter is used for all hadrons. An introduction of a separate volume parameter for the
baryons and antibaryons families may remedy the result. 
3) The problem appears on antibaryon species existing in large positive $\mu_B$, where their 
density is highly suppressed. $\;$
Therefore, their presence is expected to have a minor effect on the system.

\begin{figure}
\vspace{-1.4cm}
\centering
\includegraphics[scale=0.55,trim=0.5in 0.8in 1.in 0.2in]{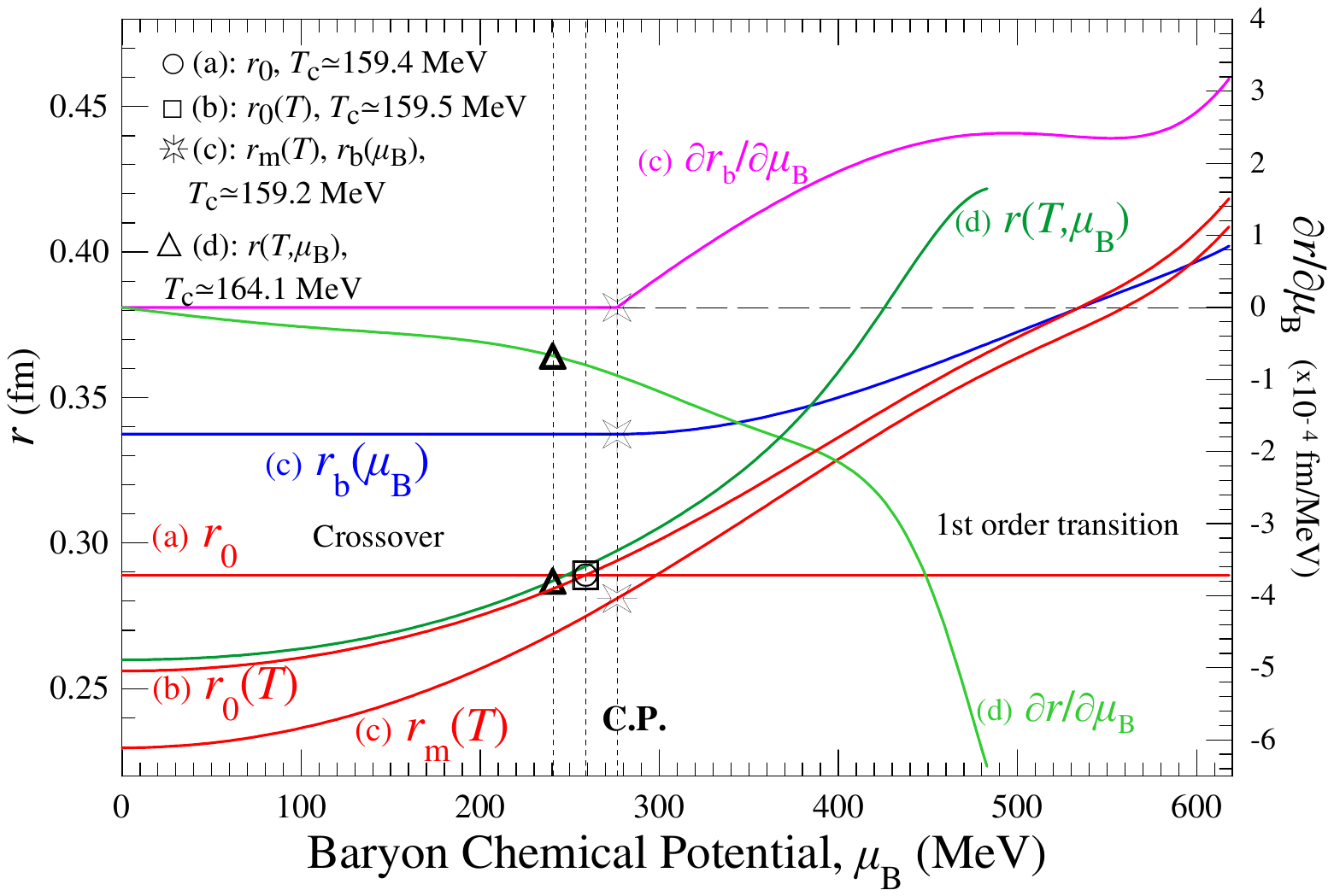}
\caption{\label{fig:rm,rb,drbdmB} {\small 
The hadronic radius $r_0$ (volume model (a)), for the calculations of Fig.~\ref{fig:Tc=159.4}, 
the hadronic radius $r_0(T)$ (volume 
model (b)), for the calculations of Fig.~\ref{fig:Tc=159.5}, the meson $r_m(T)$ and the 
baryon $r_b(\mu_B)$ radii (volume model 
(c)) for the calculations of 
Fig.~\ref{fig:Tc=159.2} and the hadronic radius $r(T,\mu_B)$ (volume 
model (d)), for the calculations of Fig.~\ref{fig:Tc=164.1} (along the respective transition 
curves), as function of the baryon chemical potential $\mu_B$ (left axis). 
Also, shown the derivative of the baryon radius 
$\partial r_b/\partial \mu_B$ for model (c) and the derivative of the common hadron 
 radius 
$\partial r / \partial \mu_B$ for model (d) with respect to the baryon chemical potential 
(right axis). All cases fulfil the additional criterion of Section \ref{sec:crit}.
The shown calculations correspond to the (vh) set.}}
\end{figure}
\begin{figure}
\centering
\includegraphics[scale=0.48,trim=0.5in 0.3in 1.in 0.2in,angle=0]{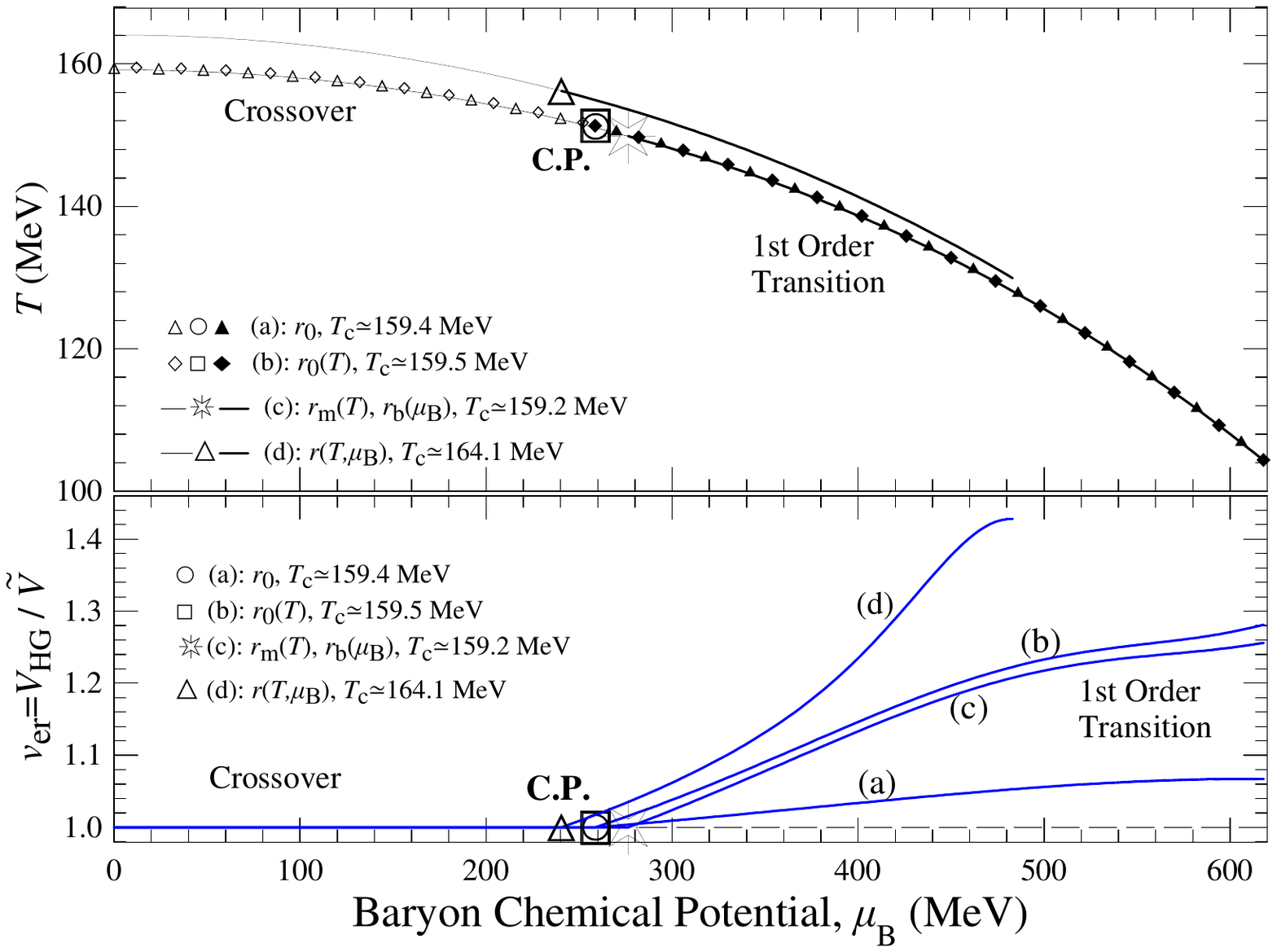}\\
\caption{\label{fig:1234T-MBVER} {\small 
{\it Lower:} 
The volume expansion ratio, $v_{er}$, as function of $\mu_B$
for the calculations of Fig.~\ref{fig:Tc=159.4},
Fig.~\ref{fig:Tc=159.5}, Fig.~\ref{fig:Tc=159.2} and Fig.~\ref{fig:Tc=164.1}
(volume models (a), (b), (c) and (d) respectively).
{\it Upper:} 
The transition curve (crossover and 1st order region) with the respective critical point for the 
calculations of 
Figs.~\ref{fig:Tc=159.4}, \ref{fig:Tc=159.5}, \ref{fig:Tc=159.2} and 
\ref{fig:Tc=164.1} in the $T,\mu_B$ plane.
The shown calculations correspond to the (vh) set.}}
\vspace{-0.4cm}
\end{figure}


We end this section by presenting
some further characteristics of
the models we analysed. We utilise
the solution for each model which further satisfies the criterion
of the equality of the densities of $q$ and $s$ quarks  
contained in mesons and baryons (described in the next section) for the (vh) set.

In Fig.~\ref{fig:rm,rb,drbdmB} we present for the models (a)-(d)
the solution on the transition curve (crossover-1st order) for the 
hadron radius. For model (c) there are two separate radii for mesons 
and baryon-antibaryons. We, also, present the solution for model (c) 
for the derivative of $r_b$ and for model (d) for the derivative of 
$r$ with respect to the baryon chemical potential on the transition curve.

In Fig.~\ref{fig:1234T-MBVER} we show for all four models of Subsections 
\ref{subsec:v.m.a}-\ref{subsec:v.m.d} the volume 
expansion ratio 
$v_{er}$ for the same values of $T_c$ as in Fig.~\ref{fig:rm,rb,drbdmB}. We, also, show the 
corresponding transition curves in the 
$T-\mu_B$ plane,  as well as, the relevant critical 
points.

\section{A criterion for the Critical Point} \label{sec:crit}

At the transition line and at low baryon density (part of the crossover region) Hadron Gas consists mainly 
of mesons\footnote{Actually at $\mu_B=$0 and high $T$ the density of baryons and antibaryons is considerable, but small compared to the mesonic density. For example, at $T=$ 158 MeV 
the density of baryons plus antibaryons is $\sim$14,5\% of the whole hadron density for the (vh) set.}. At high baryon density (part of the 1st order transition) Hadron Gas consists mainly of baryons. 
Also, the temperature of the chiral phase transition at zero baryon chemical potential is of the 
order of the mass of the lightest meson, the pion. The transition baryon chemical potential of 
the chiral phase transition at zero temperature is of the order of the mass of the lightest 
baryon, the proton. We may, therefore, postulate that the attributes of the crossover region are 
mainly due to  mesons, while the attributes of the region of the 1st order transition are mainly 
due to baryons. 

For this reason we calculate the density of quarks ($u$, $d$ named collectively as $q$ quarks 
and $s$ quarks) contained in mesons and baryons. We note that both $q$ and $s$ quarks carry 
baryon number. So, since we are seeking the critical point at positive baryon density, we have 
to take into account the total densities of both $q$ and $s$ quarks which are contained in 
hadrons, which are:
\begin{equation} \label{eq:nqsm}
n_{qs,m}=n_{HG,mesons}=n_{HG,1}+n_{HG,2}+n_{HG,5} \;,
\end{equation}
\vspace{-0.3cm}
\begin{equation} \label{eq:nqsb}
n_{qs,b}=3 n_{HG,baryons}=
3(n_{HG,3}+n_{HG,4}+n_{HG,6}+n_{HG,7}+n_{HG,8}+n_{HG,9})\;,
\end{equation}
where the index $i$ in the densities $n_{HG,i}$ refers to the respective family of hadrons.
Then an additional criterion for the position of the critical point is 
\vspace{-0.0cm}
\begin{equation} \label{eq:nqsmb}
n_{qs,m}=n_{qs,b} \Rightarrow n_{HG,mesons}=3 n_{HG,baryons} \;.
\end{equation}
As it is evident from eq.~(\ref{eq:nqsmb}) this criterion is contrasted to the 
criterion used in \cite{sm=sb}, where the 
entropy of mesons equals the entropy of baryons and leads to substantial higher values
of baryon chemical potential on the freeze-out curve.

We evaluate the densities $n_{qs,m}$ and $n_{qs,b}$ for the critical points determined through 
the models (a) ($r_0$ fixed), (b) ($r_0(T)$), (c) ($r_m(T)$ and $r_b(\mu_B)$)  and (d) ($r(T,\mu_B)$). Our results for these densities and for the (vh) set
are shown in Fig.~\ref{fig:nqs,mb}. The intersection of the two densities 
determines the position of the critical point for the (vh) set.
In the same figure we depict the positions of the critical points for the (*), (**), (lps) and (lp)
hadron sets.
Overall, for the four first hadron sets, we observe that for model (a) the critical point, which 
additionally satisfies the equality of the densities of the quarks contained in mesons and 
baryons, is located at $\mu_{B,cr} \simeq$ 251-259 MeV. The model (b) gives
almost equal $\mu_{B,cr} \simeq$ 250-259 MeV, while model (c) evaluates this position at 
$\mu_{B,cr} \simeq$ 
272-279 MeV and model (d) at $\mu_{B,cr} \simeq$ 234-240 MeV.
For the (lp) hadron set the critical point is located at $\mu_{B,cr} \simeq$ 192-202 MeV 
for volume models (a), (b) and (d). These findings should be treated cautiously, since
the (lp) hadron set contains an abundance of baryonic states 
while only a few mesonic ones more than the (**) hadron set (see Fig.~1 in \cite{lp}). 
As a consequence the point which satisfies eq.~(\ref{eq:nqsmb}) is shifted to lower baryon chemical 
potentials for the (lp) hadron set with respect to the (**) hadron set.
If we add in the (lp) set more hypothetical meson states the critical point is again shifted to
higher values of $\mu_B$. The (lps) set though, which contains fewer baryonic states than the 
(lp) set, leads to solutions close the (vh), (*) and (**) sets.

\begin{figure}[H]
\vspace{-0.6cm}
\centering
\includegraphics[scale=0.55,trim=0.5in 0.9in 1.in 0.2in,angle=0]{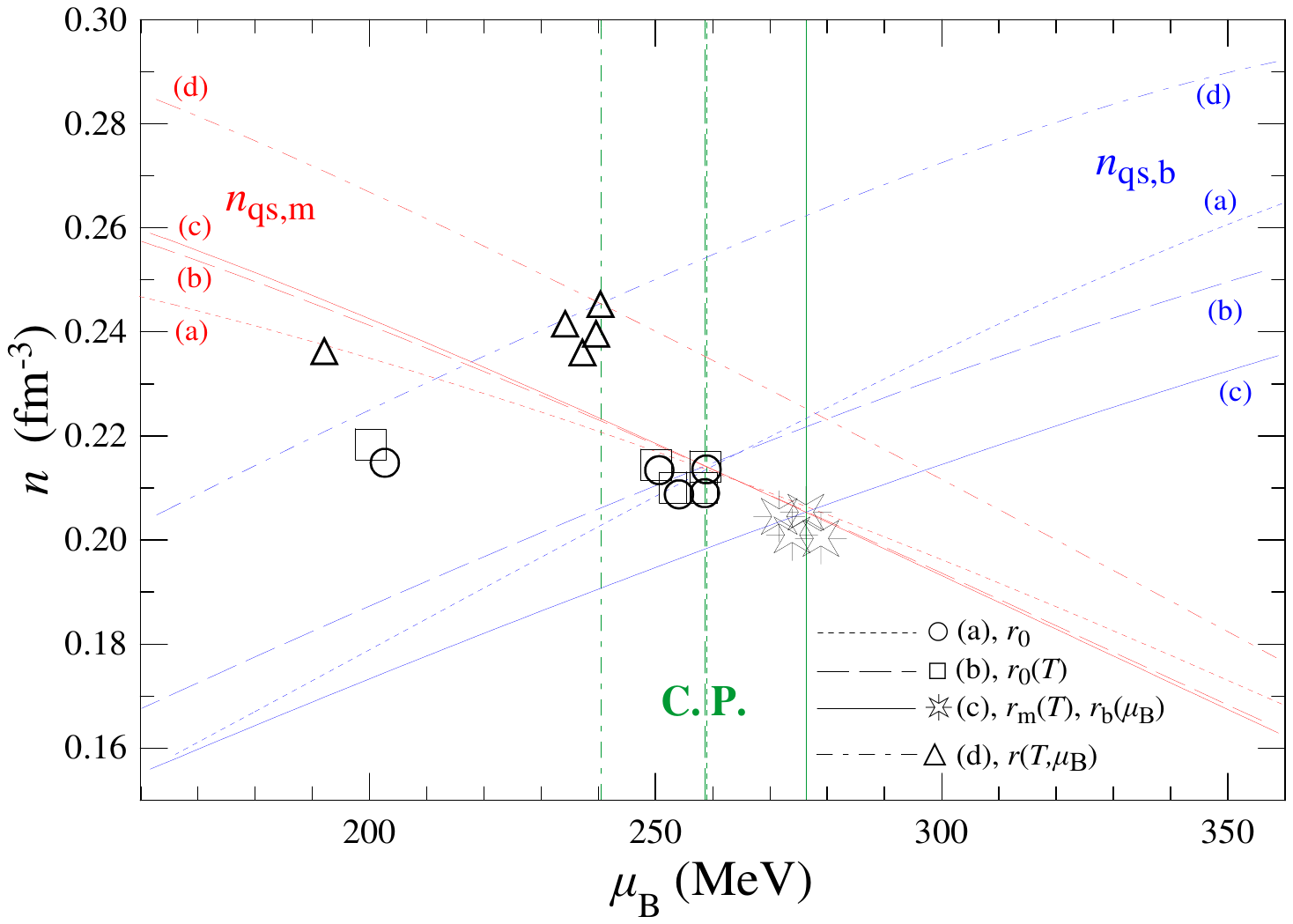}
\caption{\label{fig:nqs,mb} {\small 
The densities of quarks (up, down $q$ and strange $s$) contained in mesons $m$ and baryons $b$ 
of the Hadron Gas as function of the 
baryon chemical potential calculated for the different conditions of  Fig.~\ref{fig:cpr0} 
(model (a), dotted line), Fig.~\ref{fig:cpr0(T)} (model (b), slashed line),
Fig.~\ref{fig:cprmrb}  (model (c), continuous line) and 
Fig.~\ref{fig:cpr_TmuB}  (model (d), slashed-dotted line), as function of $\mu_B$ for the verified hadron set. The intersection of 
the two densities for each model determines a critical point which additionally fulfils 
$n_{qs,m}=n_{qs,b}$. This occurs for model (a) at $\mu_B\simeq$258.88 MeV, for model (b) at 
$\mu_B\simeq$258.54 MeV, for model (c) at $\mu_B\simeq$276.23 MeV
and for model (d) at $\mu_B\simeq$240.32 MeV. The corresponding solutions for the
critical point for the (*), (**), (lps) and (lp) hadron sets are also displayed.
The solutions with the lower baryon chemical potential for volume models (a), (b) and (d)
correspond to the (lp) set.
}}
\end{figure}

It is, also, possible to calculate and compare the densities of $q$ quarks, as well as, the 
densities of $s$ quarks alone, which are
contained in mesons and baryons. However, this entails the exact knowledge of the factors 
$c_1,~c_2$ for certain hadrons of family 2 with hidden strangeness and content 
$c_1 (u \bar{u} +d \bar{d} ) +c_2 s \bar{s}$. Also, by dealing with higher densities, that is 
inclusive densities of $q$ and $s$ quarks instead of separate densities, our results are less 
affected by any lack of knowledge of the mass spectrum of hadrons.

We remark, further, on the density of the antiquarks $\bar{q}, \bar{s}$. The density of 
antiquarks contained in mesons is always equal to the density of quarks contained in the same 
mesons. However, the density of antiquarks contained in anti-baryons is significantly less 
than the density of the quarks contained in baryons, since we are in the positive baryon density 
region. Thus, nowhere in this region the density of antiquarks contained in mesons can become 
equal to the density of antiquarks contained in anti-baryons.

\section{Further Development \& Experimental Tests} \label{sec:FD-ET}

The model we have presented assumes a chiral hadron-like phase in contact with a Hadron Gas
phase along a transition area in the phase diagram which is shrinked to a curve.
It is a minimal approach which conserves the particle numbers per family and
investigates how we can pass from crossover to the critical point and then to the first order
transition on this curve.
It relies heavily on information drawn from the HG phase, for which there is
substantial accumulated knowledge.
But it cannot accommodate the continuation of higher derivatives of the pressure 
between the two phases. 
The next step is to have a full description of the chiral phase for every value of 
temperature and independently chemical potentials.
This would require knowledge of: a) how the chiral masses vary, 
b) how the degrees of freedom gradually change from those of the hadrons to those 
of independent quarks, c) the gluonic pressure.
Such a fully constructed model should reproduce the lattice QCD calculations
for temperatures above the freeze-out curve, which, also,
could serve as a guide to fix any possible free parameters. 
For temperatures below the freeze-out curve our model is already in agreement
with the lattice QCD data (Figs.~\ref{fig:PL-AB} and \ref{fig:xB} for volume model (d)).

In this work, the chiral pion mass along the transition curve, as we move from zero to higher
baryon density, increases and finally reaches the value of the pion mass at the C.P.
So we propose the determination of the chiral pion mass from zero to higher
values of $\mu_B$ as a signal of approach to the QCD critical point.
We stress that the behaviour of the chiral pion mass  along the transition 
curve is not known {\it ad hoc}. Even for fixed mass spectrum it depends heavily on the
hadronic eigenvolumes which may not be constant. So there in no guarantee that this
quantity will keep increasing as temperature drops, or even, more importantly that it will
reach the pion mass at some value of baryon chemical potential.
So we suggest to use it as an active parameter in the statistical model
which fits the hadronic populations emerging from heavy-ion collision experiments and 
determine its value.
More specifically, the pion multiplicity measured in an experiment, $N_{\pi,ex}$, 
is fitted by an equation of the form
\begin{equation} \label{eq:fit_expion}
n_{\pi,HG}\left(T,\mu_B,\mu_S \right) V_{HG} f_1\left(T,\mu_B,\mu_S \right)
= N_{\pi,ex}\;.
\end{equation}
The quantity $f_1$ corrects the pions for the feeds from the decays of all the 
hadronic resonances.
Using similar expressions for the other experimentally measured multiplicities, the fit
determines the
thermodynamic parameters, like the temperature $T$, the necessary chemical potentials,
e.g. of baryon number $\mu_B$, of strangeness $\mu_S$ and, also, the system volume $V_{HG}$.
From eq.~(\ref{eq:HGchnBF}) we have
\begin{equation} \label{eq:ch1_HGpion}
\tilde{n}_{1}\left(T,\mu_B,\mu_S ;\tilde{m}_1\right) = 
n_{1,HG}\left(T,\mu_B,\mu_S \right) =
\frac{n_{\pi,HG}\left(T,\mu_B,\mu_S \right)}{f_2 \left(T,\mu_B,\mu_S \right)}\;,
\end{equation}
where $f_2$ is the fraction of the HG density of family 1 which is due to 
pions\footnote{For point particles $f_2=R_1$, according to eq.~(\ref{eq:Ri}).}.
Last equation assumes that we are in the crossover region, since system volume
is kept unchanged between chiral and HG phase.
Combining eqs.~(\ref{eq:fit_expion}) and (\ref{eq:ch1_HGpion}), 
we can use the following relation to fit the pion
multiplicity:
\begin{equation} \label{eq:ch1_expion}
\tilde{n}_{1}\left(T,\mu_B,\mu_S ;\tilde{m}_1\right)
f_1\left(T,\mu_B,\mu_S \right) f_2\left(T,\mu_B,\mu_S \right) V_{HG}=
N_{\pi,ex}\;.
\end{equation}
These fits with the use of eq.~(\ref{eq:ch1_expion}) for the pion multiplicity have to be 
performed for
numerous experiments on the freeze-out curve, starting from zero baryon chemical
potential and moving to higher values.
An anticipated outcome would be to extract non-negative 
values of $\tilde{m}_1$ which gradually increase as we move from zero to
 higher values of baryon 
chemical potential, peeking
at the value of the HG pion mass (at the critical point). The approach to this higher value 
will serve as a signal of a gradual nearing to the QCD critical point.
At higher values of $\mu_B$ (1st order transition), $\tilde{m}_1$ should stabilise at 
$m_{\pi}$. Of course in this region an extra parameter $\tilde{V}$ should be included
in the fit, resulting to the presence of an extra term $v_{er}=V_{HG}/\tilde{V}$
(according to eq.~(\ref{eq:pc11o})) in the {\it lhs} side of eq.~(\ref{eq:ch1_expion}).

The equilibrium of all hadronic components of HG is consistent with a
prehadronic equilibrium of a higher temperature QCD phase.
If this phase is a phase containing quarks, which will produce all hadrons, then a global
chemical equilibrium is expected.
However, if the quarks evolve, during cooling down, firstly to hadron-like quark condensates
before final hadronisation, then all the particles within a family have a
common origin, but not the hadrons in different families.
So, if all the chiral particles which are the hadronic family origins, do not achieve
full chemical equilibrium this may result in families not in full chemical equilibrium
among each other.
If separate chemical equilibrium conditions are observed for different families,
this is a good support for the hadron-like quark condensate model.
The chiral pion mass undergoes a
substantial change compared to the chiral masses of the other hadronic families
(Figs.~\ref{fig:Tc=159.4}(ii), \ref{fig:Tc=159.5}(ii), \ref{fig:Tc=159.2}(ii) and
\ref{fig:Tc=164.1}(ii)
for all the hadron volume models considered in this work).
This behaviour may signify the possibility for separate chemical conditions for 
this chiral particle.
Indeed, there are several studies where all the hadronic components are not treated as
fully equilibrated. Especially interesting for our study is that an excess of the pion 
population is observed \cite{{ex_pi1},{ex_pi2},{ex_pi3},{ex_pi4},{ex_pi5},{ex_pi6}}. 
This is connected with the introduction of a partial
equilibrium fugacity, $\gamma_q$, as an additional parameter, which is determined to be $>1$
\cite{{ex_pi1},{ex_pi2},{ex_pi3},{ex_pi4}}.
This parameter affects the light unflavoured meson families, as well as, the baryons,
 causing their increase.
But, usually its introduction to the fitted parameters leads to the reduction of the fitted 
temperature for $\gamma_q>1$. The temperature reduction compensates the increase to the 
baryons, leaving as
main effect the pion increase. Also, in other studies a separate chemical potential for the 
pions has been introduced, $\mu_{\pi}>0$ \cite{{ex_pi6},{mu_pi}}.
The additional thermodynamic parameters are used to describe the different chemical
equilibrium conditions.
However, the key issue is whether to use on the HG side separate partial equilibrium
fugacities of quarks or families. 
The interesting thing would be to detect off-equilibrium phenomena, especially for the
pion family.
This can be easily traced if firstly a fit is performed with all the multiplicities and then
another fit with all except the one in question. If the second fit is of considerably
enhanced quality, then this is a sign that the multiplicity in question is off
chemical equilibrium \cite{{Gibbs},{ex_pi7},{ex_pi8}}.
If such phenomena are observed, especially for the pion multiplicity, we can
use the chiral mass to study them further.
In the current version of our model all hadronic components of HG were assumed to be fully 
chemically equilibrated and the chiral masses for all the
families were calculated to conserve the family multiplicities at the HG phase.
However, we can do the opposite.
We can leave as a free parameter, e.g. the chiral pion mass, $\tilde{m}_1$, use 
eq.~(\ref{eq:ch1_expion}) and fit the pion 
multiplicity, regardless of its equilibration or not.
Although eq.~(\ref{eq:ch1_expion}) can be used as it stands, all the HG densities which are 
treated as
being non-chemically equilibrated must carry an additional thermodynamic parameter.
To this end, and in a first approach, eq.~(\ref{eq:ch1_HGpion}) can be transformed to
\vspace{-0.cm}
\begin{equation} \label{eq:ch1_HGpion_mup1}
\tilde{n}_{1}\left(T,\mu_B,\mu_S ;\tilde{m}_1\right) = 
n_{1,HG}\left(T,\mu_B,\mu_S, \mu_{\pi} \right) \;,
\end{equation}
where $\mu_{\pi}$ is the additional chemical potential which is added here to the pion 
family.
In eq.~(\ref{eq:ch1_HGpion_mup1}) the particle density in the chiral phase is
adjusted only through the chiral mass $\tilde{m}_1$.
In a second approach, in the {\it lhs} of eq.~(\ref{eq:ch1_HGpion_mup1}) 
we can enter $\mu_{\pi}$ as an active parameter of the chiral density.
In any case, $\mu_{\pi}$ can be fixed by the experimental data.
Then, it is possible to calculate $\tilde{m}_1$.
In the first approach, if $\tilde{m}'_1$ corresponds to the full pion chemical equilibrium
which solves eq.~(\ref{eq:ch1_HGpion_mup1}) for $\mu_{\pi}=0$, then a
decrease of $\tilde{m}_1$ below $\tilde{m}'_1$ can be associated with a pion excess
and vice-versa. In the second approach the off-equilibrium effects are shown by $\mu_{\pi}$.
Such a deviation of the pion population would point to a different origin of the pion
family, such as the chiral pion condensate, giving support to our model.
Even so, the pion chiral mass is still expected to maximize at the pion mass, signalling
the approach to the C.P. 

The above tests are carried out along the freeze out curve.
In Fig.~3 of \cite{lat_Tc2} and Fig.~15 of \cite{lat_muB2} there are multiple 
freeze-out points at adjacent low values 
of baryon chemical potential and different temperatures. If this is not due to
errors, it could suggest that at low baryon densities the freeze-out curve may
develop to a zone. In a such a case, with the same method used to determine
the chiral pion mass along the freeze-out curve, we may determine the same quantity
at different temperatures at nearly the same values of baryon chemical potential.
Then the extracted values could be compared against the corresponding values of a fully
constructed model which is described in the first paragraph of this section.

Lastly, the variation of the chiral masses of the other hadronic families, as well as
partial equilibrium phenomena, can, also, be
investigated along the freeze-out curve. This would entail to construct eqs.~like
eq.~(\ref{eq:ch1_expion}) for other families. However, the most interesting chiral mass,
after $\tilde{m}_1$, would be $\tilde{m}_3$, that of the nucleon family. 
This, in all volume models we have considered, stays close to the maximum value of the
nucleon mass and in volume model (c) it is adjusted to reach this maximum value at the
critical point, as the chiral pion mass.

\section{Conclusions} \label{sec:conclu}

We have developed an effective 
description of the QCD state for temperatures near and above the chemical freeze-out curve,
which adopts the organisation of quarks into quasi-particles with the same quark
content as hadrons but with point-like behaviour, as far as their repulsive
interaction is concerned. For 
temperatures in the region below the chemical freeze-out curve we adopt the hadronic resonance 
gas with the Bose/Fermi statistics and thermodynamically consistent volume corrections. The 
chiral phase consists of a few quark condensates, each of which shares the same quark content 
with a group of hadronic resonances with different masses. Between the two phases the particle 
number conservation is imposed as a minimum requirement of a phase transition, a condition which 
leads to the variation of the chiral masses along the transition curve.

We find that the pion particles with their partner resonances play the leading role in 
transforming the order of the transition from
higher order to first order and in developing a critical point.
The second family which may affect this transition is the family of nucleons. Thus, the exact 
knowledge of the specific spectrum of 
the rest of the hadronic particles is of secondary importance.

The position of the critical point is sensitive to the value of the radii of hadrons. For this 
reason, we use the lattice calculations with three quark flavours for the pressure and entropy 
density below the critical temperature $T_c$ and fit it with the Hadron Gas equation of state in 
order to determine the hadronic volumes. We, also, link the lattice critical temperature $T_c$ 
with the freeze-out curve which we consider as our transition line. We consider four models for the 
hadron volumes, (a) which uses the same radius for all hadrons, 
constant for all temperatures, 
(b) which uses a common hadronic radius, which depends on 
temperature, (c) which uses 
different radius values for mesons and baryons, depending on 
temperature and baryon chemical 
potential, respectively and (d) which uses a common hadronic 
radius, which depends on temperature and baryon chemical 
potential. 
Our findings are summarised in Table \ref{tab:summary}
for all five hadron sets we have used in our calculations. 
We observe that the critical point is limited by a
maximum value of chemical potential which gradually diminishes
from model (a) to (d). This maximum value is $\sim$806 MeV for model (a), 
$\sim$553 MeV for model (b), $\sim$530 MeV for model (c) and 
$\sim$477 MeV for model (d). 
If the uncertainties in $T_c$ and hadronic volumes could be lifted, then the position of the critical point could be uniquely determined.

\vspace{0.cm}
\begin{table}[!b]

\begin{center}
\begin{tabular}{|c|c|c|c|c|c|c|c|c|} \hline
Volume       & Hadron & $T_{ch,0}=T_c$ & $r_{m,cr}$  & $r_{b,cr}$  & $T_{cr}$    & $\mu_{B,cr}$ & $\mu_{s,cr}$ \\
Model        & States &   (MeV)        &   (fm)      &    (fm)     &   (MeV)     &    (MeV)     &    (MeV)     \\
\hline
\multicolumn{8}{c}{General solutions} \\ 
\hline
(a):         & (vh)   &  154.2-160.1   & 0.137-0.300 & $=r_{m,cr}$ &  78.6-160.1 & 724.6-0.0    & 109.9-0.0    \\
$r_0$        & (*)    &  153.3-159.1   & 0.129-0.300 & $=r_{m,cr}$ &  76.3-159.1 & 733.2-0.0    & 110.6-0.0    \\
             & (**)   &  152.1-158.0   & 0.106-0.300 & $=r_{m,cr}$ &  69.2-158.0 & 758.2-0.0    & 116.6-0.0    \\
             & (lps)  &  151.6-157.4   & 0.098-0.299 & $=r_{m,cr}$ &  66.8-157.4 & 766.3-0.0    &         
118.4-0.0    \\
             & (lp)   &  150.2-155.5   & 0.069-0.297 & $=r_{m,cr}$ &  54.4-155.5 & 805.5-0.0    &         
128.3-0.0    \\
\hline
(b):         & (vh)   &  155.1-163.0   & 0.253-0.297 & $=r_{m,cr}$ & 125.1-163.0 & 493.5-0.0    &  68.9-0.0     \\    
$r_0(T)$     & (*)    &  154.7-161.3   & 0.246-0.298 & $=r_{m,cr}$ & 121.6-161.3 & 516.7-0.0    &  66.6-0.0     \\
             & (**)   &  153.1-160.5   & 0.234-0.297 & $=r_{m,cr}$ & 115.4-160.5 & 553.2-0.0    &  73.3-0.0     \\
             & (lps)  &  153.0-159.5   & 0.238-0.297 & $=r_{m,cr}$ & 117.5-159.5 & 540.0-0.0    &               67.4-0.0     \\
             & (lp)   &  152.2-157.0   & 0.241-0.295 & $=r_{m,cr}$ & 124.0-157.0 & 494.3-0.0    &               59.9-0.0     \\
\hline
(c):         & (vh)   &  156.0-162.4   & 0.260-0.291 & 0.320-0.343 & 135.9-160.1 & 409.6-135.1  &  59.5-19.1   \\
$r_m(T)$,    & (*)    &  154.4-162.1   & 0.229-0.293 & 0.296-0.355 & 122.1-162.1 & 513.0-0.0    &  68.2-0.0    \\
$r_b(\mu_B)$ & (**)   &  153.6-161.2   & 0.248-0.292 & 0.317-0.355 & 130.9-161.2 & 444.0-0.0    &  59.2-0.0    \\
             & (lps)  &  152.7-160.2   & 0.223-0.293 & 0.288-0.354 & 119.1-160.2 & 529.6-0.0    &  67.5-0.0    \\
\hline
(d):         & (vh)   &  159.7-167.4   & 0.269-0.292 & $=r_{m,cr}$ & 137.6-167.4 & 411.6-0.0    &  53.8-0.0    \\
$r(T,\mu_B)$ & (*)    &  158.0-165.4   & 0.266-0.293 & $=r_{m,cr}$ & 134.8-165.4 & 428.9-0.0    &  50.7-0.0    \\
             & (**)   &  157.1-164.6   & 0.260-0.293 & $=r_{m,cr}$ & 131.3-164.6 & 453.5-0.0    &  54.0-0.0    \\
             & (lps)  &  156.2-163.4   & 0.255-0.293 & $=r_{m,cr}$ & 127.9-163.4 & 476.7-0.0    &  52.2-0.0    \\
             & (lp)   &  153.8-160.7   & 0.258-0.291 & $=r_{m,cr}$ & 134.0-160.7 & 416.8-0.0    &   48.2-0.0    \\
\hline
\multicolumn{8}{c}{Solution which additionally satisfies $n_{qs,m}=n_{qs,b}$}\\ 
\hline
(a):         & (vh)   &  159.42        & 0.2890      & $=r_{m,cr}$ & 151.23      & 258.88       &  34.93       \\ 
$r_0$        & (*)    &  158.38        & 0.2893      & $=r_{m,cr}$ & 150.92      & 250.60       &  28.66       \\
             & (**)   &  157.28        & 0.2880      & $=r_{m,cr}$ & 149.55      & 258.58       &  30.44       \\
             & (lps)  &  156.68        & 0.2882      & $=r_{m,cr}$ & 149.35      & 254.04       &  24.84       \\
             & (lp)   &  155.02        & 0.2878      & $=r_{m,cr}$ & 150.64      & 202.65       &  20.81       \\
\hline
(b):         & (vh)   &  159.53        & 0.2890      & $=r_{m,cr}$ & 151.34      & 258.54       &  34.82       \\
$r_0(T)$     & (*)    &  158.55        & 0.2892      & $=r_{m,cr}$ & 151.10      & 249.99       &  28.46       \\
             & (**)   &  157.46        & 0.2880      & $=r_{m,cr}$ & 149.73      & 257.92       &  30.22       \\
             & (lps)  &  156.89        & 0.2881      & $=r_{m,cr}$ & 149.56      & 253.26       &   24.57       \\
             & (lp)   &  155.56        & 0.2875      & $=r_{m,cr}$ & 151.22      & 200.16       &   20.12       \\
\hline
(c):         & (vh)   &  159.18        & 0.2809      & 0.3374      & 149.86      & 276.23       &  40.23       \\
$r_m(T)$,    & (*)    &  158.24        & 0.2792      & 0.3466      & 149.45      & 271.57       &  34.91       \\
$r_b(\mu_B)$ & (**)   &  157.10        & 0.2779      & 0.3455      & 148.08      & 278.85       &  36.58       \\
             & (lps)  &  156.61        & 0.2783      & 0.3444      & 148.04      & 273.81       &  30.71       \\
\hline 
(d):         & (vh)   &  164.07        & 0.2869      & $=r_{m,cr}$ & 156.21      & 240.32       &  28.89       \\
$r(T,\mu_B)$ & (*)    &  162.51        & 0.2873      & $=r_{m,cr}$ & 155.32      & 234.16       &  23.45       \\
             & (**)   &  161.78        & 0.2862      & $=r_{m,cr}$ & 154.38      & 239.56       &  24.29       \\
             & (lps)  &  160.74        & 0.2865      & $=r_{m,cr}$ & 153.67      & 237.18       &  19.34       \\
             & (lp)   &  158.33        & 0.2856      & $=r_{m,cr}$ & 154.03      & 192.07       &  18.20       \\ 
\hline
\end{tabular}
\end{center}
\setcounter{table}{1}
\caption{\label{tab:summary} {\small Summary of the critical point's thermodynamic parameters for the four models of the 
hadronic volumes and for the five hadron sets considered in this paper.}}
\end{table}

We, also, calculate the densities of the quarks which are
contained in mesons and baryons at the different critical point
locations,
motivated by the gradual increase of baryon
density and decrease 
of meson density, as we move along the transition curve towards higher values of baryon chemical 
potential and find the point where the equality between the two densities occurs.
This condition is not so affected by the hadron radius, in contrast to
the condition of the maximum chiral pion mass and so it can serve in limiting further the
solutions of the critical point.
We find that for all four considered models and for the first four hadron sets 
the solution for the critical 
point location which additionally satisfies equality of the two densities is located at 
$\mu_{B,cr}\simeq$ (234-279) MeV and $T_{cr}\simeq$(148-156) MeV. Also, in this case, the hadronic volumes acquire realistic values. In models (a), (b) and (d) where a common volume is used, the hadronic radius is $r \simeq$ 0.29 fm. In model (c)
this radius is $r_m \simeq$ 0.28 fm for mesons and $r_b \simeq$ 0.34-0.35 fm for baryons.
These findings are in agreement with the evidence of power-law fluctuations compatible with
the QCD critical point for Si+Si collisions \cite{cf_158} with freeze-out at baryon chemical
potential of $\mu_B \simeq$260 MeV \cite{fr_muB=260}.

Another observation is that the hadron set we choose to use affects mainly the value
of $T_c$, which is found to decrease as we move from the (vh) to the (lp) hadron set. 
This is important for the compatibility of the HG model with the lattice QCD
calculations. However, in general, the extracted values of the mesonic radius, baryonchemical potential 
and temperature of the critical point are not so affected by the choice of the hadron set.
This can be seen from the almost parallel structure of the relevant curves in Figs.~\ref{fig:cpr0}(i)-(iii), \ref{fig:cpr0(T)}(i)-(iii),
\ref{fig:cprmrb}(i), (iii)-(iv) and \ref{fig:cpr_TmuB}(i)-(iii), as well as, from the recorded
values in Table \ref{tab:summary} for the critical point which satisfies the 
criterion of Section \ref{sec:crit}, at least for the first four hadron sets.

In this work, first order transition is explored trough the discontinuities in the densities
of the hadronic families, whereas in \cite{lat_s} it is explored through discontinuity 
in the entropy.
To carry a similar analysis in the context of our model, the chiral phase 
pressure is needed, according to
eq.~(\ref{eq:pg}), which would introduce additional parameters and uncertainties.
Therefore, the low $\mu_B$ solutions of the C.P. which are excluded in \cite{lat_s}
cannot be excluded in our model in its present form. 

The group of the chiral QCD states, considerably fewer than the abundance of the 
hadronic states 
existing at lower temperatures,
can be taken as starting point for the effective description of the EoS at higher 
temperatures, 
where quark deconfinement is
completed. We, also, propose the determination of the 
chiral pion mass 
as a signal of approaching to the QCD critical point.
Finally, as tests of the validity of our model, we suggest the tracing of partial chemical 
equilibrium phenomena connected with different hadronic families, especially
with the pion family.


\end{document}